%% file: desqcd.tex
\documentclass[12pt]{article}
\usepackage{pstricks,pst-node}
\usepackage{amsfonts}
\usepackage{sectsty}
\newcommand\ZZ{{\mathbb Z}}
\newcommand\PP{{\mathbb P}}
\newcommand\CC{{\mathbb C}}
\newcommand\NN{{\cal N}}

\newcommand\RR{{\mathbb R}}
\newcommand\sqcdv{$\rm SQCD_5$}
\newcommand\kcs{k_{\rm cs}}
\newcommand\kwz{k_{\rm wz}}
\newcommand\wzf{\Omega_{\rm WZ}}
\newcommand\csf{\Omega_{\rm CS}}
\newcommand\tr{\mathop\mathrm{tr}\nolimits}
\newcommand\Tr{\mathop\mathrm{Tr}\nolimits}

\renewcommand\Im{\mathop\mathrm{Im}\nolimits}
\renewcommand\Re{\mathop\mathrm{Re}\nolimits}
\newcommand\diag{\mathop\mathrm{diag}\nolimits}

\newcommand\eqdef{\buildrel {\rm def}\over=} 
\newcommand\aq{\smash{\widetilde Q}\vphantom{Q}}
\newcommand\link{\Omega}
\newcommand\fdc{T}
\newcommand\ql{L}

\newrgbcolor{palered}{1 0.7 0.7}
\newrgbcolor{paleblue}{0.7 0.7 1}
\newrgbcolor{darkcyan}{0 0.6 0.6}
\newrgbcolor{brown}{0.6 0.6 0}
\newrgbcolor{orange}{1 0.5 0}
\newcommand\be{\begin{equation}}
\newcommand\ee{\end{equation}}
\newcommand\bea{\begin{eqnarray}}
\newcommand\eea{\end{eqnarray}}

\newcommand\vev[1]{\left\langle #1\right\rangle}

\newcommand\coeff[2]{{\textstyle{#1\over #2}}}
\newcommand\half{\coeff{1}{2}}

\newcommand\eq[1]{eq.~(\ref{#1})}
\newcommand\eqalign[1]{%
	\vcenter{%
		\normalbaselines \advance\baselineskip 5pt
		\advance\lineskip 5pt \tabskip=0pt
		\halign{%
			&\hfil $\displaystyle{##{}}$&
			$\displaystyle{{}##}$\hfil\cr
			#1\crcr
			}%
		}%
	}
\newcommand\eqrange[2]{eqs.~(\ref{#1}--\reftail{#2})}
\newcommand\dotline{\par\hbox to \hsize{\dotfill}\par}

\newcommand\nextline{\unskip\nobreak\hfill\break}
\makeatletter
\def\befored@t#1.#2.#3;{#1}
\def\afterd@t#1.#2.#3;{#2}
\def\refhead#1{\edef\next{\ref{#1}}\expandafter\befored@t\next..;}
\def\reftail#1{\edef\next{\ref{#1}}\expandafter\afterd@t\next..;}
\def\lsim{\mathrel{\mathpalette\@versim<}}
\def\gsim{\mathrel{\mathpalette\@versim>}}
\def\@versim#1#2{\vcenter{\offinterlineskip
        \ialign{$\m@th#1\hfil##\hfil$\crcr#2\crcr\sim\crcr } }}
\newcommand\becomes[1]{\mathchoice{\becomes@\scriptstyle{#1}}
   {\becomes@\scriptstyle{#1}} {\becomes@\scriptscriptstyle{#1}}
   {\becomes@\scriptscriptstyle{#1}}}
\def\becomes@#1#2{\mathrel{\setbox0=\hbox{$\m@th #1{\,#2\,}$}%
        \mathop{\hbox to \wd0 {\rightarrowfill}}\limits_{#2}}}
\makeatother
\psset{unit=1mm,linecolor=black,linewidth=1pt}
\def\normalbaselines{%
	\normalbaselineskip=20pt plus 0.2pt minus 0.1pt
	\baselineskip=\normalbaselineskip
	\lineskip=2pt plus 0.1pt minus 0.1pt
	\lineskiplimit=2pt
	}
\def\fixformat{%
	\normalbaselines
	\abovedisplayskip=15pt plus 5pt minus 3pt
	\belowdisplayskip=\abovedisplayskip
	\parskip=6pt plus 2pt minus 1pt
	\skip\footins=20pt plus 10pt minus 5pt
	\predisplaypenalty=5000
	\postdisplaypenalty=500
	\interlinepenalty=50
	\interdisplaylinepenalty=10000
	\flushbottom
	}
\sectionfont{\huge\rm\blue\nohang\centering}
\subsectionfont{\Large\sc\nohang\raggedright}
\makeatletter
\def\@seccntformat#1{%
	\@ifundefined{#1@cntformat}%
	{\csname the#1\endcsname\quad}% default
	{\csname #1@cntformat\endcsname}% individual control
	}
\def\section@cntformat{\thesection.\enspace}
\@addtoreset{equation}{section}
\newif\iffntmark \fntmarktrue
\renewcommand\@makefntext[1]{\noindent
	\iffntmark\llap{\@thefnmark\enspace}\fi
	#1\unskip }
\newif\ift@c \t@cfalse
\def\br@@k{\relax\ift@c\else\unskip\break\fi}
\def\brk{\protect\br@@k}
\let\t@c=\tableofcontents
\renewcommand\tableofcontents{\t@ctrue\t@c\t@cfalse}
\makeatother
\renewcommand\theequation{\thesection.\arabic{equation}}

\begin{document}
\fixformat
\include{titlepage}

\setcounter{page}{2}
\thispagestyle{empty}
\tableofcontents
\include{chapter1} % Introduction
\include{chapter2} % Semiclassical (De) Construction
\include{chapter3} % On Chern--Simons Couplings and Extra 4D Flavors
\include{chapter4} % Quantum Deconstruction of the Gauge Couplings
\include{chapter5} % Quantum Baryonic Branches
\include{chapter6} % SU(2) Examples of Phase Structures and Flop Transitions
\include{chapter7} % Deconstruction / String Universality
\include{biblio}

\end{document}

%% file: titlepage.tex
%auto-ignore
%
% title page of the CRDQ paper
%
\begin{titlepage}

\rightline{\vbox{%
    \baselineskip=15pt \tabskip=0pt
    \halign{%
	#\unskip\hfil\cr
	UTTG--13--06\cr
	hep-th/yymmnnn\cr
	}%
    }}
\vskip 2pc plus 1fil minus 0.5pc

\centerline{\Large\bf Quantum Deconstruction of 5D SQCD$\strut^\star$}
\vskip 1pc minus 0.5pc
\centerline{\large Edoardo Di Napoli$\strut^{\rm (1)}$
	and Vadim S.\ Kaplunovsky$\strut^{\rm (2)}$}
\vskip 1pc minus 0.5pc
\tabskip=0pt
\setbox0=\vtop{\halign{%
    \small\it #\hfil\cr
    \llap{\rm (1)\enspace}\relax
    University of North Carolina\cr
    Dept.\ of Physics and Astronomy\cr
    CB \#3255, Phillips Hall\cr
    Chapel Hill, NC~27599--3255, USA\cr
    \ \tt edodin@physics.unc.edu\cr
    }}
\setbox1=\vtop{\halign{%
    \small\it #\hfil\cr
    \llap{\rm (2)\enspace}\relax
    University of Texas at Austin,\cr
	Theory Group, Physics Department,\cr
	1 University Station, C1608\cr
    Austin, TX~78712--1608, USA\cr
    \ \tt vadim@physics.utexas.edu\cr
    }}
\centerline{\box0 \hskip 4pc plus 1fill \box1}

\vskip 2pc plus 1fil minus 0.5pc

\centerline{ABSTRACT}
\smallskip
We deconstruct the fifth dimension of 5D~SCQD with general numbers
of colors and flavors and general 5D Chern--Simons level;
the latter is adjusted by adding extra quarks to the 4D quiver.
We use deconstruction as a non-stringy UV completion of
the quantum 5D theory;
to prove its usefulness, we
compute quantum corrections to the \sqcdv's prepotential.
We also explore the moduli/parameter space of the deconstructed
\sqcdv\ and show that for $|\kcs|<n_c-\half n_f$ it continues
to negative values of $1/g^2_{5d}$.
In many cases there are flop transitions connecting \sqcdv\ to
exotic 5D theories such as $E_0$, and we present several examples
of such transitions.
We compare deconstruction to brane-web engineering of the same \sqcdv\
and show that the phase diagram is the same in both cases;
indeed, the two UV completions are in the same universality class,
although they are not dual to each other.
Hence, the phase structure of an \sqcdv\ (and presumably
any other 5D gauge theory) is inherently five-dimensional and
does not depends on a UV completion.
\par
\vskip 2pc plus 1fil minus 0.5pc

\begingroup
    \catcode`\@=11
    \let\@thefnmark=\relax
    \@footnotetext{%
	\nobreak
	\par\noindent \hangafter=1 \hangindent=\parindent
	{\large $\star$}\enspace
	Article based on research supported by the US National Science
	Foundation (grant PHY--0455649) and by the US Department of
	Energy (grant DE--FG02--06ER41418)
	}
\endgroup

\end{titlepage}
\newpage
\pagenumbering{arabic}

%% file: chapter1.tex
%auto-ignore
%
% Chapter 1 of the DESQCD paper
%
%%%%%%%%%%%%%%%%%%%%%%%%%%%%%%%%%%%%%%%%%%%%%%%%%%%%%%%%%%%
\section{Introduction}
\noindent{\bf Motivation:}\nextline
Five-dimensional $\NN=1$ SUSY gauge theories appear to be well
understood.
Indeed, constraints due to 8 supersymmetries combined with
gauge invariance
are so powerful that many low-energy properties of the theory
--- such as geometry of its moduli space --- can be calculated
exactly~\cite{seiberg, MS}.
But all such calculations presume
that SUSY and gauge invariance persist
on the quantum level of the 5D theory.
In other words, all our knowledge assumes some kind of
a UV completion which keeps these symmetries manifest.

In 5D all interactive field theories are non-renormalizable,
so 4D--style perturbative UV cutoffs
such as DR or covariant higher-derivative terms
are of no use.
Instead, most research into 5D gauge theories embeds them
into string or M theory as a UV completion.
For example, one may (1) compactify~M theory on a singular
Calabi--Yau threefold~\cite{CCAF, FMS, IMS, FKM, GMS},
or (2) make a web of $(p,q)$ five-branes in type~IIB string
theory~\cite{AH, AHK, DHIK, LV, KR2},
or (3) put D4--brane probes of type~IIA string in a background
of D8 branes and O8 orientifold planes~\cite{seiberg, DKV, MS}.
In such completions, one may use the full power of string/M theory
to derive the {\it global} geometry of the 5D moduli space,
including `flop' transition to different 5D phases,
sometimes involving strongly-coupled sectors with non-trivial
IR fixed points~\cite{PhasesW}.
But unfortunately, in string context it is hard to tell whether
all these phases are made of the same QFT-level degrees of freedom,
or perhaps they follow from different sectors of the string theory.
In other words, we don't know if the whole phase diagram is an
inherent property of the 5D gauge theory (regardless of a UV regulator),
or perhaps some phases are artefacts of embedding into string theory.

To resolve this issue, we need to compare phase structures of
different UV completions of the same 5D theory.
Since all `stringy' completions are dual to each other as
string/M theories, there is no use in comparing them to each
other.
Thus we need a non-stringy regulator such as lattice.
But the Euclidean 5D lattice breaks SUSY; also, it's hard
to latticize the Chern-Simons interactions of the gauge fields.
Instead, we shall use a lattice / continuum hybrid known as
{\it\blue dimensional deconstruction} \cite{ACG, HPW, ACG1, CTW, csaki}:
4 dimensions out of five remain continuous while the fifth dimension
becomes discrete.

In an earlier article~\cite{IK1}, we have used deconstruction
as a UV completion
of 5D SYM theories with maximal Chern--Simons levels $\kcs=n_c$.
In this article we extend this method to \sqcdv\ with general numbers
of colors and flavors and all allowed Chern--Simons levels.
Our main results are as follows:
(1) We develop the quantum aspects of deconstruction technology.
In particular, we show how to convert
the exact non-perturbative data (which obtain
at the 4D level of deconstructed theories) into the 5D moduli dependence
of gauge couplings, and hence into 5D phase structures.
(2) We prove universality of \sqcdv\ phase diagrams: for any choice of
$n_c$, $n_f$, and $\kcs$, the deconstructed \sqcdv\ and the string
embedding of the same 5D theory are in the same universality class and have
identical phase diagrams.
This strongly suggests that other UV regulators are also in the same
universality class and hence the phase diagrams are inherent properties of
the 5D theories.

\smallskip
\noindent{\bf Overview of 5D SQCD:}\nextline
Now that we have made out intentions clear, let us briefly introduce
two subjects that may be unfamiliar to some readers, namely
the SQCD in 5D, and the dimensional deconstruction.
We start with the the basic features of
5D gauge theories with $\NN=1$ SUSY (which in five dimensions
means 8 rather than 4 supercharges)~\cite{MS}.
First of all, there are two kinds of supermultiplets, vector and hyper:
a vector multiplet contains a gauge field $A_\mu$, a Dirac fermion
(4 complex components), and a real scalar; a hyper multiplet contains
a Dirac fermion and two complex (or 4 real) scalars.
In \sqcdv, $n_c^2-1$ vectors form an adjoint representation of $SU(n_c)$
gauge group while $n_f\times n_c$ hypers form $n_f$ fundamental
representations $\bf n_c$.
Note that there are no separate quark and antiquark multiplets; instead,
a single $\bf n_c$ of hypermultiplets contains both the quark and the
antiquark (as well as two squarks and two  antisquarks).
All Yukawa and scalar couplings of a 5D $\NN=1$ theory are related by
SUSY to the gauge coupling $g_5$; unlike in 4D, there is no independent
superpotential.
Instead, in 5D there are Chern--Simons interactions of gauge fields
and their superpartners,
\be
\eqalign{
{\cal L}_{\rm CS}\ =\ {}&
{i\,\kcs\over 24\pi^2}\,\tr\left(
        A\land F\land F\,-\,\coeff{i}{2}\,A\land A\land A\land F\,
        -\coeff{1}{10}\,A\land A\land A\land A\land A \right)\cr
&\ +{\kcs\over 8\pi^2}\,\tr\left(\Phi F_{\mu\nu}F^{\mu\nu}\right)\
        +\ \mbox{fermionic terms}\cr
}\label{CSterms}
\ee
where $\Phi$ is the adjoint scalar field.
To assure gauge invariance of the path integral, the coefficient $\kcs$
(also known as the Chern--Simons {\sl level}) is quantized:
in SYM$_5$ or \sqcdv\ with an even number of flavors, $\kcs$ must be integer;
\sqcdv\ with an odd $n_f$ needs a half-integer $\kcs$~\cite{seiberg}.

The vacuum states of 5D $\NN=1$ theories form continuous families
parametrized by two kinds of moduli: the Coulomb moduli control
the adjoint scalar VEV $\vev\Phi$ while the Higgs moduli control
the squark VEVs.
Because these scalars belong to different kinds of supermultiplets
(vector versus hyper), the two kinds of moduli do not intermix.
That is, the {\it local} geometry of the moduli space factorizes
into separate Coulomb and Higgs subspaces.
The {\it global} geometry is more complicated because non-zero squark VEVs
require some tuning of the Coulomb moduli and quark masses $m_f$.
Consequently, the overall moduli space has several branches, each with
its own Higgs and Coulomb subspaces: the Coulomb branch with $n_c-1$
independent Coulomb moduli but no Higgs moduli at all; the mesonic branches
where some Coulomb moduli are fixed to allow some squark VEVS and hence
Higgs moduli; and the baryonic branches where all Coulomb moduli are fixed.
But in all branches, the Coulomb and the Higgs moduli are completely
separated by SUSY.

Provided the UV completion of the quantum theory is manifestly supersymmetric,
the separation between the Higgs and the Coulomb moduli remains exact.
Also, there are no quantum corrections --- perturbative or non-perturbative
--- to the classical geometry of the Higgs moduli space.
This follows from promoting the gauge coupling to a background field~\cite{seiberg1}:
to do it in a supersymmetric manner,
$1/g_5^2$ must be the scalar member of a vector multiplet and therefore
cannot affect the Higgs space geometry~\cite{APS}.
As to the Coulomb moduli space geometry, the quantum corrections stop
at the one-loop level.
In terms of the prepotential,
\be
{\cal F}\ =\ {\cal F}_{\rm tree}\ +\ {\cal F}_{\rm 1-loop}\,,
\quad\mbox{exactly},
\label{Fexact}
\ee
and there are no further perturbative or non-perturbative corrections~\cite{seiberg}.
For \sqcdv,
\bea
{\cal F}_{\rm tree}(\phi_1\,\ldots,\phi_{n_c}) &=&
{1\over 2g_5^2}\sum_{i=1}^{n_c}\phi_i^2\
        +\ {\kcs\over 48\pi^2}\sum_{i=1}^{n_c}\phi^3,
        \label{Ftree}\\
\mbox{and}\quad{\cal F}_{\rm 1-loop}(\phi_1\,\ldots,\phi_{n_c}) &=&
{1\over 96\pi^2}\sum_{i,j=1}^{n_c}\left|\phi_i-\phi_j\right|^3\
        -\ {1\over 96\pi^2}\sum_{i=1}^{n_c}\sum_{f=1}^{n_f}\left|\phi_i-m_f\right|^3,\qquad
        \label{F1loop}
\eea
where the Coulomb moduli $\phi_1\,\ldots,\phi_{n_c}$ are eigenvalues
of the adjoint scalar's VEV $\vev\Phi$.
For generic values of these moduli the gauge group $SU(n_c)$ is broken to
its Cartan subgroup $U(1)^{n_c-1}$, and the gauge coupling matrix for the
abelian fields follows from the prepotential according to
\be
\left[\frac{1}{g_5^2}\right]_{ij}\
=\ \left({\rm moduli \atop metric}\right)_{ij}\
=\ \frac{\partial^2{\cal F}}{\partial\phi_i\,\partial\phi_j}\,.
\label{GFmetric}
\ee
This matrix must be positive-definite for all values of the moduli $\phi_i$,
which restricts the discrete parameters of \sqcdv\ to~\cite{MS}
\be
\eqalign{
|\kcs|\ +\ {n_f\over2}\ &\leq n_c\qquad\mbox{for}\ n_c\geq3,\cr
n_f\ &\leq 7\qquad\mbox{for}\ n_c=2.\cr
}\label{Restriction}
\ee
The abelian gauge fields have Chern--Simons interactions with each other:
\be
{\cal L}\ \supset\ \sum_{ijk}\frac{K_{ijk}}{48\pi^2}\,A_i\land F_j\land F_k
\qquad\mbox{where}\quad K_{ijk}\
=\ \frac{\partial^3{\cal F}}{\partial\phi_i\,\partial\phi_j\,\partial\phi_k}\,.
\ee
Gauge invariance requires the coefficients $K_{ijk}$ to be integer, and this
restricts the prepotential so much that there are no quantum corrections
beyond the one-loop level.

Finally, a point of terminology.
In this paper, we distinguish between the non-dynamical
{\it parameters} of the 5D theory such as $1/g^2_5$ and quark masses
and the dynamical {\it moduli} of its vacua such as $\phi_i$ or
the Higgs moduli of squark VEVs.
However, the phase diagram of the theory involves both the parameters and
the moduli; for example, in \S6.1 we shall see that an $SU(2)$ SYM theory
has a phase transition at $(8\pi^2/g_5^2)=-\phi$ rather
than at $(8\pi^2/g_5^2)=0$.
Consequently, when appropriate we shall put the parameters and the moduli
into a combined parameter/moduli space.

\smallskip
\noindent{\bf Overview of Deconstruction:}\nextline
And now we turn to the dimensional deconstruction.
Most generally, the deconstruction relates simple
gauge theories in spaces of higher dimension to more complicated theories
in fewer dimensions of space:  The extra dimensions of space are
`deconstructed' into quiver diagrams of the `theory space'~\cite{ACG,
  HPW}. 
In simple cases, deconstruction is a three-step procedure:
First, one discretizes the extra dimension(s) --- say, the $x^4$ space
coordinate of a $4+1$ dimensional theory --- 
into a lattice of small but finite spacing~$a$.
On this lattice, the $A_{0,1,2,3}$ components of the gauge field reside
on lattice nodes, while the $A_4$ component is realized via unitary
matrices $U_\ell=\mathop{\mbox{path-ordered~exp}}
\left(i\int_{a\ell}^{a(\ell+1)}\!A_4\,dx^4\right)$ residing on links.
Second, one reinterprets the lattice as a quiver diagram describing
a complicated 4D field theory with a large number of gauge group
factors (one per lattice site) with equal couplings $g_4^{(\ell)}\equiv
g_4=g_5/\sqrt{a}$.
The link variables $U_\ell$ become 4D non-linear sigma models $\link_\ell$
with $F_\pi=1/(ag_4)$ and transforming in bi-fundamental representations
$(\overline{\square}_\ell,\square_{\ell+1})$ of the gauge group
$\prod_\ell SU(n)_{\,\ell}$.
Finally, one adds degrees of freedom to make the theory renormalizable
in 4D; this includes promoting non-linear sigma models $\link_\ell$
to linear sigma models, or perhaps realizing them as techni-pions of some
kind of technicolor (with a separate technicolor group for each $\link_\ell$).
The resulting 4D gauge theory can often be summarized by a quiver diagram
--- hence the name {\it\blue quiver theory} --- for example
\be
\psset{unit=1cm,linewidth=1pt,arrowscale=1.5}
\def\segment#1{%
        \rput{#1}(0,0){%
                \pscircle*(2,0){0.1}
                \pscircle(1.732,-1){0.1}
                \psarc{<-}(0,0){2}{333}{357}
                \psarc{<-}(0,0){2}{303}{327}
                }
        }
\begin{pspicture}[0.5](-2,-2)(+10,+2)
\segment{90}
\segment{30}
\segment{330}
\segment{270}
\segment{210}
\pscircle*(-1.732,+1){0.1}
\psarc[linestyle=dotted]{<-}(0,0){2}{93}{147}
\pscircle*(4,+1){0.1}
\rput[l](5,+1){color group factor}
\pscircle(4,0){0.1}
\rput[l](5,0){technicolor group factor}
\psline{->}(3.5,-1)(4.5,-1)
\rput[l](5,-1){techniquark or antiquark}
\end{pspicture}
\label{YMquiver}
\ee
for the 5D YM theory deconstructed in \cite{ACG}.
In order to have a finite number of 4D fields, the quiver should have
a finite size $\ql$ --- in 5D terms, this means that the deconstructed dimension
$x^4$ is also compactified on a large circle of length $2\pi R=\ql a$ ---
but eventually one may take the $\ql\to\infty$ limit  and  recover
the uncompactified 5D physics.
In this limit, the lattice spacing~$a$ remains finite and serves as the
UV regulator:
For energies $E\ll(1/a)$ the physics is 5D but for $E\gsim(1/a)$ it becomes 4D.

Dimensional deconstruction of supersymmetric theories breaks half
of supercharges for every discretized dimension~\cite{csaki, csaki2}; in particular,
for $\NN=1$ theories in 5D, deconstructing one dimension breaks
4 out of 8 supercharges.
Fortunately, the 4 supercharges which remain unbroken act as custodial
symmetries of the complete 5D SUSY in the low-energy effective theory.
Indeed, for $E\ll(1/a)$ the fifth dimension is effectively continuous,
and if the effective 5D theory happen to have $SO(4,1)$ Lorentz symmetry
as well as any SUSY at all, then it must have the complete 5D SUSY algebra
with all 8 supercharges.
This is not as easy as it sounds because the $SO(4,1)$ Lorentz symmetry
of the continuum limit is far from automatic in partially latticized
theories.
Instead, one often needs to fine-tune the lattice action ($i.\,e.$, the
4D Lagrangian of the quiver theory) to make sure that all light particle
species have the same light speed in the discretized direction.
However, once this is achieved, the recovery of all 8 supercharges
in the 5D continuum limit is automatic~\cite{csaki}.

From the 4D point of view, the quiver theory has 4 exact supercharges,
which means that some properties of the theory are holomorphic
and can be calculated exactly, including all the non-perturbative effects.
Such properties include the phase structure of the theory, and also
moduli dependence of gauge couplings for the massless vector fields.
The basic idea of {\it quantum deconstruction} is to interpret these data
in 5D terms; this allows {\sl practical} use of the
dimensional deconstruction as a UV completion of quantum 5D theories.

\smallskip
\noindent{\bf Outline:}\nextline
This article is organized as follow.
In the next section (\S2) we deconstruct \sqcdv\ with quarks at the
semiclassical level of analysis.
Instead of following the 3-step procedure outlined above ---
discretize, re-interpret, and make renormalizable --- we work
in reverse.
That is, we start with a quiver diagram, build a corresponding
4D, $\NN=1$, $[SU(n_c)]^\ql$ gauge theory, and then show that
it indeed deconstructs the 5D~SQCD.
Specifically, we verify that the classical vacua of the 4D
quiver theory correspond  to the classical vacua of the 5D SQCD,
and for each vacuum, the spectrum of light 4D  particles agrees
with the Kaluza--Klein reduction
(on a latticized circle of length $2\pi R=\ql a$) 
of the 5D gluons, quarks and their superpartners.

In \S3 we deconstruct the Chern--Simons interactions
of the 5D gauge fields.
We show how to control the Chern--Simons level $\kcs$ by adding
extra quark flavors to the 4D theory.
The extra flavors do not have any light modes and thus do not
deconstruct any 5D particles; instead, they decouple at the
5D threshold $E=(1/a)$.
But integrating out those quarks leaves behind quantum
corrections to the low-energy Lagrangian; in 5D terms,
such corrections lower $\kcs$ by the number $\Delta F$
of extra flavors.
Altogether, we end up with $F=n_f+\Delta F$ 4D flavors
where $n_f$ is the number of 5D flavors, and $\Delta F$ is used
to set
\be
\kcs\ =\ n_c\ -\ {n_f\over2}\ -\Delta F .
\label{UltimateCS}
\ee
Interestingly, the values of $\kcs$ which may be deconstructed
in this way are precisely the values allowed by \eq{Restriction}.

In \S4 we deconstruct the 5D gauge coupling and their moduli dependence.
We start with the Coulomb phase of the 4D quiver where the
massless gauge bosons belong to $[U(1)]^{n_c-1}\subset \mathop{\rm diag}
\left[SU(n_c)^\ql\right]$~\cite{paper1}; the couplings of these abelian bosons are
encoded in a hyperelliptic spectral curve,
which was computed in~\cite{KSdN}.\footnote{%
        In \cite{KSdN}, we studied the $[SU(n_c)]^\ql$ quiver theory
        from the 4D point of view.
        We analyzed the quiver's chiral ring, which summarizes
        its exactly calculable holomorphic data.
        In this paper, we use these data to obtain the 5D properties
        of the deconstructed \sqcdv.
        }
This curve has moduli and parameters, and our first task is to map
them to 5D moduli $\phi_i$ and parameters $m_f$ and $h=(8\pi^2/g_5^2)$.
Then, we take the decompactification limit $\ql\to\infty$ while
all the moduli and parameters remain fixed.
In this limit, the spectral curve simplifies (as long as $g_5$ is
weak enough), which helps us to evaluate the abelian gauge coupling
matrix $\tau_{ij}$.
We find that $\Im\tau_{ij}=\ql a\times{}$a finite matrix
$[4 \pi g_5^{-2}(\phi)]_{ij}$, which we interpret as the deconstructed
5D abelian coupling matrix; this corresponds to 4D massless fields
being zero modes of the 5D fields compactified on a circle of length
$2\pi R=\ql a$.

Although the 4D quiver theory has only 4 supercharges, the deconstructed
5D gauge couplings are consistent with a prepotential, which confirms
SUSY enhancement to 8 supercharges in the continuum 5D limit.
Moreover, the prepotential turns out exactly as in \eqrange{Fexact}{F1loop},
which shows that dimensional deconstruction indeed works at the quantum level:
the 5D loop corrections follow from loop and instantonic corrections
in 4D (see also ~\cite{csaki2}).

The deconstructed tree-level 5D coupling $h=(8\pi^2/g_5^2)$ is also
affected by the 4D quantum corrections.
We find that the allowed range of $h$ depends on the Chern--Simons level
of the deconstructed \sqcdv:  For $\kcs=\pm(n_c-\half n_f)$, $h$ runs from
$+\infty$ down to a finite lower limit $h_{\rm min}=\pm\half\sum_f m_f$;
but for other values of $\kcs$ there is no lower limit and $h$ can take
any value between $+\infty$ and $-\infty$.
The negative values of $h$ do not make sense in terms of ordinary
\sqcdv; instead, they corresponds to exotic strongly-coupled phases of
the 5D theory~\cite{MS, AH}.

In \S5, we discuss quantum corrections to  baryonic Higgs branches of
the deconstructed \sqcdv.
(The corrections are to the Coulomb moduli and parameters of such branches
rather than to the Higgs moduli.)
We also find that 5D theories with $|\kcs|\leq(n_f/2)$ have
exotic Higgs branches at strong  coupling: the $h$ parameter must
be fixed at $h=h_b$ where $h_b=O(m_f)\ll(1/a)$.
In particular, SYM theories ($n_f=0$) with $\kcs=0$ have exotic Higgs
branches at $h=0$.
The physical nature of the exotic Higgs branches is unclear
from the 5D point of view,
but in 4D they are simply baryonic branches which involve some of the
$\Delta F$ extra quark flavors.
At weak coupling, these quarks are heavy (mass${}=O(1/a)$) and decouple
from the 5D physics, but in the strongly coupled quiver they develop
zero modes, hence a baryonic branch for $h=h_b\ll(1/a)$.
String/M implementations of 5D SYM with $|\kcs|\leq(n_f/2)$
have similar exotic Higgs branches~\cite{IMS} at fixed $h=h_b(m_f)$, and alas
their physical nature is just as unclear from the 5D QFT point of view
as in dimensional deconstruction.

In \S6 we present four examples of deconstructed \sqcdv\ theories
and study their $h<0$ phases.
For $h<0$, the 4D theory is strongly coupled at the $\rm 4D\to 5D$
threshold $E=(1/a)$, but thanks to unbroken $\NN=1$ SUSY, the
spectral curve of the quiver is exactly calculable despite the
strong coupling.
This allows us to deconstruct the $h<0$ regime of the 5D theory
just as easily as the weakly-coupled $h>0$ regime.
Sometimes, the two regimes are separated by a phase transition:
Although in 4D there is only one Coulomb phase because the
spectral curve is holomorphic, the decompactification limit
$\ql\to\infty$ of the spectral curve may be
different for $h<0$ than for $h>0$, and that
leads to different phases in 5D.

For simplicity, all our examples have $n_c=2$.
The first example (\S6.1) has $n_f=0$ and $\Delta F=1$
while the second example (\S6.2) has $n_f=0$ and $\Delta F=2$.
For $h>0$, they deconstruct SYM theories with different
5D vacuum angles ($\theta=\pi$ for $\Delta F=1$
and $\theta=0$ for $\Delta F=2$).
But the $h<0$ regimes of the two models are very different:
the $\Delta F=1$ model has two distinct Coulomb phases
--- the SYM phase and the $E_0$ phase --- separated by a flop
transition, while the $\Delta F=2$ model has only one Coulomb phase,
but it also has an exotic Higgs branch.

In \S6.3 we present two more $SU(2)$ models, this time with $n_f=2$
and $\Delta F=0$ or 1.
For simplicity, we restrict our analysis to equal quark masses
(modulo sign) for the two 5D flavors.
Nevertheless, we find several distinct phases, both Coulomb and Higgs.

All 4 models are presented in much detail, which makes for a rather
looong section~\S6.
But the main result can be stated in once sentence:
{\it in all examples, the deconstructed \sqcdv\ has exactly the same
phase diagram as the string-theoretical UV completion of the same
5D theory.}
The readers who are not interested in technical details may skim
this section and focus only on the phase
diagrams themselves: they appear on pages \pageref{D0t:Diagram3},
\pageref{D0:Diagram2}, \pageref{D2:Diagram2}--\pageref{D2:Diagram4},
\pageref{D2:Diagram6}, \pageref{D2t:CoulombPhases},
and \pageref{D2t:HiggsPhases}.

Finally, in \S7 we show that for all $n_c$, $n_f$, and $\kcs$,
the deconstructed \sqcdv\ and the string-theoretical implementation
of the same 5D theory via a brane web are {\it always} in the same
universality class.
In particular, they always have similar phase diagrams and similar
prepotentials ${\cal F}(\phi_1,\ldots,\phi_{n_c};h;m_1,\ldots,m_{n_f})$.
However, the two UV completions are {\it not dual} to each other
and become dissimilar outside of the zero-energy limit.

Instead, the 5D universality between deconstruction and brane webs
is similar to the 4D universality between SQCD and
MQCD~\cite{witten97, HSZ, BIKSY}.
In fact, our proof is based on the 4D universality:
We start with deconstructed \sqcdv, treat it as a 4D $[SU(n_c)]^\ql$
quiver theory, and take its M-theory counterpart ---
the M5 brane spanning the 4D Minkowski space times the quiver's
spectral curve.
This M theory is not dual to the deconstructed \sqcdv, but it's
in the same universality class.

And then we show that the $\ql a\to\infty$ limit of the M5 brane
is dual to a type~IIB $(p,q)$ 5--brane web,
and moreover this web implements the very \sqcdv\
we have started from.
And since duality implies universality (but not the other way around),
it follows that the brane web is in the same universality class as
the deconstructed theory; but they are not dual to each other.

To summarise, in this paper we show how to use dimensional deconstruction
as a UV completion of a 5D SUSY gauge theory such as \sqcdv.
We show how to extract 5D quantum effects such as loop corrections to
the prepotential from the 4D loop and instantonic effects --- which can
be calculated exactly thanks to the unbroken $\NN=1$ SUSY in 4D.
We show how to deconstruct the 5D phase diagrams, including the
non-classical $h<0$ phases.
And we show that at the end of the day, the dimensional deconstruction
is in the same universality class as string-theoretical UV completions
of the same 5D theory.

%%%%%%%%%%%%%%%%%%%%%%%%%%%%%%%%%%%%%%%%%%%%%%%%%%%%%%%%%%%
 

%% file: chapter2.tex
%auto-ignore
%
% Chapter 2 of the DESQCD paper
%
%%%%%%%%%%%%%%%%%%%%%%%%%%%%%%%%%%%%%%%%%%%%%%%%%%%%%%%%%%%%%%%%%%%%%
\section{Semiclassical (De) Construction}
In this section, we perform a {\it reverse deconstruction}
--- at the semiclassical level of analysis ---
of \sqcdv\ with arbitrary numbers of colors and flavors.
That is, we start with a quiver diagram, build a 4D $\NN=1$
gauge theory, and then show that it indeed deconstructs the 5D~SQCD.
Specifically, we verify that the classical vacua of the 4D quiver
theory correspond to the classical vacua of the 5D SQCD, and for each vacuum,
the spectrum of light 4D particles agrees with the Kaluza--Klein reduction
(on a latticized circle of length $2\pi R=\ql a$) of the 5D gluons, quarks
and their superpartners.

We start with quiver diagrams of general form
\be
\psset{unit=4mm,linecolor=black}
\def\site{%
    \pscircle*[linecolor=green]{1}\relax
    \rput{*0}(0,0){\large $n_c$}\relax
    \psline{->}(-1.8,+0.45)(-0.9,+0.45)\relax
    \psline{->}(-1.8,-0.45)(-0.9,-0.45)\relax
    \psline{->}(-1.8,-0.15)(-0.99,-0.15)\relax
    \psline{->}(-1.8,+0.15)(-0.99,+0.15)\relax
    \psline{<-}(+1.8,+0.45)(+0.9,+0.45)\relax
    \psline{<-}(+1.8,-0.45)(+0.9,-0.45)\relax
    \psline{<-}(+1.8,-0.15)(+0.99,-0.15)\relax
    \psline{<-}(+1.8,+0.15)(+0.99,+0.15)\relax
    \rput(-2,0){\Large\{}\relax
    \rput(+2,0){\Large\}}\relax
    \rput{*0}(-2.9,0){$n^{}_f$}\relax
    \rput{*0}(+2.9,0){$n^{}_f$}\relax
    }
\begin{pspicture}[](-14,-14)(+14,+14)
    \psset{linewidth=1pt,linecolor=red}
    \rput{0}(+10,0){\site}
    \rput{45}(+7.07,+7.07){\site}
    \rput{315}(+7.07,-7.07){\site}
    \rput{90}(0,+10){\site}
    \rput{270}(0,-10){\site}
%    \rput{135}(-7.07,+7.07){\site}
    \rput{225}(-7.07,-7.07){\site}
    \rput{180}(-10,0){\site}
    \psset{linewidth=1.5pt,linecolor=blue}
    \psarc{<-}{10}{50.5}{84.5}
    \psarc{<-}{10}{5.5}{39.5}
    \psarc{<-}{10}{320.5}{354.5}
    \psarc{<-}{10}{275.5}{309.5}
    \psarc{<-}{10}{230.5}{264.5}
    \psarc{<-}{10}{185.5}{219.5}
    \psarc[linestyle=dotted]{<-}{10}{95.5}{174.5}
\end{pspicture}
\label{quiver}
\ee
Physically, each green circle of this diagram corresponds to
a simple $SU(n_c)_\ell$ factor of the net 4D gauge group
\be
G_{\rm 4D}\ =\  \prod_{\ell=1}^\ql  SU(n_c)^{}_\ell
\label{G4D}
\ee
while the red and blue arrows denote the chiral superfields:
\be
\psset{unit=0.6mm,linecolor=red,linewidth=1pt}
\eqalign{
\vcenter{\hbox to 9mm{%
        \psline{->}(0,6)(10,6)
        \psline{->}(0,4)(10,4)
        \psline{->}(0,2)(10,2)
        \psline{->}(0,0)(10,0)
        \hfil\Large\rm\}%
        }}\,
    \mbox{quarks}\ Q_\ell^f\ &
=\ ( {\bf\square}_{\,\ell} ),\quad f=1,2,\ldots,n_f^{},\cr
\vcenter{\hbox to 9mm{%
        \psline{<-}(0,6)(10,6)
        \psline{<-}(0,4)(10,4)
        \psline{<-}(0,2)(10,2)
        \psline{<-}(0,0)(10,0)
        \hfil\Large\rm\}%
        }}\,
    \mbox{antiquarks}\ \aq^f_\ell\ &
=\ ( \overline{\bf\square}_{\,\ell} ),\quad f=1,2,\ldots,n_f^{},\cr
\vcenter{\hbox to 25mm{%
        \psline[linecolor=blue,linewidth=1.5pt]{->}(0,0)(40,0)
        }}\,
    \mbox{bi-fundamental~link~fields}\ \link^{}_\ell\ &
=\ (\square_{\,\ell+1}, \overline\square_{\,\ell}),\cr
}\label{Fields}
\ee
where $\ell=1,2,\ldots,\ql $ is understood modulo
quiver size~$\ql $.
{}From the 4D point of view, $\ql $ is a fixed parameter of the  theory,
but for the deconstruction purposes we must later take the $\ql \to\infty$
limit in order to recover the un-compactified 5D physics.

Similar to many other deconstructed theories, the quiver~(\ref{quiver})
can be obtained by orbifolding a 4D theory with a much larger but simple gauge
group and higher SUSY,
namely $\NN=2$ SQCD with $\ql \times n_c$ colors (but only $n^{}_f$
flavors)~\cite{DM, AHCKKM}:
A $\ZZ_\ql $ twist removes the extra supercharges and reduces the gauge
symmetry from $SU(\ql \times n_c)$ down to
\be
S([U(n_c)]^\ql )\ =\ [SU(n_c)]^\ql \times [U(1)]^{\ql -1}.
\label{orbigroup}
\ee
However, the abelian photons of the orbifold theory suffer from
triangular anomalies and therefore must be removed from the
effective low-energy theory.
In string orbifolds such removal is usually accomplished via the
Green--Schwarz terms \cite{GP,IRU},
but at the field theory level we simply discard the abelian factors
of the orbifolded symmetry~(\ref{orbigroup}) and interpret the nodes
(green circles) of the quiver diagram~(\ref{quiver}) as purely non-abelian
$SU(n_c)_\ell^{}$ factors.

To complete the 4D quiver theory we must define its tree-level couplings.
The orbifolding procedure gives us two types of couplings inherited from the
`original' $\NN=2$ SQCD, namely the same gauge coupling $g_\ell\equiv g$
for all the $SU(n_c)_\ell^{}$ factors of the quiver,
and the Yukawa coupling $\gamma$ of the ``hopping'' superpotential
\be
W_{\rm hop}\
=\ \gamma\sum_{\ell=1}^\ql \sum_{f=1}^{n^{}_f}\Bigl(
        \aq^f_{\ell+1} \link^{}_\ell Q^f_{\ell}\
        -\ \mu^{}_f\, \aq^f_\ell Q^f_\ell
        \Bigr)
\label{Whop}
\ee
which makes the quark fields propagate in the discretized $x^4$ direction.
Classically $\gamma=g$ because of $\NN=2$ SUSY of the ``mother theory'';
in lattice terms, this equality assures that the quarks and the gluons have
equal light speeds.
Besides the couplings, we also have quark masses $\mu_f^{}$.
Formally, we may derive them from the orbifolded 5D quark masses,
but we shall see momentarily that the relation between the 4D and the 5D
quark masses of the deconstructed theory is more complicated.

Finally, to make the deconstruction work, we need
the O'Raifeartaigh superpotential
\be
W_\Sigma\ =\ \beta\sum_{\ell=1}^\ql  s_\ell^{}\times
\Bigl(\det(\link_\ell^{})\,-\,v^{n_c}\Bigr)
\label{Wsigma}
\ee
where $s_\ell^{}$ are singlet fields (one for each $\ell=1,2,\ldots,\ql $)
not shown on the quiver diagram~(\ref{quiver}).
These singlets and the O'Raifeartaigh terms~(\ref{Wsigma}) do not
follow from the orbifolding:
We simply add them by hand at the same time as we remove
the abelian factors from the orbifolded gauge group~(\ref{orbigroup}).
The purpose of this modification is to turn each bi-linear link field
$\link_\ell$ into an $SL(n_c,\CC)$ linear sigma model where {\sl on-shell}
\be
\det(\link_\ell)\ \equiv\ v^{n_c}\ =\ \mbox{const}.
\label{SigmaModel}
\ee
Note that $SL(n_c,\CC)$ is the complexified $SU(n_c)$ group manifold,
and this is precisely what we want for the link variables of a supersymmetric
$SU(n_c)$ gauge theory on the lattice.
In 5D terms,
\be
\link_\ell(x)\ =\ v\times
\mathop{\hbox{\Large\rm exp}}\limits_{\rm Path\atop ordered}\left(
         \int\limits_{a\ell}^{a(\ell+1)}\!\!\!dx^4
         \Bigl(iA_4(x)+\phi(x)\Bigr)\right)\
+\ \mbox{fermionic terms}
\label{Dmap}
\ee
where $\phi(x)$ is the scalar superpartner of the 5D vector field $A_\mu(x)$,
$\mu=0,1,2,3,4$.

Having defined the 4D quiver theory, we must now verify that it indeed
deconstructs the 5D SQCD.
At the semi-classical level of analysis of this section, this means
verifying that: (A) the vacuum field configurations of the 5D and the 4D
theory correspond to each other according to the field map~(\ref{Dmap}), and
(B) for each vacuum, the spectrum of light 4D particles follows from
the Kaluza--Klein reduction of the 5D particles
on a latticized circle of length $2\pi R=\ql a$~\cite{ACG, HPW}.
That is, for each 5D quark or gluon (or a superpartner) we must have a series
of $\ql $ 4D particles with similar quantum numbers and 4D masses given by
\be
M_4^2\ =\ m_5^2\ +\ P_4^2\ +\ O(am_5^3,a^2P_4^4,\ldots\,)
\label{Reduction}
\ee
where
\be
 P_4\ =\ \frac{2\pi k}{\ql a}\ +\ p_0,\qquad k=0,1,2,\ldots\ \mbox{mod}\ \ql 
\label{Pquantization}
\ee
is the quantized momentum in the $x^4$ direction and the quantization shift
$p_0$  allows for Wilson lines, {\it etc.}
Furthermore, all the {\it light} 4D particles ($M_4\ll(1/a)$) must belong
to such Kaluza--Klein series, although the {\it heavy} 4D particles
($M_4\gsim(1/a)$) do not have to have 5D counterparts.

We begin with the simplest 5D vacuum state with unbroken $SU(n_c)$
where all the gluons are massless while the quarks have
their bare masses~$m^{}_f$.
According to the field map~(\ref{Dmap}),
$\vev\phi=\vev{A_4}=0$ in 5D translates to the 4D link field VEVs
\be
\vev{\link_\ell^{}}\ \equiv\ v\times
\mbox{\large\bf 1}_{n_c\times n_c}
\label{SymmetricVacuum}
\ee
which break the 4D gauge symmetry~(\ref{G4D}) down to the `diagonal'
\be
SU(n_c)_{\rm diag}^{}\ =\ \diag\left[\prod_\ell SU(n_c)_\ell^{}\right]
\label{DiagonalSU}
\ee
while the rest of the 4D vector fields acquire masses
\be
M^2_4(k)\ =\ g^2|v|^2\times 4\sin^2\frac{\pi k}{\ql } .
\ee
This spectrum indeed matches \eq{Reduction} for $m_5=0$,
provided we identify the lattice spacing as
\be
a\ =\ \frac{1}{g|v|}\,.
\label{Vspacing}
\ee

Similarly, the quark mass spectrum also has deconstructive form
{\sl for flavors with bare 4D masses \blue $\mu_f^{}$ near $v$}.
Indeed, consider the mass matrix for the quarks $Q^f_\ell$ and
the antiquarks $\aq^f_\ell$ of a fixed flavor~$f$ but all $\ell=1,\ldots,\ql $:
The exact eigenvalues of this matrix are given by
\be
M_4^2(k)\ =\ |\gamma|^2\times \left| v e^{2\pi ik/\ql }\,-\,\mu^{}_f\right|^2,
\qquad k=1,\ldots,\ql ,
\ee
and {\blue for $\mu^{}_f$ near $v$},
this spectrum does have deconstructive form (\ref{Reduction}) where
\bea
m_5 &=& |\mu^{}_f|\,-\,|v|, \label{QuarkMass} \\
p_0 &=& |\gamma v|\times\arg(\mu^{}_f/v ), \label{FlavorWL} \\
\mbox{and}\ a &=& \frac{1}{|\gamma v|}\,.
\label{Qspacing}
\eea

Clearly, in order to satisfy both eqs.~(\ref{Vspacing}) and~(\ref{Qspacing})
we must have equal gauge and Yukawa couplings $g=|\gamma|$.
In a quantum theory, this means {\it fine-tuning} the non-holomorphic
K\"ahler parameters
of the quiver theory such that the renormalized physical couplings satisfy
\be
g_{\rm phys}\ =\ |\gamma|_{\rm phys}\quad\mbox{\red exactly},
\label{Ggamma}
\ee
or in non-perturbative terms, in the very low energy limit $E\ll v/\ql$,
the effective theory
(the diagonal $SU(n_c)$ with an adjoint field $\Phi$ and several quark flavors)
should be $\NN=2$  supersymmetric.
Without this condition, the deconstructed theory would have quarks and gluons
with different effective speeds of light in the $x^4$ direction.
This is a common problem in lattice theories with some continuous dimensions
({\it eg.} Hamiltonian lattice theories with continuous time
but discrete space),
and the common solution is fine-tuning of the lattice parameters.
For the problem at hand, the fine tuning of the quiver theory involves
the K\"ahler parameters (such as coefficients of the quarks' and antiquarks'
kinetic-energy Lagrangian terms) and does not
affect any of the holomorphic properties of the quiver such as its chiral ring.
Consequently, in the following we shall simply assume that the couplings
are fine-tuned according to \eq{Ggamma} and focus on other issues.

Next, consider the Coulomb branch of the \sqcdv\ moduli space where
the squarks have zero VEVs but$\vev\phi\neq 0$.
Generically, the $\vev\phi$ matrix has $n_c$ distinct eigenvalues
$(\phi_1,\ldots,\phi_{n_c})$,
the $SU(n_c)$ gauge symmetry is broken down to its Cartan subgroup
$(U(1))^{n_c-1}$ and the off-diagonal gluons $G_{ij}$ have masses
\be
m_5[G_{ij}]\ =\ \phi_i\,-\,\phi_j\,.
\label{Vmass5D}
\ee
At the same time, the quarks have color- and flavor-dependent masses
\be
m_5[q^{i,f}_{}]\ =\ m^{}_f\,-\,\phi_i\,.
\label{Qmass5D}
\ee
Similarly, the 4D quiver theory has Coulomb branch vacua with zero squark VEVs
but non-trivial link VEVs $\vev{\link_\ell}\neq v\times 1_{n_c\times n_c}$.
Combining the D--term constraints for all the $SU(n_c)_\ell$ gauge groups
\be
\link_\ell^\dagger\link^{}_\ell\ -\ \link_{\ell-1}^{}\link_{\ell-1}^\dagger\
\propto\ \mbox{\large\bf 1}_{n_c\times n_c}\quad\forall\ell,
\ee
with the F--term constraints~(\ref{SigmaModel}), we find that all the
$\vev{\link_\ell}$ matrices must be equal to each other modulo
an $\ell$-dependent gauge transform.
Moreover, we may diagonalize all the matrices at once and set
\be
\forall\ell:\ \vev{\link_\ell}\
=\ v\times\mathop{\rm diag}\left(
    e^{a\varphi_1},e^{a\varphi_2},\ldots,e^{a\varphi_{n_c}}
    \right)
\label{CoulombVEV}
\ee
for some complex numbers $(\varphi_1,\varphi_2,\ldots,\varphi_{n_c})$
satisfying $\sum_i\varphi_i=0$.
According to the field map~(\ref{Dmap}),
this corresponds to the 5D VEVs
\be
\vev\phi\,+\,i\vev{A^4}\
=\ \mathop{\rm diag}(\varphi_1,\varphi_2,\ldots,\varphi_n),
\ee
$i.\,e.,\ \phi_i=\Re\varphi_i$ are the 5D scalar VEVs of the Coulomb branch
while the $A^4_i=\Im\varphi_i$ are Wilson lines of the diagonal gauge
fields around the compactified $x^4$ dimension.

Generically, all the $\varphi_i$ are distinct and the 4D gauge symmetry
is broken all the way down to the Cartan $(U(1))^{n_c-1}_{\rm diag}$ subgroup
of the diagonal $SU(n_c)$.
The mass matrix of the remaining 4D gauge fields has eigenvalues
\be
M_4^2[G_{ij}^{(k)}]\
=\ g^2|v|^2\times\left|e^{2\pi ik/\ql }e^{a\varphi_i}\,-\,e^{a\varphi_j}\right|^2,
\label{Vmass4D}
\ee
and it is easy to see that for $\phi_i,\phi_j\ll(1/a)$ and $|k|\ll \ql $ this
4D spectrum has the deconstructed form~(\ref{Reduction})
where the lattice spacing $a$ is exactly as in \eq{Vspacing}, the 5D mass
\be
m_5[G_{ij}]\ =\ \Re(\varphi_i\,-\,\varphi_j)\ \equiv\ \phi_i\,-\,\phi_j
\ee
is in perfect agreement with the 5D formula~(\ref{Vmass5D}),
and the $P_4$ quantization shift $p_0$ is precisely the appropriate Wilson line
\be
p_0[G_{ij}]\ =\ \Im(\varphi_i\,-\,\varphi_j)\ \equiv\ A^4_i\,-\,A^4_j\,.
\ee
At the same time, the 4D quarks have mass eigenvalues
\be
M_4^2[Q^{i,f}_{(k)}]\
=\ |\gamma|^2\times\left|v e^{2\pi ik/\ql }e^{a\varphi_i}\,-\,\mu^{}_f\right|^2
=\ \frac{1}{a^2}\times\left|e^{2\pi ik/\ql }e^{a\varphi_i}\,-\,e^{am^{}_f}\right|^2
\label{Qmass4D}
\ee
where in the second equality we use \eq{Qspacing} for the lattice spacing~$a$
and {\sl define} complex 5D masses $m_f^{}$ according to
\be
m^{}_f\ \buildrel \mbox{def}\over= \ \frac{1}{a}\,\log\frac{\mu^{}_f}{v}\,.
\label{MassDef}
\ee
In the deconstruction limit, the real parts $\Re(m^{}_f)$ of
these complex masses act as the bare 5D quark masses,
while their imaginary parts $\Im(m^{}_f)$ 
correspond to Wilson lines of the flavor symmetries, {\it cf.}\ \eq{FlavorWL}.
Indeed, if we restrict to $m^{}_f\ll (1/a)$ ($i.\,e.,\ \mu^{}_f$ near $v$),
$\varphi_i\ll(1/a)$ and $|k|\ll \ql $,
then the quark mass spectrum~(\ref{Qmass4D})
has the deconstructed form~(\ref{Reduction}) where the 5D masses and
the Wilson lines are exactly as for the Coulomb branch of the \sqcdv:
\bea
m_5[Q_{}^{i,f}] &=& \Re(m^{}_f\,-\,\varphi_i)\ =\ \Re m^{}_f\, -\,\phi_i\,,\\
p_0[Q_{}^{i,f}] &=& \Im(m^{}_f\,-\,\varphi_i)\ =\ \Im m^{}_f\, -\,A^4_i\,.
\eea

Besides the Coulomb branch, the moduli space of \sqcdv\ has mesonic
Higgs branches \underline{iff} some of the quark masses are degenerate.
On such branches, some squark fields have non-zero VEVs,
some of the $\phi_i$ eigenvalues are fixed, and
the surviving gauge theory has reduced rank $r<(n_c-1)$.
For example, for $m_1=m_2$ there is a mesonic Higgs branch
with non-zero squark VEVs $\vev{q^{i,f}_\alpha}$ for $i=1$ and $f=1,2$;
the hypermultiplet components (indexed by $\alpha,\beta$)
are constrained by the D-terms to satisfy
$$
\refstepcounter{equation}
D^t\ =\ \eta^t_{\alpha\beta}\vev{q^{1,1}_\alpha}\vev{q^{1,1}_\beta}^*\
+\ \eta^t_{\alpha\beta}\vev{q^{1,2}_\alpha}\vev{q^{1,2}_\beta}^*\
=\ 0\quad\mbox{for}\ t=1,2,3.
\label{Dterm5D}\eqno(\theequation)\rlap{\strut\footnotemark}
$$
\footnotetext{%
        In 5D, $\NN=1$ SUSY there are three D terms forming a triplet
        of the $SU(2)_R$ symmetry.
        Consequently, for any broken gauge symmetry there are
        three D-term constraints $D^t$, $t=1,2,3$:
        The $\eta^t_{\alpha\beta}$ matrices in \eq{Dterm5D} represent the
        action of the $SU(2)_R$ symmetry on the hypermultiplet
        components $q_\alpha$.
        }
On this branch, the $\phi_1$ eigenvalue is frozen at $\phi_1=m_1=m_2$
while the remaining eigenvalues $\phi_2,\ldots,\phi_{n_c}$ remain free
(except for the $\sum_i\phi_i=0$ constraint);
for generic $\phi_2,\ldots,\phi_{n_c}$ the gauge symmetry
is Higgsed down to $(U(1))^{n_c-2}$ while the remaining gluons have masses
\be
m_5^2[G_{ij}]\
=\ (\phi_i-\phi_j)^2\ 
+\ \left(\delta_{i,1}\,+\,\delta_{j,1}\,
        -\,\coeff{2}{n_c}\,\delta_{i,1}\delta_{j,1}\right)
\times\frac{g_5^2}{4}\sum_{f,\alpha}|\vev{q^{1,f}_\alpha}|^2 .
\label{Vmasses5Dhiggs}
\ee

Likewise, the quiver theory has a mesonic Higgs branch whenever
$\mu_f^\ql =\mu_{f'}^\ql \neq 0$.
Indeed, let $\mu_1^\ql =\mu_2^\ql \neq0$ and let us freeze the $(ve^{a\varphi_1})^\ql $
link eigenvalue at the same value, or equivalently let
\be
\varphi_1\ =\ m_1\ +\ \frac{2\pi ik_1}{\ql a}\ =\ m_2\ +\ \frac{2\pi i k_2}{\ql a}
\label{MesonicOrigin}
\ee
for some integers $k_1$ and $k_2$.
At this point, the scalar potential has flat directions for the squark
and antisquark modes
\be
\vev{Q^{i,f}_\ell}\ =\ e^{2\pi i k^{}_f\ell/\ql }\times Q^{i,f}_{},\qquad
\vev{\aq^{i,f}_\ell}\ =\ e^{2\pi i k^{}_f(1-\ell)/\ql }\times \aq^{i,f}_{},
\label{SquarkVEVs4D}
\ee
of the color $i=1$ and the flavors $f=1,2$ only,
subject to F--term and D--term constraints
\be
\sum_f Q^{1,f}\tilde Q^{1,f}\ =\ 0,\qquad
\sum_f(|Q^{1,f}|^2\,-\,|\tilde Q^{1,f}|^2)\ =\ 0.
\label{DFterm4D}
\ee
These squark VEVs Higgs the $(SU(n_c))^\ql $ gauge symmetry down to
the $(SU(n_c-1))^\ql $,
which is further broken by the link VEVs $\vev{\link_\ell}$
down to a subgroup of the diagonal $SU(n_c-1)$.
For generic values of the un-frozen eigenvalues
$\varphi_2,\ldots,\varphi_{n_c}$,
the unbroken gauge symmetry is $U(1)^{n_c-2}$,
and all the remaining gauge fields have masses
\be
\eqalign{
M_4^2[G_{ij}^{(k)}]\ = &
{}\ g^2|v|^2\times\left|
        e^{2\pi i k/\ql }\,e^{a\varphi_i}\,-\,e^{a\varphi_j}
        \right|^2\cr
&+\ \left(
        \delta_{i,1}\,+\,\delta_{j,1}\,
        -\,\coeff{2}{n_c}\,\delta_{i,1}\delta_{j,1}
        \right)
    \times\frac{g^2}{2}\sum_f\left( |Q^{1,f}|^2+|\aq^{1,f}|^2\right).\cr
}\label{Vmasses4Dhiggs}
\ee

Clearly, this mesonic Higgs branch of the quiver theory deconstructs
the similar branch of the \sqcdv:
Its root ($i.\,e.$, the point where it connects to the Coulomb branch)
 is at the same place $\Re\varphi_1=\Re m_1=\Re m_2$, and
the F/D term constraints~(\ref{DFterm4D}) match
the 5D constraints~(\ref{Dterm5D}) once we repackage
\be
(Q,\aq^\dagger)^{i,f}\ \mapsto\ \sqrt{a}\,q^{i,f}_\alpha .
\label{Qnorm}
\ee
(The factor $\sqrt{a}$ here translates between the 4D and the 5D canonical
normalizations of the quark fields.)
Finally, the 4D mass spectrum~(\ref{Vmasses4Dhiggs}) has the deconstructed form
(\ref{Reduction}) where the 5D masses are exactly as in \eq{Vmasses5Dhiggs},
provided we translate squark VEVs according to \eq{Qnorm} and identify the 5D
gauge coupling according to the classical deconstruction formula \cite{ACG}
\be
g_5^2\ =\ ag^2 .
\ee

Further coincidences among the quark masses of \sqcdv\ allows for richer
mesonic Higgs branches with more squark VEVs, more frozen $\phi_i$ eigenvalues
({\it eg.}, $\phi_1=m_1=m_2$, $\phi_2=m_3=m_4$) and a lower rank of
the unbroken gauge symmetry.
The 4D quiver theory with multiple coincidences among $\mu_f^\ql $ has similar
Higgs branches,
and the deconstruction works so similarly to the above that we don't
need to repeat the argument.

Instead, let us consider the baryonic Higgs branch which exists
for $n^{}_f\geq n_c$ when $n_c$ of the quark masses add to zero,
{\it eg.,} $m_1+m_2+\cdots m_{n_c}=0$.
On this branch, the $\phi$ field is completely frozen at
$\phi_i=m_i\ \forall i=1,\ldots,n_c$, all the quarks with $i=f$ develop
similar VEVs
\be
\vev{q^{i,f}_\alpha}\ =\ \delta^{i,f}\times q_\alpha
\label{BaryonVEV5D}
\ee
and the gauge symmetry is completely Higgsed down.
The particle spectrum comprises a single massless hypermultiplet
$q$ (the baryonic modulus),
$n_c(n^{}_f-n_c)$ short hypermultiplets $q^{i,f}$ with masses
\be
m_5[q^{i,f}]\ =\ \phi_i\,-\,m^{}_f\ =\ m_i\,-\,m^{}_f\quad
\mbox{for}\ f>n_c\ \mbox{only},
\label{Qmasses5Dbaryon}
\ee
plus $n_c^2-1$ long vector multiplets with masses
\be
m_5^2[G_{ij}]\ =\ \coeff12 g_5^2\sum_\alpha|q_\alpha|^2\ +\ (\phi_i-\phi_j)^2.
\label{Vmasses5Dbaryon}
\ee

The quiver theory also has a baryonic Higgs branch when $n^{}_f\geq n_c$
and the product of some $n_c$ 4D masses equals to $v^{n_c}_{}$.
Indeed, let $\mu_1^\ql \times\mu_2^\ql \times\cdots\times\mu_{n_c}^\ql =v^{\ql n_c}$
or equivalently $m_1+m_2+\cdots m_{n_c}=0$,
and let us freeze all the link eigenvalues at
\be
\varphi_i\ =\ m_i\ +\ \frac{2\pi i k_i}{\ql a}\quad\forall i=1,2,\ldots,n_c\,.
\label{BaryonicOrigin}
\ee
At this point, the quark mass matrix due to superpotential~(\ref{Whop})
has zero modes for all $i=f$, which allows non-zero VEVs
\be
\vev{Q^{i,f}_\ell}\ =\ \delta^{i,f}\times e^{2\pi i k_i\ell/\ql }\times Q^i,\qquad
\vev{\aq^{i,f}_\ell}\ =\ \delta_i^f\times e^{2\pi i k_i(1-\ell)/\ql }\times\aq^i,
\ee
subject to the D term constraint
\be
\mbox{same}\ (|Q^i|^2\,-\,|\aq_i|^2)\ \forall i,
\label{Dterm4Dbaryon}
\ee
and the F-term constraint
\be
\frac{\partial W}{\partial\link^{i,i}_\ell}\
=\ \gamma Q^i\aq^i\
-\ \frac{\beta v^{n_c}\,\vev{s_\ell}}{v e^{a\varphi_i}}\ =\ 0.
\label{Fterm5Dbaryon}
\ee
The simplest solutions to these constraints are either same
$Q^i\equiv Q\ \forall i$ and $\aq^i\equiv0$ (baryonic VEVs only)
or {\it vice verse} same $\aq^i\equiv\aq\ \forall i$  and $Q^i\equiv 0$
(antibaryonic VEVs only),
but thanks to the singlet fields $s_\ell$ enforcing the determinant
constraints~(\ref{SigmaModel}),
there are other solutions where both baryonic and antibaryonic VEVs
are present at the same time while $\vev{s_\ell}\equiv s\neq 0$.
In such solutions
\be
Q^i\times\aq^i\ =\ \mbox{const}\times e^{-a\varphi_i},
\ee
and the color dependence of the right hand side goes away
in the deconstruction limit of $\varphi_i=m_i\ll(1/a)$.
Consequently, the squark VEVs become simply
\be
\vev{Q^{i,f}_\ell}\
=\ \delta^{i,f}\times e^{2\pi i k_i\ell/\ql }\times Q,\qquad
\vev{\aq^{i,f}_\ell}\
=\ \delta^{i,f}\times e^{2\pi i k_i(1-\ell)/\ql }\times\aq
\label{BaryonicVEVs}
\ee
for some arbitrary pair $(Q,\aq)$ of complex VEVs ---
which obviously deconstruct the 1 hypermultiplet VEV $\vev{q_\alpha}$
of the 5D theory.

Furthermore, all the gauge symmetries of the 4D quiver theory 
are Higgsed down and the vector multiplets acquire masses
\bea
M^2_4[G_{ij}^{(k)}] &
=& g^2(|Q|^2+|\aq|^2)\ +\ g^2|v|^2\times\left|
        e^{2\pi i k/\ql }\,e^{a\varphi_i}\,-\,e^{a\varphi_j}
        \right|^2
    \label{Vmasses4Dbaryon}\\
&\approx& g^2(|Q|^2+|\aq|^2)\ +\ (\phi_i-\phi_j)^2\ +\ P_4^2
%    \label{Vmasses4DbaryonDeco}
\eea
where the approximation on the second line applies in the deconstructive
limit of $\varphi_i,m_i,P_4\ll(1/a)$.
These 4D masses are in obvious agreement with the 5D vector masses
(\ref{Vmasses5Dbaryon}), so all we need to check is the supermultiplet
structure.
In 4D, $\NN=1$ SUSY, the Higgs mechanism eats one chiral multiplet for each
vector multiplet which becomes massive, thus each
$G_{ij}^{(k)}$ vector eats one linear combination of
the three chiral multiplets $\link_\ell^{ij}$, $Q_\ell^{ij}$
and $\aq_\ell^{ij}$ with similar charges.
Meanwhile, the other two linear combinations of these three multiplets
acquire masses via the Yukawa couplings, and thanks to
$\gamma=g$ and $\varphi_i=m_i$, they end up with exactly the same masses
(\ref{Vmasses4Dbaryon}) as the vector fields.
Altogether, this gives us complete $\NN=2$ massive long  multiplets --- or
equivalently long 5D massive vector multiplets reduced to 4D.

Finally, the remaining chiral fields of the quiver theory comprise a massless
pair $(Q,\aq)$ which deconstructs the baryonic hypermultiplet modulus of the
5D theory, plus massive quarks and antiquarks with flavors $f>n_c$.
These quarks have masses
\bea
M^4_2[q_{i,f}^{(k)}] &
=& |\gamma V|^2\times\left| e^{2\pi i k/\ql }\,e^{am_i}\,-\,e^{am^{}_f}\right|^2
        \quad\mbox{for}\ f>n_c\ \mbox{only}\\
&\approx& (m_i-m^{}_f)^2\ +\ P_4^2\quad
        \mbox{for}\ m_i,m^{}_f,P_4\ll(1/a)
\eea
and clearly deconstructs the 5D short hypermultiplets $q_{i,f}$ with
masses~(\ref{Qmasses5Dbaryon}).

This completes our classical analysis of the deconstructed \sqcdv.
At the quantum level of the 4D quiver theory, calculating the mass spectra
for various vacua of the theory becomes much more difficult --- indeed,
the state-of-the-art $\NN=1$ technology does not allow for
exact non-perturbative calculation of physical masses.
Instead, our quantum analysis of the deconstructed~\sqcdv\ will focus on the
exactly calculable holomorphic properties of the 4D theory such as the
moduli dependence of its gauge couplings.
We shall return to this issue in section~4.

%% file: chapter3.tex
%auto-ignore
%
% Chapter 3 of the DESCQD paper
%

\section{On Chern--Simons Couplings \brk and Extra 4D Flavors}
The quarks we have studied in the previous section were light
compared to the lattice cutoff of the deconstructed 5D theory:
$|m^{}_f|\ll(1/a)$, hence according to \eq{MassDef} $\mu^{}_f\approx v$.
{}From the quiver point of view, there are no 4D restrictions on the
bare quark masses and one may freely add extra flavors with $\mu^{}_f\gg v$ or
$\mu^{}_f\ll v$.
However, all such extra flavors are very heavy from the
5D point of view ($|m^{}_f|\gsim(1/a)$) and decouple from the continuum limit
of the 5D theory.
In fact, the extra flavors with $\mu^{}_f\gg v$ decouple from the
4D quiver theory above the deconstruction threshold, so there is really
no point in considering them any further.
On the other hand, the extra flavors with $\mu^{}_f\ll v$ are very much
present at the deconstruction threshold, and even through they ultimately
decouple from the low-energy 5D theory,
the 5D couplings receive quantum corrections
from integrating out such extra flavors.
Specifically, the extra flavors affect the Chern--Simons level of the
deconstructed 5D theory:
\be
\kcs\ =\ n_c\ -\ \#\{f:\,\mu^{}_f\ll v\}\
-\ \coeff12\#\{f:\,\mu^{}_f\sim v\}
\label{ChernSimons}
\ee
The purpose of the present section is to prove this formula for $n_c\geq 3$.

Let us start with a special case where all flavors have $\mu^{}_f\equiv0$
and the quarks $Q_\ell$ and antiquarks $\aq_\ell$ at the same site $\ell$
uncouple from each other.
Hence, there is no interaction between different link fields~$\link_\ell$
via quarks and antiquarks: Each $\link_\ell$ couples to the specific pair
of $Q_\ell$ and $\aq_{\ell+1}$, and they don't couple
to any other link~$\link_{\ell'}$.
Therefore, we may treat each link~$\link_\ell$ as a separate $SU(n_c)$
sigma model and calculate its Wess--Zumino interactions without any concern
for the other link fields $\link_{\ell'}$.

The Wess-Zumino interactions are topological and they depend only on the way
the chiral fermion transform under symmetries of the sigma model, so
let us consider the fermions which couple to
the non-linear scalar field~$\link_\ell$:
In component field formalism,
\be
{\cal L}_{\rm Yukawa}[\link_\ell]\ =
g\tr\Bigl( \link_\ell^\dagger\lambda_{\ell+1}\Psi^\link_\ell\ 
        -\ \Psi^\link_\ell\lambda_\ell\link^\dagger_\ell\Bigr)\
+\ \gamma\tr(\Psi^{\aq}_{\ell+1}\link_\ell\Psi^Q_\ell)\
+\ \beta v^{n_c}\,\Psi^s_\ell\tr(\Psi^\link_\ell\link_\ell^{-1})\quad
+\ \rm H.~c.,
\label{Yukawas}
\ee
where $\lambda_\ell,\lambda_{\ell+1}$ are the gauginos and
the $\Psi^\link_\ell$, $\Psi^{\aq}_{\ell+1}$, $\Psi^Q_\ell$, and $\Psi^s_\ell$
are the fermionic members of the appropriate chiral multiplets;
no other fermions couple to the~$\link_\ell$ sigma model.
The $\link_\ell$ sigma model has a chiral symmetry $SU(n_c)_L\times SU(n_c)_R
\equiv SU(n_c)_{\ell+1}\times SU(n_c)_\ell$.
Under the right-hand symmetry $U\in SU(n_c)_\ell$, the scalar field $\link_\ell$
transforms according to $\link_\ell\to\link_\ell\times U^\dagger$ while
the fermionic transformation rules follow from the invariance of the Yukawa
Lagrangian~(\ref{Yukawas}).
Specifically,
\be
\!\Psi^\link_\ell\,\to\,\Psi^\link_\ell U^\dagger,\quad\!
\lambda_\ell\,\to\,U\lambda_\ell U^\dagger,\quad\!
\lambda_{\ell+1}\,\to\,\lambda_{\ell+1}\,,\quad\!
\Psi^{\aq}_{\ell+1}\,\to\,\Psi^{\aq}_{\ell+1},\quad\!
\Psi^Q_\ell\,\to\,U\Psi^Q_\ell\,,\quad\!
\Psi^s_\ell\,\to\,\Psi^s_\ell.\!
\ee
{}From the chiral $SU(n_c)_R\equiv SU(n_c)_\ell$ point of view,
the $\Psi^\link_\ell$ amounts
to $n_c$ species of antiquarks each transforming according
to $\tilde\psi\to \tilde\psi\times U^\dagger$,
while the $\Psi^Q_\ell$ packs $n^{}_f$ species of quarks each
transforming according to $\psi\to U\times\psi$.
For $n_c\neq n^{}_f$ this is a chiral transform,
hence the Wess--Zumino action
\be
S_{\rm WZ}\ =\ \kwz\times\int\limits_{\RR^4}\!\!\wzf(\link_\ell)
\label{WZaction}
\ee
where $\wzf$ is the universal Wess-Zumino 4--form\footnote{%
        The name ``Wess--Zumino form'' is often used for the 5--form
        $d\wzf$ rather than the 4--form $\wzf$ itself.
        Unlike the 4--form, the 5--form is manifestly chirally symmetric,
        and it's also a much simpler function of the non-linear
        scalar field~$\link_\ell$.
        The action~(\ref{WZaction}) can be written in terms of the 5--form
        integrated over 5 dimensions: the ordinary four, plus an auxiliary
        fifth dimension.
        But that fifth dimension has absolutely nothing to do with the
        deconstructed fifth dimension of the \sqcdv, so to avoid
        dimensional confusion, we use the 4D form
        of the Wess-Zumino action in this paper.
        },      
and
\be
\kwz\ =\ n_c\ -\ n^{}_f\,.
\ee

As explained in \cite{SkibaSmith,IK1}, the Wess--Zumino couplings of the
link fields deconstruct the Chern--Simons coupling of the 5D gauge fields:
Let $\wzf(\link_\ell,A^\mu_\ell,A^\mu_{\ell+1})$ be a gauged WZ 4--form
for the link field $\link_\ell$ and the (4D) gauge fields $A^\mu_\ell$
and $A^\mu_{\ell+1}$ under which it is charged; then in the continuum
5D limit $a\to0$,
\be
\sum_\ell\int\limits_{\RR^4}\!\!\wzf(\link_\ell,A^\mu_\ell,A^\mu_{\ell+1})\
=\ \int\limits_{\RR^5}\!\!\csf(A^\mu_{\rm 5D})\quad +\ O(a)
\label{CScoupling}
\ee
where
\be
\csf\ =\ \frac{i}{24\pi^2}\,
\tr\left(A\land F\land F\,-\,\coeff{i}{2} A\land A\land A\land F\,
-\,\coeff{1}{10} A\land A\land A\land A\land A\right)
\label{NACS}
\ee
is the Chern--Simons 5--form.
The coefficient $\kcs$ of the Chern--Simons coupling~(\ref{CScoupling}) is
quantized; in light of the above,
\be
\kcs\ =\ \kwz\ =\ n_c\ -\ n^{}_f\,.
\label{CSlevel}
\ee
Note that the effective low-energy theory in the 5D continuum limit is a pure
SYM whereas all the quarks which were present in the 4D theory have decoupled
at the deconstruction threshold $E\sim(1/a)$.
Nevertheless, thanks to those decoupled quarks, the Chern--Simons level of the
SYM theory is lowered from $\kcs=n_c$ for the quark-less quiver we have studied
in~\cite{IK1} down to $\kcs=n_c-n^{}_f$.

The general case which allows $\mu^{}_f\neq 0$ is more complicated:
The quark masses relate fermions $\Psi_\ell^Q$ and $\Psi_\ell^{\aq}$
which couple to different sigma models and we no longer have isolated
sigma models with separate Wess--Zumino couplings.
Hence, instead of a direct deconstruction of the 5D Chern--Simons
coupling~(\ref{NACS}), we assume it exists at some level $\kcs$
and calculate this level by taking the very-low-energy limit $E\ll(1/\ql a)$.
In this limit, the 5D theory is dimensionally reduced to 4D (without the
Kaluza--Klein excitations), the $A^4$ component of the vector field becomes
a scalar, and the CS coupling becomes a field-dependent set of $\Theta$ angles:
\be
{\cal L}_{\rm cs}^{\rm 4D}\
=\oint\limits_{x^4}\!\kcs\csf(A^\mu_{\rm 5D})\
=\ \frac{i\kcs \ql a}{8\pi^2}\,\Tr\left(A^4\,F\land F\right)\
=\ \frac{i}{16\pi^2}\sum_i(\kcs \ql aA^4_i)\,F_i\land F_i
\ee
where in the last equality here we have restricted the 4D gauge fields
to the abelian $F_i$ ($i=1,2,\ldots,n_c$, $\sum_iF_i=0$)
which remain massless after the Wilson lines $\ql aA^i_4$ break the $SU(n_c)$
down to the $U(1)^{n_c-1}$.

{}From the quiver point of view, the
\be
\Theta_i\ =\ \kcs\times \ql a\, A^4_i
\label{CStheta}
\ee
are $\Theta$ angles which arise from the
field-dependent masses of the charged fermions.
The Adler--Bardeen theorem provides an exact formula:
\be
\Theta_i\ =\ -\sum_{q} q_i^2\,\arg\det(M_{q})
\label{AdlerBardeen}
\ee
where $q$ is the array of abelian charges $(q_1,q_2,\ldots,q_{n_c})$
and $M_q$ is the mass matrix of fermions with the same charges $q$.
Our task therefore is to evaluate this formula and show that 
the $\Theta_i$ angles indeed have form~(\ref{CStheta})
for the Chern--Simons level~$\kcs$ specified in~\eq{ChernSimons}.

In the eigen-basis of the Wilson lines,
the fermionic mass matrix pairs the quarks $(\Psi^Q_\ell)^{i,f}$
with the antiquarks $(\Psi^{\aq}_{\ell'})^{i,f}$
and the gauginos $(\lambda_\ell)^{ij}$ with the link
fermions $(\Psi^\link_{\ell'})^{ji}$:
\bea
{\cal L}^{\rm fermion}_{\rm mass} &=&
\sum_{i,f}\sum_{\ell,\ell'} M_{\ell,\ell'}[Q^{i,f}]\times
        (\Psi^Q_\ell)^{i,f}\,(\Psi^{\aq}_{\ell'})^{i,f} \nonumber \\
&&+\sum_{i,j}\sum_{\ell,\ell'} M_{\ell,\ell'}[\lambda^{ij}]\times
        (\lambda_\ell)^{ij}\,(\Psi^\link_{\ell'})^{ji} ,\\
\mbox{where}\ M_{\ell,\ell'}[Q^{i,f}] &=&
\gamma v e^{a\varphi_i}\times\delta_{\ell+1,\ell'}\
    -\ \gamma\mu_f^{}\times\delta_{\ell,\ell'}\, \\
\mbox{and}\ M_{\ell,\ell'}[\lambda^{ij}] &=&
\left(gv e^{a\varphi_i}\right)^*\times\delta_{\ell,\ell'+1}\
    -\ \left(gv e^{a\varphi_j}\right)^*\times\delta_{\ell,\ell'}\,.
\eea
The determinants of the mass matrices with respect to the quiver
indices~$\ell,\ell'$ are completely straightforward:
\bea
\det M[Q^{i,f}] &=&
\pm\gamma^\ql \left( v^\ql \,\exp(\ql a\varphi_i)\ -\ \mu_f^\ql \right),
\label{Qdeterminant} \\
\det M[\lambda^i_{\,j}] &=&
(gv^*)^\ql \left( \exp(\ql a\varphi_i^*)\ -\ \exp(\ql a\varphi_j^*) \right) .
\label{Gdeterminant}
\eea
Taking into account the abelian charges of the gauginos, their
combined contribution to the $\Theta_i$ angle~(\ref{AdlerBardeen})
amounts to
\be
[\Theta_i]_\lambda\ =\ -\sum_{j\neq i}\arg\left(
        \det M[\lambda^{ij}]\times\det M[\lambda^{ji}]\right)
\ee
Assuming for simplicity that the 5D scalar $\phi_i$ have zero VEVs
and only the Wilson lines break the $SU(n_c)$ --- thus $\varphi_j=iA^4_j$
--- we have
\be
\det M[\lambda^{ij}]\times\det M[\lambda^{ji}]
=\ 4(gv^*)^{2\ql }\,\sin^2\frac{\ql a}{2}(A^4_i-A^4_j)\times \exp(-i\ql a(A^4_i+A^4_j)),
\ee
and therefore
\be
[\Theta_i]_\lambda\
= \sum_{j\neq i}\left( \ql a(A^4_i+A^4_j)\,+\,2\ql \arg(v)\right)\
=\ n_c\times \ql aA^4_i\ +\ \rm const.
\ee

Now consider the quark mass determinant~(\ref{Qdeterminant}).
In the large $\ql $ limit,
\be
\det M[Q^{i,f}]\ \approx\ \gamma^\ql \cases{
\pm v^\ql \,\exp(i\ql a A^4_i) & when $|v|>|\mu^{}_f|$,\cr
\mp \mu_f^\ql  & when $|v|<|\mu^{}_f|$,\cr }
\ee
hence the quark contribution to the $\Theta_i$ angle~(\ref{AdlerBardeen})
\be
\eqalign{
[\Theta_i]_Q\ &
=\ -\sum_f^{|\mu^{}_f|<|v|}\left( \ql a A^4_i\,+\,\ql \arg(\gamma v)\right)\
    -\sum_f^{|\mu^{}_f|>|v|} \ql \arg(\gamma\mu^{}_f)\cr
&=\ \mbox{const}\ -\ (\ql a A^4_i)\times\#\{f: |\mu^{}_f|<|v|\} .\cr }
\ee
Totaling the quark and the gaugino contributions, we arrive at
\be
\Theta_i\ =\ \kcs'\times(\ql a A^4_i)\ +\ \rm const,
\ee
in full agreement with the dimensionally reduced Chern--Simons
coupling~(\ref{CStheta}) for
\be
\kcs'\ =\ n_c\ -\ \#\{f: |\mu^{}_f|<|v|\} .
\label{StrictCS}
\ee

Although the above Chern--Simons level $\kcs'$ is not quite as in
\eq{ChernSimons}, the discrepancy involves only quarks with
4D masses $\mu^{}_f\approx v$.
The problem lies in our taking the low-energy limit too literally
and hence integrating out any 4D fermionic mode which is not
exactly massless, including all of the quark modes.
Consequently, the resulting Chern--Simons level~$\kcs'$ corresponds
to the low-energy limit of the \sqcdv\ from which all the quarks
have been integrated out.
Thus,
\be
\kcs'\ =\ \kcs\ +\ \sum_f \coeff12\mathop{\rm sign}(\Re m^{}_f)
\ee
where $\kcs$ refers to the 5D theory which has light quarks only
(the 5D masses $|m^{}_f|\leq m_{\rm max}\ll(1/a)$), and only such light quarks
appear in the $\sum_f$.
In terms of the 4D masses $\mu^{}_f=ve^{am^{}_f}$ ({\it cf.}\ \eq{MassDef}),
this means
\be
\kcs'\ =\ \kcs\ +\ \coeff12\,\#\{f:\ |v|<|\mu^{}_f|<|v|+m_{\rm max}\}\
-\ \coeff12\,\#\{f:\ |v|-m_{\rm max}<|\mu^{}_f|<|v|\} .
\ee
Finally, comparing this formula to \eq{StrictCS} we arrive at
\be
\kcs\ =\ n_c\ -\ \#\{f:\ |\mu^{}_f|<|v|-m_{\rm max}\}\
-\ \coeff12\,\#\{f:\ |v|-m_{\rm max}<|\mu^{}_f|<|v|+m_{\rm max}\}
\label{CSexact}
\ee
which is exactly what we have promised in \eq{ChernSimons}
(but now have restated in a more precise manner).

Eq.~(\ref{CSexact}) tells us that deconstructing \sqcdv\ with a given
Chern--Simons level may require more quark flavors in the 4D quiver theory
then are present in 5D.
To avoid notational confusion, let $n^{}_f$ henceforth refer to the number
of 5D quark flavors with masses $|m|\ll(1/a)$ while $F$ denotes the total
number of 4D flavors of the quiver.
According to \eq{CSexact}, we need
\be
F\ =\ n^{}_f\ +\ \Delta F\quad \mbox{where}\
\Delta F\ =\ n_c\ -\ \frac{n^{}_f}{2}\ -\ \kcs \,;
\label{Flavors}
\ee
for $f=1,2,\cdots,n^{}_f$ the 4D quark masses should be set to
$\mu^{}_f=v e^{am^{}_f}$ according to \eq{MassDef},
but for $f>n^{}_f$ we want $\mu^{}_f\ll v$;
for simplicity, we let $\mu^{}_f=0$ for $f=(n^{}_f+1),\ldots,F$.

Note that quantum consistency of \sqcdv\ requires integer $\kcs$
when $n^{}_f$ is even but half-integer $\kcs$ when $n^{}_f$ is odd.
Also, positivity of the moduli-dependent gauge
couplings~(\ref{Fexact}--\reftail{GFmetric}) for $h>0$ requires
\be
|\kcs|\ +\ \frac{n^{}_f}{2}\ \leq\ n_c \,.
\label{FlavorLimit}
\ee
In terms of \eq{Flavors}, these rules translate to $\Delta F$ being a
non-negative integer (good, since otherwise deconstruction would be impossible)
and $F\leq 2n_c$.
Since each $SU(n_c)_{\,\ell}$ gauge group of the quiver couples to the total of
$(n_c+F)$ chiral fields in the $(\square+\overline{\square})_\ell$
representation, this means that the quiver theory should be asymptotically free,
or at least asymptotically finite.
This is good for the quiver as a UV completion of the 5D theory, but it also
means strong quantum corrections in the IR limit of the quiver theory.
In \S6, we shall see that when such quantum corrections become strong enough,
the deconstructed \sqcdv\ may have a flop transition to a different 5D phase.

We conclude this section with a few words about \sqcdv\ theories with only
two colors~\cite{MS}.
The $SU(2)$ group does not have a cubic invariant, hence the 5D Chern--Simons
coupling does not exist for $n_c=2$.
Instead, there is a $\ZZ_2$ topological invariant and hence a vacuum angle
$\theta$ which takes 2 discrete values $0$ and $\pi$ (modulo $2\pi$).
It would be interesting to deconstruct this vacuum angle directly from the
quantum quiver theory, but here we prefer a simpler derivation:
Let us realize an $SU(2)$ quiver theory with $F$ 4D flavors as the
mesonic Higgs branch of an $SU(3)$ quiver with $F'=F+2$.
This gives us two ways to deconstruct the 5D theory with $n_c=2$:
We may first deconstruct an $SU(3)$ theory with $n'_f=n^{}_f+2$, then Higgs
the 5D theory down to $SU(2)$ (which eats up the two extra flavors), hence
\be
[\theta]_{SU(2)}\ =\ \pi\times[\kcs]_{SU(3)}\
=\ \pi\times\left[n'_c\,-\,\coeff12 n'_f\,-\,\Delta F\right]_{SU(3)}\
=\ \pi\times\left[n_c\,-\,\coeff12 n^{}_f\,-\,\Delta F\right]_{SU(2)}
\ee
Alternatively, we may first Higgs the $SU(3)$ down to $SU(2)$ in 4D and
deconstruct afterward, but the end result should be the same, thus
\be
\blue \mbox{for}\ n_c=2,\quad
\theta\ =\ \pi\times\Delta F\ -\ \frac{\pi}{2}\times\ n^{}_f\quad
\mbox{modulo}\ 2\pi .\black \label{VacuumAngle}
\ee
Note that for odd $n^{}_f$ this angle takes values $\pm\pi/2$ instead of $0$
or $\pi$, but this is OK since the real vacuum angle obtains only after
after integration out of the 5D fermions, thus
\be
\bar\theta\ =\ \theta\ +\ \frac\pi2\sum_f\mathop\mathrm{sign}(\Re m^{}_f)
=\ \pi\times\biggl(\Delta F\,+\,\#\{f:m^{}_f<0\}\biggr)
\ee
which indeed takes values $0$ and $\pi$ for any $n^{}_f$.

Another peculiarity of the \sqcdv\ with $n_c=2$ is that it allows up to 7 quark
flavors instead of usual limit $n^{}_f\leq 2n_c$ for $n_c\geq 3$.
The $SU(2)$ quiver theory however loses asymptotic freedom
and becomes UV-divergent and IR-trivial for $n^{}_f>4$.
Consequently, at the quantum level, the quiver~(\ref{quiver}) fails to
deconstruct \sqcdv\ with $n_c=2$ and $n^{}_f=5$, 6 or 7.
We suspect such theories can be deconstructed in terms of more complicated
quivers and we hope to present them in a future publication, but in this article
we shall henceforth assume $F\leq 2n_c$ even for $n_c=2$.

%% file: chapter4.tex
%auto-ignore
%
% chapter 4 of the DESCQD paper
%

\section{Quantum Deconstruction of \brk the Gauge Couplings}
The deconstructed \sqcdv\ has only 4 exact supersymmetries, hence at the
non-perturbative level of analysis only the holomorphic features of the
4D quiver are {\em exactly} calculable.
{}From the deconstruction point of view, the most important holomorphic
feature is the moduli dependence of the abelian gauge couplings
$\tau_{ij}(\varphi)$ for the Coulomb branch of the quiver's moduli space.
Whereas the unbroken gauge symmetry $(U(1))^{n_c-1}\subset SU(n_c)_{\rm diag}$
of the quiver deconstructs the
Kaluza--Klein reduction of unbroken 5D symmetry on a circle of length
$2\pi R=\ql a$, in the large quiver limit $\ql \to\infty$ we should have
\be
2\pi\Im\tau_{ij}(\varphi)\
\equiv\ \left[\frac{8\pi^2}{g_4^2(\varphi)}\right]_{ij}\
=\ \ql a\times \left[\frac{8\pi^2}{g_5^2(\phi)}\right]_{ij}\ +\ O(1) .
\label{g45}
\ee
In this section, we shall see that this is indeed the case, and furthermore
the 5D gauge couplings on the right hand side of \eq{g45} are exactly as in
\eqrange{Fexact}{GFmetric} for the un-deconstructed 5D SQCD.

In a separate article~\cite{KSdN} we have analyzed the entire chiral ring
of the 4D quiver theory~(\ref{quiver});
for the present purposes, let us simply state  without proof the key
results which are relevant for the gauge couplings.
First of all, the Seiberg--Witten spectral curve of the quiver
is the Riemann surface of the quadratic equation
\be
Y^2\ -\ Y\times P(X)\ +\ (-1)^F\alpha B(X)\ =\ 0
\label{SWcurve}
\ee
where $P(X)$ and $B(X)$ are polynomials of respective degrees $n_c$ and $F$,
and $\alpha$ is a constant parametrizing the non-perturbative quantum effects.
Specifically, $\alpha$ originates at the diagonal instanton level of the
$[SU(n_c)]^\ql $ quiver, meaning one instanton of the diagonal
$SU(n_c)_{\rm diag}$ gauge group,
or equivalently, one instanton in each of the $SU(n_c)_\ell$ factor.
For $F<2n_c$,
\be
\alpha\ =\ \left((-\gamma)^F\Lambda^{2n_c-F}\right)^\ql 
\label{Adef}
\ee
where $\Lambda$ is the usual dimensional transmutant of the asymptotically
free 4D gauge coupling $g$
(note same $g_\ell\equiv g$ for all $\ell$
hence same $\Lambda_\ell\equiv\Lambda$).
%and the $\pm$ sign is $(-1)^{F(\ql -1)}$.
In the asymptotically-flat case of $F=2n_c$,
$$
\refstepcounter{equation}
\alpha\ \approx\ \left[\gamma^{2n_c}\exp(2\pi i\tau_{\rm UV})\right]^\ql \
\approx\ \left[\exp\left(
        i\theta\,-\,\frac{8\pi^2}{g_{\rm phys}^2}
        \right)\right]^\ql .
\label{F2nc}\eqno(\theequation)\rlap{\strut\footnotemark}
$$
\footnotetext{%
        The second equality here follows from \eq{Ggamma} for the
        renormalized gauge and Yukawa couplings, and the approximation
        is ignoring the threshold effects.
        Instead, we use the {\sl massless} renormalization group
        equations for the $g_{\rm ren}(E)$ and $\gamma_{\rm ren}(E)$
        and then impose $g_{\rm ren}(E)=\gamma_{\rm ren}(E)$ for some
        low-energy normalization point~$E$.
        In terms of the running kinetic energy factors $Z_\link(E)$,
        $Z_Q(E)$ and $Z_{\aq}(E)$ for the charged fields, the renormalized
        Yukawa and gauge couplings are given by
        \bea
        \gamma_{\rm ren}^2(E) &=&
        \frac{|\gamma_{\rm holomorphic}|^2}{Z_\link(E)\,Z_Q(E)\,Z_{\aq}(E)}\,,
                \nonumber \\
        \noalign{\noindent and\par}
        g^{2n_c}_{\rm ren}(E)\times
                \exp\left(\frac{8\pi^2}{g^2_{\rm ren}(E)}\right) &=&
        \frac{\exp(2\pi\Im\tau_{\rm UV})\times(E/\mbox{cutoff})^{2n_c-F}}
                {(Z_\link(E))^{n_c}\,(Z_Q(E)\,Z_{\aq}(E))^F}\,.\nonumber
        \eea
        Substituting $F=2n_c$ and combining the two equations, we obtain
        $$
        \left(\frac{\gamma_{\rm ren}}{g_{\rm ren}}\right)^{2n_c}\times
        \exp\left(-\frac{8\pi^2}{g^2_{\rm ren}}\right)\
        \equiv\ \left|\gamma^{2n_c}_{\rm hol}\,\exp(2\pi\tau_{\rm UV})\right|
        $$
        at all renormalization scales.
        Eq.~(\ref{F2nc}) follows from this formula once we identify
        $g_{\rm phys}=g_{\rm ren}(E)$ and $\gamma_{\rm phys}=\gamma_{\rm ren}(E)$
        for some renormalization point $E$ and apply \eq{Ggamma}.
        }
As to the polynomials,
\be
B(X)\ =\ \prod_{f=1}^F\left(X\,-\,\mu_f^{}\right)\
=\ X^{\Delta F}\times \prod_{f=1}^{n_f^{}}\left(X\,-\,\mu_f^{}\right)
\label{Bdef}
\ee
parametrizes the bare 4D quark masses $\mu_f^{}$ of the theory, and
\be
P(X)\ =\ \prod_{i=1}^{n_c}\left(X\,-\,\varpi_i\right)
\label{Pdef}
\ee
parametrizes the Coulomb moduli space of the quiver.
Since this space has only $n_c-1$ independent moduli, the roots $\varpi_i$
of $P(X)$ are subject to one constraint, namely
\be
\prod_{i=1}^{n_c}\varpi_i\ \equiv\ V^{\ql n_c}\ =\ \mbox{const},
\label{ProductConstraint}
\ee
where
\bea
V^{\ql n_c} &=&
(v_1^{n_c})^\ql \,+\,(v_2^{n_c})^\ql\quad\mbox{for}
\label{Vformula}\\
v_1^{n_c}\ +\ v_2^{n_c} &=&
v^{n_c}_{}\quad\mbox{and}\quad
v_1^{n_c}\ \times\ v_2^{n_c}\
=\ \Lambda^{2n_c-F}\,(\gamma\mu_1)(\gamma\mu_2)\cdots(\gamma\mu_F).
\label{v1v2}
\eea
Qualitatively,
\be
\eqalign{
V\ &=\ v,\ \mbox{exactly,} &
\quad\mbox{for}\ \Delta F\,&>\,0,\cr
\mbox{but}\ V\ &=\ v\ +\ O(\Lambda^{2n_c-F}) &
\quad\mbox{for}\ \Delta F\,&=\,0.\cr
}\label{Vqual}
\ee

The gauge coupling matrix $\tau_{ij}$ follows directly from the
spectral curve~(\ref{SWcurve}), but in order to study its dependence
on the deconstruction-appropriate moduli $\varphi_j=\phi_j+iA^4_j$
we must first {\sl define} the $\varphi_j$ in a gauge-invariant way.
Classically, there is a simple definition in terms of eigenvalues
of the quiver-ordered product $\link_\ql \link_{\ql -1}\cdots\link_2\link_1$
of the bi-fundamental link fields:
\be
\mathop{\rm eigenvalues}\Bigl[ \link_\ql \link_{\ql -1}\cdots\link_2\link_1\Bigr]\
=\ \left( \left[v e^{a\varphi_1}\right]^\ql , \left[v e^{a\varphi_2}\right]^\ql ,
        \ldots,\left[v e^{a\varphi_{n_c}^{}}\right]^\ql \right),
\ee
{\it cf.}\ \eq{CoulombVEV}.
Or equivalently, we may define the resolvent function and look for its poles:
\be
T(X)\ \eqdef\ \Tr{1\over X\,-\,\link_\ql \cdots\link_1}\
\buildrel {\rm cla}\over= \,
\sum_{i=1}^{n_c}{1\over X\,-\,[v\exp(a\varphi_i)]^\ql }\,.
\label{resolvent}
\ee
Unfortunately, in the quantum theory the resolvent is defined as
\be
T(X)\ =\ \vev{\Tr{1\over X\,-\,\link_\ql \cdots\link_1}}\quad
\mbox{instead of}\quad \Tr{1\over X\,-\,\vev{\link_\ql }\cdots\vev{\link_1}}
\ee
and consequently it has branch cuts instead of poles;
specifically,
\be
T(X)\ =\ \frac{\partial_X Y}{Y}\
=\ {1\over\sqrt{P^2-4(-1)^F\alpha B}}\left( \partial_X P\
        -\ {2(-1)^F\alpha\over P+\sqrt{P^2-4(-1)^F\alpha B}}\,\partial_X B\right).
\label{resolventeq}
\ee
However, in the weak coupling limit $\Lambda\to 0$, the branch cuts become very
short and can be approximated as poles located at the roots $\varpi_i$
of the $P(X)$ polynomial:
\be
\Lambda\to 0\ \Longrightarrow\ \alpha B\ll P^2\ \Longrightarrow\
T(X)\ \approx\ \frac{\partial_X P}{P}\
=\sum_{i=1}^{n_c}\frac{1}{X\,-\,\varpi_i}\,,
\ee
which immediately suggests the definition
\be
\varphi_i\ \eqdef\ \frac{1}{a}\log\frac{\root \ql  \of{\varpi_i}}{v}\
\Longleftrightarrow\
P(X)\ =\ \prod_{i=1}^{n_c}\left(X\,-\,v^\ql \exp(\ql a\varphi_i)\right) .
\label{phidef0}
\ee
Or rather
\be
\varphi_i\ \eqdef\ \frac{1}{a}\log\frac{\root \ql  \of{\varpi_i}}{V}\
\Longleftrightarrow\
P(X)\ =\ \prod_{i=1}^{n_c}\left(X\,-\,V^\ql \exp(\ql a\varphi_i)\right)
\label{phidef}
\ee
in order to map the quantum-corrected
moduli constraint~(\ref{ProductConstraint})
onto classical-like trace condition $\sum_i\varphi_i=0$.

Outside the week coupling limit, $T(X)$ generally has branch cuts
of finite length;
however, at a point where the Coulomb branch of the
quiver's moduli space joins a mesonic Higgs branch,
one of the branch cuts does degenerate into a pole.
In terms of the Seiberg--Witten spectral curve, this happens
when a root of $P(X)$ coincides with a double root of $B(X)$,
{\it eg}.\ $\varpi_1=\mu_1^\ql =\mu_2^\ql $:
At this point, the quadratic equation~(\ref{SWcurve}) factorizes as
\be
Y\ =\ (X-\varpi_1)\times \widetilde Y,\qquad
\widetilde Y^2\ -\ \widetilde Y\times\frac{P(X)}{(X-\varpi_1)}\
+\ \frac{(-1)^F\alpha B(X)}{(X-\mu^\ql _{1,2})^2}\ =\ 0,
\label{HiggsOrigin}
\ee
where the second \eq{HiggsOrigin} describes a hyperelliptic curve
of reduced genus ($g=n_c-2$ instead of $g=n_c-1$) ---
which corresponds to the reduced
rank of the Higgs branch's gauge symmetry $(U(1))^{n_c-2}$ ---
and the resolvent
\be
T(X)\ =\ \frac{\partial_X Y}{Y}\
=\ \frac{1}{X-\varpi_1}\ +\ \frac{\partial_X \widetilde Y}{\widetilde Y}
\ee
has a pole at $X=\varpi_1$ on both sheets of the Riemann surface.

{}From the 5D point of view, this point of the moduli space
corresponds to $\varphi_1=m_1=m_2$.
Therefore, the $\rm 4D\leftrightarrow 5D$ map of moduli
and masses should have $\varphi_i=m_f^{}$ {\sl exactly} when $\varpi_i=\mu_f^\ql$,
regardless of any quantum corrections.
Hence, the classical mass formula~(\ref{MassDef}) calls for defining the 5D
Coulomb moduli according to \eq{phidef0}, weak coupling or strong coupling.
Alternatively, we may rescale $v\to V$ in both mass and moduli maps;
this gives us \eq{phidef} for the moduli --- and hence $\sum_i\varphi_i=0$
--- while the 5D quark masses are given by
\be
m_f^{}\ \eqdef\ \frac{1}{a}\log\frac{\mu_f^{}}{V}\,.
\label{massdef}
\ee

In light of the definitions (\ref{phidef}) and (\ref{massdef}), it is
convenient to rescale the $X$ and $Y$ coordinates of the spectral curve
by appropriate powers of $V$ and rewrite the curve as
\begingroup\blue\bea
y^2\ -\ y\times p(x) &+& e^{-\ql aS}\,b(x)\ =\ 0
        \label{SW}\\
\noalign{\noindent where}
p(x) &=& \prod_{i=1}^{n_c}\left(x\,-\,e^{\ql a\,\varphi_i^{}}\right),
        \label{pdef}\\
b(x) &=& (-1)^F x^{\Delta F}\times\prod_{i=1}^{n_f^{}}
        \left(x\,-\,e^{\ql am_f^{}}\right),
        \label{bdef}\\
\mbox{and}\quad e^{-\ql aS} &=& \frac{\alpha}{V^{\ql (2n_c-F)}}
        \nonumber\\
&\hbox to 1em{\hss\rput{90}{$\Longleftrightarrow$}\hss}&\nonumber\\
S &\eqdef& \frac{1}{a}\times\cases{
\log\frac{V^{2n_c-F}}{(-\gamma)^F \Lambda^{2n_c-F}} & for $F<2n_c$,\cr
        \noalign{\vskip 8pt}
        \frac{8\pi^2}{g^2}\,-\,i\theta & for $F=2n_c$.\cr
        }\qquad\label{Sdef}
\eea\endgroup
In these notations, \eqrange{SW}{bdef} describe the spectral curve
of a 5D theory compactified on a circle of length $2\pi R=\ql a$ without any
indication that the compactified dimension is discrete;
all the 4D aspects of the deconstructed \sqcdv\ are `hiding'
in the definitions (\ref{phidef}), (\ref{massdef})
and (\ref{Sdef}) of the 5D moduli and parameters.
In other words, the details of the deconstruction {\sl decouple} from
the spectral curve~(\ref{SW}--\reftail{bdef}) of the compactified \sqcdv.

Thanks to this decoupling, the spectral curve has all the symmetries
of the 5D theory, even when the 4D quiver theory does not respect them
at the non-holomorphic level.
From the deconstruction point of view, such enhanced symmetries
of the 4D spectral curve acts as {\sl custodial symmetries} of the
5D symmetries.
For example, consider the $\bf C$ symmetry
which acts on spectral curves according to
\be
x\,\to\,{1\over x}\,,\qquad y\,\to\,{y\over(-x)^{n_c}}\,,
\label{Csym:XY}
\ee
\be
\Delta F\,\to\,2n_c-n^{}_f-\Delta F,\qquad
\varphi_j\,\to\,-\varphi_j\,,\qquad
m^{}_f\,\to\,-m^{}_f\,,
\label{Csym:4dp}
\ee
\be
S\ \to\ S\ -\,\sum_{f=1}^{n_f^{}} m_f^{}\,,
\label{Csym:S}
\ee
or equivalently
\be
H\ \eqdef\ S\ -\ \half\sum_{f=1}^{n_f^{}} m_f^{}
\quad\mbox{is invariant.}
\label{Hdef}
\ee
\par\noindent
Note that $\bf C$ is not a symmetry of the quiver theories themselves
but only of their spectral curves, or more accurately, it's a symmetry of
the family of such spectral curves with variable $m_f^{}$ and $\Delta F$
parameters (but fixed $n_c$ and $n_f$).
Nevertheless, the very existence of this 4D symmetry implies a 5D
symmetry which acts according to \eq{Csym:4dp} translated into 5D terms, namely
\be
\kcs\,\to\,-\kcs\,,\qquad
\phi_j\,\to\,-\phi_j\,,\qquad
A^\mu_j\,\to\,-A^\mu_j\,,\qquad
m^{}_f\,\to\,-m^{}_f\,.
\label{Csym:5dp}
\ee
In other words, $\bf C$ acts as a {\sl custodial symmetry of the 5D charge
conjugation symmetry}.

\medskip
\centerline{\blue\large $\star\qquad\star\qquad\star$}
\smallskip

After all these preliminaries, let us consider the decompactification
limit $\ql a\to\infty$ of the spectral curve~(\ref{SW}--\reftail{bdef}).
In this limit, the moduli $\varphi_i^{}$ and the parameters $S$ and
$m_f^{}$ remain fixed, hence the $x$ and $y$ coordinates of the curve
scale as
\be
x\ =\ \exp(\ql a\times \xi),\qquad y\ =\ \exp(\ql a\times\eta)
\label{XiEta}
\ee
for fixed $\xi$ and $\eta$.
Hence,
\be
\vcenter{\openup 5pt \ialign{%
        \hfil $\displaystyle{#(x)}$\ &
        $\displaystyle{{}\sim\ \exp\Bigl(\ql a\times O(#)\Bigr)}$\hfil &
        \quad except for $\xi$ very near one of the $#\,$,\hfil\cr
        p & \xi,\varphi & \varphi_i^{}\cr
        b & \xi,m & m_f^{}\cr
        }}
\ee
and the ratio
\be
\frac{e^{-\ql aS}\,b(x)}{p^2(x)}\
\sim\ \exp\left[ -\ql a\times\Bigl(S+O(\xi,\varphi,m)\Bigr)\right]\
\label{BPratio}
\ee
generally becomes (for $\ql a\to\infty$) either extremely large or extremely
small, depending on $S$.
For sufficiently large $S$ ($i.\,e.$, for sufficiently weakly coupled
quiver theory, {\it cf.}\ \eq{Sdef}), the exponent on the right hand
side of \eq{BPratio} is generally negative, and therefore
\be
e^{-\ql aS}\, b(x)\ \ll p^2(x)\quad
\mbox{except for}\ x\ \mbox{very near one of the}\ e^{\ql a\varphi_i^{}} .
\label{polyratio}
\ee
Consequently, the roots $x_1,x_2,\ldots,x_{2n_c}$ of the discriminant
\be
D(x)\ =\ p^2(x)\ -\ 4e^{-\ql aS}\,b(x)
\ee
cluster in tight pairs near the roots $e^{\ql a\varphi_i^{}}$ of $p(x)$.
Assuming for simplicity a generic point of the Coulomb moduli space where
all the $\varphi_i$ are distinct, we have
\be
x_{2i-1},x_{2i}\ \approx\ e^{\ql a\varphi_i^{}}\,\pm\ d_i\quad(i=1,\ldots,n_c)
\label{pairs}
\ee
where
\be
d_i\ =\ 2e^{-\ql aS/2}\times\frac{\sqrt{b(e^{\ql a\varphi_i})}}
        {p'(e^{\ql a\varphi_i})}\
\ll\ (e^{\ql a\varphi_j}-e^{\ql a\varphi_i})\ \forall j\neq i.
\label{dformula}
\ee
{}From the spectral curve's point of view, the discriminant roots
$x_1,\ldots,x_{2n_c}$ are branching points of the Riemann surface~(\ref{SW})
over the $X$ plane,
and their clustering into tight pairs means that in a suitable basis of
the abelian gauge fields of the theory, all gauge couplings are weak.
Specifically, the abelian coupling matrix is given \cite{IK1} by
\be
\left.\vcenter{
    \openup 2\jot \ialign{
        $\displaystyle{\tau_{#}}$\enspace\hfil &
        $\displaystyle{{}=\ \frac{i}{2\pi}\log #}$\hfil\cr
        i\neq j & \frac{e^{\ql a\varphi_i}\,e^{\ql a\varphi_j}}
                {(e^{\ql a\varphi_i}-e^{\ql a\varphi_j})^2}\cr
        i=j & \frac{(e^{\ql a\varphi_i})^2}{(d_i/2)^2} \cr
        }
    }\right\}\
\mbox{modulo}\ \ZZ,
\ee
and all we need to do now is to calculate the $d_i$ according to \eq{dformula}.

To avoid compactification artefacts due to finite size $\ql a$ of the
deconstructed dimension, we strengthen the assumption of the $\varphi_i$
being all distinct and assume that the differences are larger than the
compactification scale:
\be
\forall i\neq j:\ \left|\Re(\varphi_i-\varphi_j)\right|\
\gg\ \frac{1}{\ql a}\,,
\ee
hence $\forall i\neq j$
 either $e^{\ql a\varphi_i}\gg e^{\ql a\varphi_j}$ or vice verse
$e^{\ql a\varphi_i}\ll e^{\ql a\varphi_j}$.
This allows us to approximate
\be
\eqalign{
\left( e^{\ql a\varphi_i}\,-\,e^{\ql a\varphi_j}\right)\ &
\approx\ \max\left({+e^{\ql a\varphi_i}}, {-e^{\ql a\varphi_j}}\right)\cr
&=\ \pm\exp\left(\frac{\ql a}{2}\Bigl(
        \varphi_i+\varphi_j
        +\left\lfloor\varphi_i-\varphi_j\right\rfloor
        \Bigr)\right)\cr
}\label{phipprox}
\ee
where $\lfloor\varphi_i-\varphi_j\rfloor$ denotes $(\varphi_i-\varphi_j)\times
\mathop\mathrm{sign}\Re(\varphi_i-\varphi_j)$.\footnote{%
        For a real number $\alpha$, $\lfloor\alpha\rfloor\equiv|\alpha|$,
        but for complex numbers we need a new notation.
        However, for any complex number $\beta$,
        $\Re(\lfloor\beta\rfloor)=\left|\Re(\beta)\right|$.
        }
Likewise, we assume no coincidences between the moduli $\varphi_i$
and the masses $m^{}_f$, hence
\be
\left( e^{\ql a\varphi_i}\,-\,e^{\ql am^{}_f}\right)\
\approx\ \pm\exp\left(\frac{\ql a}{2}\Bigl(\varphi_i+m^{}_f+
                \left\lfloor\varphi_i-m^{}_f\right\rfloor\Bigr)\right).
\label{mapprox}
\ee
Thanks to approximations (\ref{phipprox}) and (\ref{mapprox}), we have
\bea
\tau_{i\neq j}(\varphi) &\approx &
\frac{-i\ql a}{2\pi}\times\left\lfloor\varphi_i\,-\,\varphi_j\right\rfloor ,
        \label{quivertauij} \\
e^{\ql a\varphi_i}\times p'(e^{\ql a\varphi_i}) &\approx &
\pm \exp\left(
        \frac{\ql a}{2}\left( n_c\, \varphi_i\
        +\,\sum_k\lfloor\varphi_i-\varphi_k\rfloor
        \right)\right), \\
b(e^{\ql a\varphi_i}) &\approx & \pm\exp\left(
        \frac{\ql a}{2}\left(
             (2\Delta F+n^{}_f)\,\varphi_i\,+\sum_f m_f^{}\,
             +\,\sum_f\lfloor\varphi_i-m^{}_f\rfloor
             \right)
        \right),\qquad\\
\noalign{\smallskip\noindent hence\par}
\tau_{i=j}(\varphi) &\approx&
\frac{i}{2\pi}\log\frac{(e^{\ql a\varphi_i}\, p'(e^{\ql a\varphi_i}))^2}
        {e^{-\ql aS}\, b(e^{\ql a\varphi_i})} 
        \vrule width 0pt depth 20pt\label{quivertauii} \\
&\approx & \frac{i\ql a}{2\pi}\times\left(\eqalign{
        H\ &+\ \left( n_c\,-\,\Delta F\,-\,\frac{n^{}_f}{2}\right)\times\varphi_i \cr
        &+\,\sum_k\left\lfloor\varphi_i-\varphi_k\right\rfloor\ 
                -\ \frac12\sum_f\left\lfloor\varphi_i-m^{}_f\right\rfloor\cr
        }\right)\ \mbox{modulo}\ \frac{\ZZ}{2}\,,\qquad \nonumber\\
\noalign{\smallskip\noindent where\par}
H & \eqdef & S\ -\ \half\sum_{f=1}^{n_f^{}}m_f^{}\,, \label{Hdef2}
\eea
and all approximations become exponentially good in the large quiver
limit $\ql \to\infty$.

According to eqs.\ (\ref{quivertauij}) and (\ref{quivertauii}),
the entire abelian coupling matrix $\tau_{ij}(\varphi)$ of the quiver
is proportional to $\ql a$, exactly as promised in \eq{g45}.
Furthermore, the deconstructed 5D gauge couplings of the 
\sqcdv's Coulomb branch
\bea
\left[\frac{8\pi^2}{g_5^2}\right]_{i\neq j} &=&
-\left|\phi_i-\phi_j\right| \label{deconGij} \\
\noalign{\noindent and\par\smallskip}
\left[\frac{8\pi^2}{g_5^2}\right]_{i=j} &=&
\Re(H)\ +\ \kcs\times\phi_i\ +\,\sum_k|\phi_i-\phi_k|\
        -\ \frac12\sum_f|\phi_i-\Re(m^{}_f)| \label{deconGii}\qquad
\eea
are consistent with a prepotential~\cite{seiberg, MS}:
\bea
\left[\frac{1}{g_5^2}\right]_{ij} &=&
\frac{\partial^2{\cal F}}{\partial\phi_i\,\partial\phi_j},
\label{GFdeco}\\
8\pi^2{\cal F} &=&
\sum_{i=1}^{n_c}\left({h\over 2}\,\phi_i^2\,+\,{\kcs\over6}\,\phi_i^3\right)\
        +\ {1\over12}\sum_{i,j=1}^{n_c}\left|\phi_i-\phi_j\right|^3\
        -\ {1\over12}\sum_{i=1}^{n_c}\sum_{f=1}^{n_f}\left|\phi_i-m_f\right|^3.\qquad
\label{Fdeco}
\eea
This indicates SUSY extension from 4 supercharges in 4D to 8 supercharges
in the continuum limit of the fifth dimension,
which is a major ingredient of dimensional deconstruction.\footnote{%
        By themselves, eqs.~(\ref{GFdeco}) do not prove SUSY extension
        in the continuum limit.
        A complete proof would require calculating the K\"ahler function
        of the moduli fields and checking that the $\phi_i$ have the
        same metric as the gauge fields, and also that the Higgs moduli have
        the right metric and the right speed of light in the deconstructed
        dimension.
        Alas, we don't know how to calculate the non-perturbative K\"ahler
        function, so all we can do is hope that we may fine-tune it to
        agree with the extended SUSY.
        On the other hand, the gauge couplings are exactly calculable
        and cannot be fine-tuned, and that's why checking eqs.~(\ref{GFdeco})
        is so important:
        If there is a prepotential, we may fine-tune the K\"ahler function
        to complete the SUSY extension; but if there is no prepotential,
        fine-tuning would not help.
        }
And most importantly, {\blue\it the deconstructed prepotential~(\ref{Fdeco})
is exactly as in \eqrange{Fexact}{F1loop} for the
un-deconstructed \sqcdv} (with a `stringy' UV regulator which preserves
all 8 supersymmetries), provided we identify the tree-level
Chern--Simons coefficient as $\kcs=n_c-\frac12 n^{}_f-\Delta F$
according to \eq{Flavors}, and the
tree-level inverse gauge coupling $h=(8\pi^2/g_5^2)$
as $h=\Re(H)$.\footnote{%
        Note that the tree level gauge coupling is invariant
        under the charge conjugation $\bf C$.
        That's why the same $H$ appears in both eqs.~(\ref{Hdef})
        and~(\ref{Hdef2}).
        }

We conclude this section by establishing the limits (if any) of the $h$
parameter space of the deconstructed \sqcdv.
Going back to \eq{Sdef}, we see that for $F<2n_c$, $S$ --- and hence $H$ ---
depends on $V$, and according to \eq{Vqual} $V$ suffers from non-perturbative
corrections when $\Delta F=0$.
In the large quiver limit, \eq{Vformula} becomes
\be
V\ \becomes{\ql \to\infty}\ \max(v_1,v_2),
\label{Vlimit}
\ee
hence according to \eq{v1v2}
\be
|V|^{2n_c}\ \geq\ \left|v_1^{n_c}\times v_2^{n_c}\right|\
=\ \left|\Lambda^{2n_c-F}\gamma^F\,\mu_1\mu_2\cdots\mu_F\right|
\ee
regardless of the tree-level parameter $v$, and therefore for $\Delta F=0$,
\be
\left|\frac{V^{2n_c-F}}{\gamma^F\Lambda^{2n_c-F}}\right|\
\geq\ \prod_{f=1}^{F=n_f}\left|\frac{\mu_f^{}}{V}\right|\
\Longrightarrow\ \Re(S)\ \geq\,\sum_{f=1}^{n_f^{}}m_f^{}\,.
\ee
Hence, in light of eqs.~(\ref{Flavors}) and (\ref{Hdef2}),
\be
\blue\mbox{for}\ \kcs\,=\, +\left(n_c\,-\,\coeff12n^{}_f\right),\quad
h\,=\,\Re(H)\ \geq\ \frac12\sum_{f=1}^{n^{}_f}\Re(m^{}_f) .\black
\label{Hlimit1}
\ee
On the other hand, for $\Delta F>0$ there is no lower limit on the magnitude
of $V$; instead $V=v$ without any quantum corrections whatsoever.
Hence, allowing for arbitrarily large or small $\Lambda/v$ ratios, we find
that
\be
\blue \mbox{for}\ |\kcs|\ <\ \left(n_c\,-\,\coeff12n^{}_f\right),\quad
\mbox{all}\ h\,>\,-\infty\ \mbox{are allowed}.
\black\label{Hnolimit}
\ee
Finally, for $F=2n_c$ \eq{Sdef} does not depend on $V$ but only on the
asymptotically-finite coupling $g^2$ of the 4D gauge theory $[SU(n_c)]^\ql $,
hence $\Re(S)\geq0$ and therefore,
\be
\blue \mbox{for}\ \kcs\ =\ -\left(n_c\,-\,\coeff12n^{}_f\right),\quad
h\ \geq\ -\frac12\sum_{f=1}^{n^{}_f}\Re(m^{}_f) .
\black\label{Hlimit2}
\ee
Note that despite completely different 4D origins of
the limits (\ref{Hlimit1}) and (\ref{Hlimit2}), in 5D these limits are related
to each other by the charge conjugation~$\bf C$.

Formul\ae\ (\ref{Hlimit1}--\reftail{Hlimit2}) describe
the complete range of the $h$ parameter of deconstructed \sqcdv\
for different values of quark masses and Chern--Simons levels.
However, for
\be
h\ <\ h_0\ =\ \frac12\sum_{f=1}^{n^{}_f}\left|\Re(m^{}_f)\right|
\label{h0}
\ee
the 5D inverse gauge couplings (\ref{deconGij}--\reftail{deconGii}) become
negative at the origin $\phi_i\equiv 0$ of the dynamical moduli space.
Physically, this is quite impossible from both 5D and 4D points of view;
indeed, the very existence of a spectral curve such as (\ref{SWcurve})
guarantees that the $\Im\tau_{ij}(\varphi)$ matrix is positive definite%
\footnote{%
        By abuse of terminology, we call the $n_c\times n_c$ matrix
        $\Im\tau_{ij}$ ``positive definite'' when we mean
        $$
        \sum_{i,j}\Im\tau_{ij}\,F_iF_j\ >\ 0\quad\forall F_i\
        \mbox{such that}\ \sum_i F_i\ =\ 0.
        $$
        }
for all moduli $\varphi_i$.
In 4D, this apparent paradox goes away when we remember that before deriving
\eqrange{deconGij}{deconGii} we assumed a ``sufficiently large'' $\Re(S)$
to assure that the branching points $x_1,\ldots,x_{2n_c}$
of the Riemann surface~(\ref{SW}) come in close pairs~(\ref{pairs}), and
it is precisely this assumption which fails for $h\leq h_0$.
(Indeed, the inequalities~(\ref{dformula}) fail precisely when
the right hand sides of eqs.~(\ref{deconGii}) are no longer positive.)
In the $h<h_0$ regime, the branching points $x_1,\ldots,x_{2n_c}$
are arranged differently, and in \S6 we shall see
how such re-arrangements correspond to different phases
of the 5D theory, separated from the ``ordinary \sqcdv'' phase
by flop transitions~\cite{PhasesW} in the parameter/moduli space.

%% file: chapter5.tex
%auto-ignore
%
% chapter 5 of the DESQCD paper
%
\section{Quantum Baryonic Branches}
In stringy UV completions of 5D SYM theories with $\kcs=0$
($\theta=0$ for $SU(2)$) there is a peculiar Higgs branch, which connects
to the Coulomb branch at the superconformal point $h=0,\ \phi_i=0\,\forall
i$~\cite{MS, AH}.
In this section we shall see that the deconstructed $\rm SYM_5$ --- as well
as some \sqcdv\ theories --- have similar Higgs branches.
In quiver terms, they are exotic baryonic branches, where by exotic we mean
that the baryonic VEVs involve 4D flavors with $\mu=0$ instead of
the 5D flavors with $\mu=Ve^{am}$.
Classically, such exotic branches do not exist, and even in the quantum
theory they show up only at a particular value of the coupling parameter $H$,
namely $H=0$ for $n_f=0$ and $\Delta F=n_c$.

But before we delve into quantum baryonic branches of the quiver, let
us briefly review the Higgs branches of 5D SYM from the stringy point of view.
A rather graphic picture of such branches obtains  via the $(p,q)$
5--brane web construction of type~IIB superstring~\cite{AHK}.\footnote{%
        Briefly, there is a bunch of 5--branes, each spanning the
        5D coordinates $X^0,\ldots,X^4$ and a segment of a real straight
        line in the $(X^5,X^6)$ plane; the segments form a web.
        A brane with $\rm(D,NS)$ charges $(p,q)$ is oriented
        according to $X^5+iX^6=(p+\tau_s q)\times\rm real+const$;
        where $\tau_s$ is the complex type~IIB coupling; this condition
        provides for 8 unbroken supercharges in 5D.
        The brane joints in the web are governed by the zero-force
        condition, which is equivalent to the $(p,q)$ charge conservation.
        Please see \cite{AHK,LV,KR2} for more details.
        }
For $h>0$ the $SU(n_c)$ SYM webs look like
\be
\psset{unit=0.4cm,linewidth=0.2,linecolor=blue}
\begin{pspicture}[0.5](-18,-5)(+18,+5)
\rput(-10,0){%
    \psline(-8,-5)(-6,-3)(-5,-1)(-5,+1)(-6,+3)(-8,+5)
    \psline(+8,-5)(+6,-3)(+5,-1)(+5,+1)(+6,+3)(+8,+5)
    \psline(-6,-3)(+6,-3)
    \psline(-5,-1)(+5,-1)
    \psline(-5,+1)(+5,+1)
    \psline(-6,+3)(+6,+3)
    }
\rput(+10,0){%
    \psline(-8,-5)(-3.5,-0.5)(-3.35,-0.2)(-3.35,+0.2)(-3.5,+0.5)(-8,+5)
    \psline(+8,-5)(+3.5,-0.5)(+3.35,-0.2)(+3.35,+0.2)(+3.5,+0.5)(+8,+5)
    \psline(-3.5,-0.5)(+3.5,-.5)
    \psline(-3.35,-0.2)(+3.35,-.2)
    \psline(-3.35,+0.2)(+3.25,+.2)
    \psline(-3.5,+0.5)(+3.5,+.5)
    }
\end{pspicture}
\label{BaB:BraneWebs}
\ee
where the left web corresponds to the Coulomb branch with distinct $\phi_i$
and the right web to the unbroken $SU(n_c)$ point for $\phi_i=0\,\forall i$:
a stack of $n_c$ coincident horizontal brane segments
gives rise to non-abelian $SU(n_c)$ gauge symmetry.
The inverse 5D coupling~$h$ is proportional to the length of the stack, which
depends on relative positions of the external legs.
The directions of those external legs depend on the Chern--Simons level:
for $\kcs=0$ the legs diagonally across from each other are parallel, and
for $h=0$ they line up in straight lines.
Hence, when all the brane boxes in the middle of the web
collapse to a point for $\phi_i=0\,\forall i$, the
external legs can re-connect as two intersecting infinite branes: 
\be
\psset{unit=0.4cm,linewidth=3pt,linecolor=blue}
\begin{pspicture}[](-17,-5)(+17,+5)
\rput(-12,0){%
    \psline(-5,-5)(-3,-3)(-2,-1)(-2,+1)(-3,+3)(-5,+5)
    \psline(+5,-5)(+3,-3)(+2,-1)(+2,+1)(+3,+3)(+5,+5)
    \psline(-3,-3)(+3,-3)
    \psline(-2,-1)(+2,-1)
    \psline(-2,+1)(+2,+1)
    \psline(-3,+3)(+3,+3)
    }
\psline[linecolor=black,linewidth=1.5pt,arrowscale=1.8]{->}(-8,0)(-4,0)
\rput(0,0){%
    \psline(-5,-5)(0,0)(-5,+5)
    \psline(+5,-5)(0,0)(+5,+5)
    \pspolygon[fillcolor=red,fillstyle=solid](-.3,-.3)(+.3,-.3)(+.3,+.3)(-.3,+.3)
    }
\psline[linecolor=black,linewidth=1.5pt,arrowscale=1.8]{->}(+4,0)(+8,0)
\rput(+12,0){
    \psline(-5,-5)(+5,+5)
    \psline[border=2pt](-5,+5)(+5,-5)
    }
\end{pspicture}
\label{BaB:BraneWebSplit}
\ee
Pulling the two reconnected branes apart (in a direction perpendicular to
both branes) corresponds to the Higgs branch of the 5D SYM.
Note that this branch exists only for $\kcs=0$: for other Chern--Simons
levels the external legs  cannot reconnect because they are not parallel.

In M theory construction of 5D SYM theories there are similar Higgs branches
for $\kcs=0$ only (or $\theta=0$ for $SU(2)$).
To obtain an $SU(n_c)\ \rm SYM_5$, we compactify M theory on a Calabi--Yau
sixfold with a conical singularity of the CY, where the cone's base is
a $Y^{p,q}$ Sazaki--Einstein space with $p=n_c$ and $q=\kcs$~\cite{GMSW, HKW}.
Or rather, the superconformal point obtains from such a singularity
while the Coulomb branch corresponds to resolutions of
its K\"ahler structure.
For $q=0$ (and only $q=0$) we may also deform the complex structure
of the singularity, which requires keeping the K\"ahler structure unresolved;
in field theory this corresponds to a Higgs branch with frozen Coulomb
parameters/moduli $h=0$ and $\phi_i=0\,\forall i$.
The reason cones over $Y^{p,0}$ are special in this way is that they are
$\ZZ_p$ orbifolds of the conifold, whose complex structure has one deformation
mode; this mode is $\ZZ_p$ invariant $\forall p$,
and so it's inherited by the orbifolds.
Other cones with $q\neq0$ have rigid complex structures which cannot
be deformed; consequently, SYM theories with $\kcs\neq0$ do not
have Higgs branches. 

The goal of this section is to deconstruct the Higgs branch of SYM with
$\kcs=0$ --- as well as similar Higgs branches in some strongly-coupled
\sqcdv\ theories --- via exotic baryonic branches of the quiver theory.
But first, we need to look at the ordinary baryonic branches of a quiver
with generic $n_c$, $n_f$, and $\Delta F$.
As discussed at the end of \S2, classical baryonic branches have squark
and antisquark VEVs as in \eq{BaryonicVEVs} (modulo a flavor symmetry),
and their existence requires $n_c$ zero modes
in the quarks' mass matrix, thus \eq{BaryonicOrigin} for the
Coulomb moduli $\varphi_i$.
Actually, (\ref{BaryonicOrigin}) is an overdetermined system of
$n_c$ equations for $n_c-1$ independent moduli,
hence a baryonic branch exists only if $m_1+\cdots+m_{n_c}=0$
(modulo $2\pi i/\ql a$), or in 4D terms, if
$\mu_1^\ql\times\cdots\times\mu_{n_c}^\ql=V^{\ql n_c}$.
Of course this is all modulo a flavor symmetry, so in general a baryonic
branch with flavors $f_1,\ldots,f_{n_c}$ exists if and only if
\be
\prod_{f\in\rm Baryon}\mu^\ql_f\ =\ V^{\ql n_c}.
\label{BaB:Classical}
\ee
Obviously this condition excludes 4D flavors with $\mu=0$, so classical
baryonic branches involve only the 5D flavors with $\mu=Ve^{am}\neq0$.

In the quantum theory we cannot look at individual squark VEVs;
instead, we analyze VEVs of gauge-invariant chiral operators.
In our 4D paper~\cite{KSdN} we found that the {\it off-shell} chiral ring
of the quiver theory contains a whole zoo of baryon-like operators, but
in the {\it on-shell} ring, they are all related to each other via equations
of motion (AKA Konishi anomaly equations).
Consequently, for each choice of $n_c$ distinct flavors, there is at most
one independent baryonic VEV (and likewise, one antibaryonic VEV).
And similar to the classical theory, such (anti)baryonic VEVs over-determine
the Coulomb moduli, although \eq{BaB:Classical} is corrected by instanton
effects.
The exact constraint is best stated in terms of the spectral
curve~(\ref{SWcurve}) of the quiver: A baryonic branch exists when the
curve has no branch cuts at all; instead, it factorizes according to
\be
Y^2\,-\,P(X)\times Y\,+\,(-1)^F\alpha B(X)\ 
=\ \Bigl(Y\,-\,B_1(X)\Bigr)\times\Bigl(Y\,-\,(-1)^F\alpha B_2(X)\Bigr)\ =\ 0
\label{BaB:curve}
\ee
where
\be
B_1(X)\ =\ \prod_{f\in\rm Baryon}(X-\mu^\ql)\quad
\mbox{of degree}\ =\ n_c
\label{BaB:Factor1}
\ee
encodes masses of flavors involved in the baryonic VEV, and
\be
B_2(X)\ =\ \prod_{f\not\in\rm Baryon}(X-\mu^\ql)\quad
\mbox{of degree}\ =\ F\,-\,n_c
\label{BaB:Factor2}
\ee
encodes masses of the remaining flavors.
Factorization of the spectral curve implies that the link resolvent $T(X)$
({\it cf.}\ \eqrange{resolvent}{resolventeq}) has poles instead of
branch cuts:
\be
T(X)\ \eqdef\ \vev{\Tr\frac{1}{X\,-\,\link_\ql\cdots\link_1}}\
=\ {1\over Y}{\partial Y\over\partial X}\
=\ \left\{\vcenter{\openup 1\jot \ialign{
    $\displaystyle{\sum_{#}\frac{1}{X\,-\,\mu^\ql_f}}$\quad\hfil &
    on the #\hfil\cr
    f\in\rm Baryon & physical sheet,\cr
    f\not\in\rm Baryon & unphysical sheet.\cr
    }}\right. %\}
\label{BaB:Tres}
\ee
Note that the poles differ between the two Riemann sheets, and consequently
the Coulomb moduli $\varpi_i=V^\ql e^{\ql a\varphi_i}$ do not match the
poles; instead, \eq{BaB:curve} implies
\be
P(X)\equiv\prod_{i=1}^{n_c}(X-\varpi_i)\
=\ B_1(X)\ +\ (-1)^F\alpha\times B_2(X).
\label{BaB:Pcon}
\ee
Physically, the first term on the right hand side reproduces the classical
\eq{BaB:Classical} for the moduli of a baryonic branch, while the second
term is the quantum correction; it arises at the one-diagonal-instanton level
(one instanton in the $SU(n_c)_{\rm diag}$,
$i.\,e.$ one instanton in each $SU(n_c)_\ell$ factor).
Eq.~(\ref{BaB:Tres}) over-determines the moduli: in order to maintain
the product constraint~(\ref{ProductConstraint}), the masses must satisfy
\be
\prod_{f\in\rm Baryon}\mu_f^\ql\
+\ (-1)^{F}\alpha\times\prod_{f\not\in\rm Baryon}\mu_f^\ql\
=\ V^{\ql n_c} .
\label{BaB:Masses}
\ee
Unlike its classical analogue~(\ref{BaB:Classical}), this formula involves
masses of all $F$ flavors rather than just the $n_c$ flavors involved
in the baryonic VEVs.
In particular, it involves flavors with $\mu=0$, if any, and this gives
rise to two distinct types of baryonic branches: they either contain
none of the $\mu=0$ flavors, or else they contain all of them at once.

Ordinary baryonic branches are of the first type: all $n_c$ flavors
have $\mu=Ve^{am}\neq 0$ and are visible in 5D.
For $\Delta F>0$, \eq{BaB:Masses} for such branches reduces to the classical
formula~(\ref{BaB:Classical}), or in 5D terms $\sum_{f\in\rm Baryon}m_f=0$.
However, the moduli $\varphi_i$ suffer quantum corrections according to
\be
\prod_{i=1}^{n_c}\left(x\,-\,e^{\ql a\varphi_i}\right)\
=\ \prod_{f\in\rm Baryon}\left(x\,-\,e^{\ql am_f}\right)\
  +\ (-1)^Fe^{-\ql aS}\times x^{\Delta F}\times\
    \prod_{f\not\in\rm Baryon}\left(x\,-\,e^{\ql am_f}\right).
\label{BaB:phicon1}
\ee
In the decompactification limit $\ql a\to\infty$, the second term 
on the right hand side becomes negligible if the coupling is weak enough;
in this regime we recover the classical constraint~(\ref{BaryonicOrigin}).
On the other hand, for strong coupling the second term remains important
even in the 5D limit; we shall see an example of such quantum shift
of the baryonic branch in ~\S6.3.

Baryonic branches of the second type are exotic:
they involve $\mu=0$ flavors, indeed all $\Delta F$ of them,
plus $(n_c-\Delta F)$ 5D flavors to complete the baryon.
Clearly, such branches require
$F=n_f+\Delta F\geq n_c$ but $\Delta F\leq n_c$;
in 5D terms this amounts to a constraint on the Chern--Simons level:
\be
\begingroup\blue
\mbox{Exotic baryons exist only for}\ |\kcs|\ \leq {n_f\over2}\,.
\endgroup
\label{BaBE:kcs}
\ee
In particular, {\blue in SYM theories with $n_f=0$, the exotic baryonic
branches exist only for $\kcs=0$ ($\theta=0$ for $n_c=2$).}

For an exotic baryonic branch, \eq{BaB:Masses} depends on the
coupling $\alpha$ as much as on the masses $\mu_f$;
solving it for $\alpha$ gives us
\be
(-1)^F\alpha\
=\ \left[\frac{V^{n_c}}{\prod\limits_{f\not\in\rm Baryon}\mu_f^{}}\right]^\ql ,
\ee
or in terms of $S$ and $H$ parameters
({\it cf.}\ eqs.~(\ref{Sdef}) and~(\ref{Hdef}))
\be
S\ =\,\sum_{f\not\in\rm Baryon}^{\rm 5D\ only} m_f\ \Longrightarrow\
H\ =\,\sum_{f\not\in\rm Baryon}^{\rm 5D\ only}{m_f\over2}\
-\,\sum_{f\in\rm Baryon}^{\rm 5D\ only}{m_f\over2}\,.
\label{BaBE:H}
\ee
In particular, for a SYM theory with $\kcs=0$, the exotic branch requires
$H=0$.
For other values of the coupling $H$ --- either too weak or too strong ---
exotic baryons cannot develop VEVs.
This behavior is in perfect agreement with the Higgs branch of the brane
web~(\ref{BaB:BraneWebSplit}): To pull the web apart in the out-of-plane
direction, the external legs of the web must be perfectly aligned,
which means $h=0$ exactly, no more and no less.

The $\varphi_i$ moduli of an exotic baryonic branch are constrained by
\be
\prod_{i=1}^{n_c}\left(x\,-\,e^{\ql a\varphi_i}\right)\ 
=\ x^{\Delta F}\times\prod^{\rm 5d}_{f\in\rm Baryon}
        \left(x\,-\,e^{\ql am_f^{}}\right)\
+\ (-1)^F\,e^{-\ql aS}\prod^{\rm 5d}_{f\not\in\rm Baryon}
        \left(x\,-\,e^{\ql am_f^{}}\right),
\ee
or in light of \eq{BaBE:H},
\be
\prod_{i=1}^{n_c}\left(x\,-\,e^{\ql a\varphi_i}\right)\ 
=\ x^{\Delta F}\times\prod^{\rm 5d}_{f\in\rm Baryon}
        \left(x\,-\,e^{\ql am_f^{}}\right)\
  +\ (-1)^F\prod^{\rm 5d}_{f\not\in\rm Baryon}
        \left(x\times e^{-\ql am_f^{}}\,-\,1\right).
\label{BaBE:phicon2}
\ee
Note that both terms on the second line here are of comparable magnitudes
and both remain important in the decompactification limit.
In particular, for $n_f=0$ and $\Delta F=n_c$ (the SYM theory with $\kcs=0$)
the first term becomes $x^{n_c}$ while the second term is $\pm1$.
Together, they provide for
\be
\prod_{k=1}^{n_c}\left(x\,-\,e^{\ql a\varphi_k}\right)\ =\ x^{n_c} \pm\ 1\
\Longrightarrow\ \ql a\varphi_k\ =\ 2\pi i\times k\quad\mbox{or}\quad
2\pi i\times(k-{1\over2}),\quad k=1,\ldots,n_c,
\ee
or in 5D terms, all $\phi_k=0$.
Together with the $h=0$ condition, this brings us to the superconformal
point where the exotic baryonic Higgs branch meets the Coulomb branch.
Again, this is in perfect agreement with the string theory:
in terms of the brane web~(\ref{BaB:BraneWebSplit}), aligning the external legs
(setting $h=0$) is not enough, one must also collapse all the cycles
(by setting all $\phi_k=0$) before pulling the web apart ($i.\,e.$,
turning on a hypermultiplet VEV).

In \cite{KSdN} we rejected the exotic baryonic branches because of
their link resolvents: according to \eq{BaB:Tres}, when a baryonic
branch involves a $\mu=0$ flavor, $T(X)$ has a pole at $X=0$ on the
physical sheet.
This implies that the product $\link_\ql\cdots\link_1$ of link fields
has a zero eigenvalue --- or rather $\Delta F$ zero eigenvalues, judging
by the residue of the pole at $X=0$ --- and we thought that to be
impossible since all the $\link_\ell$ matrices are invertible.
In retrospect, that was a mistake.

Indeed, in the quantum theory $\vev{\det(\link_\ql\cdots\link_1)}\neq
\prod_\ell\vev{\det(\link_\ell)}$, and in \cite{KSdN}
we have actually calculated the quantum corrections to
$\vev{\det(\link_\ql\cdots\link_1)}$ arising from
instantons in the diagonal $SU(n_c)_{\rm diag}$
as well as instantons individual $SU(n_c)_\ell$ factors.
But somehow, we overlooked the possibility that the quantum corrections
may cancel the classical contribution and lead to
$\vev{\det(\link_\ql\cdots\link_1)}=0$ despite invertibility of the individual
$\link_\ell$ matrices.
For an example of such cancellation, consider a quiver with $F=n_c$,
in which case
\be
\vev{\det\Bigl(X\,-\,\link_\ql\cdots\link_1\Bigr)}\
=\ P(X)\ -\ (-1)^{n_c}\times\alpha
\ee
({\it cf.}\ \cite{KSdN} for details), and hence
\be
\vev{\det(\link_\ql\cdots\link_1)}\ =\ V^{\ql n_c}\ -\ \alpha\
=\ V^{\ql n_c}\times\left(1\,-\,e^{-\ql aH}\right)\
\to\ 0\quad\mbox{for}\ H=0.
\ee
More generally, for a quiver with $F>n_c$ the characteristic
polynomial of the link product is given by
\be
\vev{\det\Bigl(X\,-\,\link_\ql\cdots\link_1\Bigr)}\
=\ \mathop{\rm Polynomial\ part\ of}\left[
        \frac{P(X)\,+\,\sqrt{P^2(X)-4(-1)^F\alpha B(X)}}{2}\right] .
\label{BaB:DetCorr}
\ee
When the quiver has a baryonic branch ---ordinary or exotic ---
and the spectral curve factorizes according to \eq{BaB:curve},
\eq{BaB:DetCorr} reduces to
\be
\vev{\det\Bigl(X\,-\,\link_\ql\cdots\link_1\Bigr)}\
=\ \cases{B_1(X) & on the physical sheet,\cr
        (-1)^F\alpha B_2(X) & on the unphysical sheet,\cr
        }
\label{BaB:determinant}
\ee
in perfect agreement with the link resolvent~(\ref{BaB:Tres}) of
a baryonic branch.
In particular, for an exotic branch,
the determinant~(\ref{BaB:determinant}) has precisely $\Delta F$ zero
eigenvalues, which agrees with $T(X)$ having a pole of residue $\Delta F$
at $X=0$ on the physical sheet.

To summarize, quiver theories with $\Delta F\leq n_c\leq \Delta F+n_f$
(which correspond to \sqcdv\ with $|\kcs|\leq{n_f\over2}$) have exotic
baryonic branches involving $\mu=0$ flavors.
Such branches exist only for specific values of the coupling $H$
as well as of the Coulomb moduli $\varphi_i$; in particular, for
the SYM with $\kcs=0$ (or $\theta=0$ for $n_c=2$), the exotic baryonic
branch grows out of the 5D superconformal point $h=0,\ \phi_i=0\,\forall i$.
In string implementations of the same 5D SYM, the existence of a Higgs
branch at the superconformal point is well-known, but its baryonic nature
is a novel result.

%% file: chapter6.tex
%auto-ignore
%
% chapter 6 of the DESCQD paper
%
\section{$SU(2)$ Examples of Phase Structures \brk and Flop Transitions.}
In this section we present four examples of deconstructive quiver theories
and study their phases for $h={8\pi^2\over g^2_{\rm 5d}}>0$ and $h<0$.
In 4D terms, negative $h$ means $|\Lambda|>|V|$ and hence strongly coupled
$SU(2)_\ell$ factors of the quiver theory at the $\rm 4D\to5D$ threshold $1/a$.
However, thanks to  unbroken $\NN=1$ SUSY in 4D, the holomorphic spectral
curve~(\ref{SW}) remains non-perturbatively exact despite the strong coupling,
and thus may be used for deconstructing the exotic phases of 5D theories.

Our presentation here is quite detailed, which makes for a rather looooong
section.
So let us state our main result upfront:
{\blue in all examples, the deconstructed \sqcdv\ has exactly the same
phase diagram as the string-theoretical UV completion of the same 5D theory}.
And now, the readers who are not interested in technical details may skip
over many formulae in this section and just look at the phase diagrams
themselves; they appear on pages \pageref{D0t:Diagram3}, \pageref{D0:Diagram2},
\pageref{D2:Diagram2}--\pageref{D2:Diagram4}, \pageref{D2:Diagram6},
\pageref{D2t:CoulombPhases}, and \pageref{D2t:HiggsPhases}.

For simplicity, all our examples have $n_c=2$ and hence Coulomb branches with
only one abelian gauge coupling, and also only one independent modulus
$\varphi=\varphi_2=-\varphi_1$; without loss of generality we assume
$\phi\equiv\Re\varphi>0$.
In 4D, the quiver theory with $n_c=2$ has an elliptic rather than hyperelliptic
spectral curve~(\ref{SW}), with four branching points $x_1,x_2,x_3,x_4$
located at roots of the quartic equation
\be
D(x)\ \equiv\  \bigl(x^2\,-\,2\cosh(\ql a\varphi)\times x\,+\,1\bigr)^2\,
-\,4(-1)^F e^{-\ql aS}\times x^{\Delta F}
\prod_{f=1}^{n_f}\left(x-e^{\ql am_f}\right)
=\ 0.
\label{FourBP}
\ee
The 4D abelian gauge coupling $\tau$ --- or rather its invariant
\be
j(\tau)\ \equiv\ j\left(\textstyle{a\tau+b\over c\tau+d}\right)\
=\ e^{-2\pi i\tau}\ +\ 744\ +\ \mbox{a convergent power series in}\
	e^{+2\pi i\tau}
\label{Jdef}
\ee
under electric-magnetic duality ---
follows from the cross-ratio
\be
\chi\ =\ \frac{(x_1-x_4)(x_2-x_3)}{(x_1-x_2)(x_3-x_4)}
\label{CrossRatio}
\ee
of the branching points according to
\be
j(\tau)\ =\ -256\,\frac{(\chi^2+\chi+1)^3}{\chi^2(\chi+1)^2}\,.
\label{JCR}
\ee
A finite 5D coupling corresponds to $\Im\tau\propto \ql a$ and hence exponentially
large $j$; in terms of the cross ratio~(\ref{CrossRatio}), this means
$\chi\to0$, $\chi\to-1$, or $\chi\to\infty$, with
\be
\Re\log\left({16\over\chi}\,,\ {16\over\chi+1}\,,\ \mbox{or}\ (16\chi)\right)\
=\ {\ql a\times\fdc\over2}\qquad\mbox{where}\quad
\fdc\,\eqdef\,\frac{8\pi^2}{g^2_{5d}[U(1)]}\,.
\label{Finite5Dcoupling}
\ee

Our first two examples have $n_f=0$ in 5D, but the 4D quivers do have
$\mu=0$ quarks, $\Delta F=1$ in the first example (\S6.1)
and $\Delta F=2$ in the second example (\S6.2).\footnote{%
	There is also a distinct theory with $\Delta F=0$;
	it was studied in much detail in \cite{IK1}.
	But this theory has $h\ge 0$ only, and it does not have
	any phase transitions at all.
	}
In 5D, both examples yield SYM theories for $h>0$;
according to \eq{VacuumAngle}, the first SYM has vacuum angle $\theta=\pi$
while the second has $\theta=0$.
But we shall see that the $h<0$ regimes of the two examples are very
different:
the $\Delta F=1$ model has two distinct Coulomb phases  --- the SYM phase
and the $E_0$ phase --- separated by a flop transition,
while the $\Delta F=2$ model has only one Coulomb phase, but it also has
a Higgs phase (which we deconstruct as the exotic baryonic branch).
In string theory constructions, these two 5D theories are properly known as
the $D_0$ (for $\theta=0$)   and the $\tilde D_0$ (for $\theta=\pi$), although
they are often called $E_1$ and $\tilde E_1$ after their respective
superconformal limits at $h=\phi=0$ \cite{MS}.
In this article however, we call them $D_0$ and $\tilde D_0$ because we
focus on deconstructing the non-conformal Coulomb and Higgs phases of
the two theories.

In \S6.3 we present two more $SU(2)$ models, this time with two quark
flavors in 5D.
For simplicity, we restrict our analysis to equal masses (modulo sign) for
the two flavors.
The first model has $m_1=-m_2$ and $\Delta F=1$ while the second has
$m_1=m_2$ and $\Delta F=0$;
we present them together in \S6.3 because their spectral curves are dual to
each other.
However, the quiver theories themselves are not dual, and even their
moduli spaces are not quite dual.
In particular, the $\Delta F=1$ model has more Higgs branches than the
$\Delta F=0$ model.

%Again, there is a superconformal limit (called $E_3$ in Seiberg's notations
%\cite{seiberg,MS})
%for $h=m_1=m_2=\phi=0$, but we focus on the non-conformal 5D physics
%elsewhere in the moduli space, Coulomb or Higgs.

%%%%%%%%%%%%%%%%%%%%%%%%%%%%%%%%%%%%%%%%%%%%%%%%%%%%%%%%%%%%%%%%%%%%%
% section 5.1
% D_0-tilde model.
%
\subsection{The $\widetilde D_0$ Model: $n_c=2$, $n_f=0$, $\Delta F=1$.}
We begin with the flavorless $SU(2)$ model with $\theta=\pi$,
or in 4D quiver terms, $n_c=2$, $n_f=0$, and $\Delta F=1$.
The spectral curve of this model has branching points at roots of
the discriminant
\be
D(x)\ 
\equiv\ \bigl(x^2-2\cosh(\ql a\varphi)\times x+1\bigr)^2\ +\ 4e^{-\ql aH}\times x\
=\ 0.
\label{D0t:Disc}
\ee
There is no simple general formula for these roots, but in the $\ql a\to\infty$
limit there are simple approximations for various regimes of $h=\Re(H)$ and
$\phi=\Re(\varphi)$.
As a warm up exercise, let us start with the $h>0$ regime and reproduce the
Seiberg's formula~\cite{seiberg} for $SU(2)$:
\be
\fdc\equiv\frac{8\pi^2}{g^2_{5d}[U(1)]}\ =\ 2h\ +\ (8-n_f)\times\phi .
\label{SeibergFla}
\ee
For $h>0$ and also $h,\phi\gg{1\over \ql a}$, the four roots of~\eq{D0t:Disc}
lie approximately at
\be
x_{1,2}\ \approx\ e^{+\ql a\varphi}\ \pm\ 2i\,e^{-\ql a(H+\varphi)/2},\qquad
x_{3,4}\ \approx\ e^{-\ql a\varphi}\ \pm\ 2i\,e^{-\ql a(H+3\varphi)/2}
\label{D0t:Roots:positive}
\ee
and have a large cross-ratio~(\ref{CrossRatio})
\be
\chi\ \approx\ {-1\over 16}\times e^{\ql a(H+4\varphi)}
\ee
Hence, according to \eq{Finite5Dcoupling}
\be
\fdc\ =\ 2h\ +\ 8\phi,
\label{D0t:posH}
\ee
in perfect agreement with the Seiberg's formula~(\ref{SeibergFla}).

Another useful cross-check of the $h>0$ phase
is to reproduce the $SU(2)$ restoration in 5D for $\phi\to0$.
In 4D terms, classical $SU(2)$ restoration happens for $\ql a\varphi=0$
or $\pi i$ (modulo $2\pi i$);
the two allowed values of the Wilson line are due to $W^\pm$ particles
having charges $\pm2$ in fundamental (quark) charge units.
Moreover, quantum effects in a classically-unbroken $SU(2)$ SYM split
a single singularity into a close pair of Seiberg--Witten points
where magnetic monopoles or dyons become massless and $j(\tau)$ has a pole.
In Seiberg--Witten terms, $\tr(({\rm adjoint\ scalar)}^2)$ corresponds to
$(\ql a\varphi)^2$ or $(\ql a\varphi-\pi i)^2$
while the strong-coupling scale $\Lambda^4_{\rm SW}$ of the 5D theory
compactified to 4D ($i.\,e.$, the diagonal $SU(2)$ of the quiver)
corresponds to $\pm e^{-\ql aH}$.
Hence, we expect $j(\tau(\varphi))$ to have poles at
\be
(\ql a\varphi)^2\ =\ \pm O\left(e^{-\ql aH/2}\right)
\qquad\mbox{and}\qquad
(\ql a\varphi-\pi i)^2\ =\ \pm O\left(e^{-\ql aH/2}\right)
\label{SWpairs}
\ee
In terms of the spectral curve, a pole of $j$ means that
two of the branching points $x_1,x_2,x_3,x_4$ collide with each other.
To see how it happens in the our  model, let's take the
$\sinh^2(\ql a\varphi)\to0$ limit of \eq{D0t:Disc}.
In this limit, all four branching points cluster around $\pm1$ (depending
on the sign of $\cosh(\ql a\varphi)$), so to resolve the situation, we shift
and rescale
\bea
x &=& x'\times\sinh(\ql a\varphi)\ +\ \cosh(\ql a\varphi),\\
D'(x) &=& \frac{D(x)}{\sinh^4(\ql a\varphi)}\nonumber\\
 &\approx& \bigl(x^{\prime2}-1\bigr)^2\
	\pm\ 4\frac{e^{-\ql aH}}{\sinh^4(\ql a\varphi)}\,.
\eea
The rescaled discriminant has a double root when the second term on the last
line above equals to $-1$, or in $\varphi$ terms when
\be
(\ql a\varphi)^2\ \approx\ \pm2i\, e^{-\ql aH/2}\qquad {\rm or}\qquad
(\ql a\varphi-\pi i)^2\ \approx\ \pm2\, e^{-\ql aH/2} ,
\label{D0t:SWpairs}
\ee
in perfect agreement with \eq{SWpairs}.\footnote{%
	Note that the $\varphi$ moduli space in 4D is a half-cylinder:
	besides the $\varphi\equiv\varphi+(2\pi i/\ql a)$ redundancy of the
	Wilson line, we also identify $\varphi\equiv-\varphi$ because
	of the symmetry between the two eigenvalues $\varphi_1=-\varphi_2$.
	This space has two $\ZZ_2$ orbifold singularities
	at $\ql a\varphi=0$ or $\pi i$,
	and the proper single-valued coordinates near these points are
	respectively $(\ql a\varphi)^2$ and $(\ql a\varphi-\pi i)^2$.	
	Therefore, eq.~(\ref{D0t:SWpairs}) describes
	four singularities rather then eight.
	}
Thus, we conclude that {\blue the $\phi\to0$ limit in the $h>0$ regime
of the quiver theory properly deconstructs the $SU(2)$ restoration in 5D.}

Now let us consider the negative $h$ regime of the theory.
In this regime, the roots of \eq{D0t:Disc} form three different patterns
depending on the ratio of $\phi$ to $-h$:
\begin{itemize}
\item
For $\phi>(-h)>0$, the pattern is similar to the $h>0$ regime,
and the roots lie at
\be
x_{1,2}\ \approx\ e^{+\ql a\varphi}\ \pm\ 2i\,e^{-\ql a(H+\varphi)/2},\qquad
x_{3,4}\ \approx\ e^{-\ql a\varphi}\ \pm\ 2i\,e^{-\ql a(H+3\varphi)/2}.
\label{D0t:Roots:posHPhi}
\ee
Consequently, the cross ratio~(\ref{CrossRatio}) is
$\chi=\frac{-1}{16}\,e^{\ql a(H+4\varphi)}$, and the 5D inverse coupling is
\be
\fdc\ =\ 2h\ +\ 8\phi ,
\label{D0t:posHPhi}
\ee
exactly as for $h>0$.

\item
For $\phi<(-h)$ but $\phi>(-h/3)>0$, the pattern is slightly different:
\bea
x_{1,2} &\approx& e^{+\ql a\varphi}\ \pm\ 2i\,e^{-\ql a(H+\varphi)/2},\nonumber\\
\mbox{but}\ x_3 &\approx& -2\,e^{-\ql a(H+2\varphi)},\label{D0t:Roots:negHPhi}\\
\mbox{and}\ x_4 &\approx& -\half\,e^{+\ql aH}.\nonumber
\eea
In this case, the cross ratio is $\chi=\frac{i}{8}\,e^{\ql a(3H+9\varphi)/2}$,
and hence the 5D inverse coupling is
\be
\fdc\ =\ 3h\ +\ 9\phi.
\label{D0t:negHPhi}
\ee

\item
Finally, for $0<\phi<(-h/3)$ we have a very different pattern of one
small root and three large roots equidistant from each other:
\bea
x_{1,2,3} &\approx&
-\root3\of{4}\,e^{2\pi ik/3}\times e^{-\ql aH/3}\
	+\ \coeff23 \,e^{\ql a\varphi},\nonumber\\
&&\quad k=1,2,3,\label{D0t:Roots:negative}\\
x_4 &\approx& -\coeff14 \,e^{+\ql aH}.\nonumber
\eea
For this pattern, the cross-ratio is $\chi\approx e^{-2\pi i/3}\,
-\,2^{1/3}3^{-1/2}i\,e^{\ql a(3\varphi+H)/3}$, and hence
\be
j(\tau)\ \approx\ 512 e^{+\ql a(H+3\varphi)}\ \ll\ 1.
\ee
Such small $j$ indicates strong rather than weak 4D gauge coupling:
$\tau$ asymptotes to a self-dual point $e^{2\pi/3}$ (a corner of the
Teichmuller space) and stops depending on $\varphi$ as long as
$\phi<(-h/3)$.
In 5D terms, such strong $\tau$ deconstructs to
\be
\fdc\ =\ \frac{\sqrt{3}/(4\pi)}{\ql a}\ \approx\ 0
\label{D0t:negative}
\ee
$i.\,e.$, $g^2_{5D}[U(1)]=O(\ql a)$ and becomes infinite in the decompactification
limit.
\end{itemize}

Let us summarize the various regimes of the deconstructed
$\fdc(\phi,h)$ behavior in one picture:
\be
\psset{xunit=2cm,yunit=1.5mm,runit=1mm,linewidth=1pt}
\psset{linecolor=black,fillstyle=solid}
\def\jjj#1#2{\rput(#1,0){%
	\psline[linewidth=0.5pt](0,0)(0,-1)
	\rput[t](0,-2){$#2$}
	}}
\begin{pspicture}[](0,-4)(5,42)
\psline[linewidth=0.5pt]{->}(0,0)(4,0)
\rput[l](4.1,0){$\phi$}
\psline[linewidth=0.5pt]{->}(0,0)(0,38)
\rput[b](0,38.5){$\fdc$}
\psline[linecolor=blue](0,6)(4,38)
\pscircle*[linecolor=blue](0,6){1}
\rput[rb](3,31){$\rm\blue slope=8$}
\rput[lb](4,38){\blue\psframebox[linewidth=0.5pt,framearc=0.2]{$h>0$}}
\jjj{0}{0}
\jjj{1}{|h|/3}
\jjj{3}{|h|}
\psline[linecolor=red](0,0)(1,0)(3,18)(4,26)
%\psline[linecolor=red,linestyle=dotted,linewidth=2pt](0,0)(1,0)
\pscircle[linecolor=red,fillcolor=white](0,0){1}
\pscircle[linecolor=red,fillcolor=white](1,0){1}
\pscircle[linecolor=red,fillcolor=white](3,18){1}
\rput[lt](3.5,21){$\rm\red slope=8$}
\rput[lt](2,8){$\rm\red slope=9$}
\rput[lb](4,26){\red\psframebox[linewidth=0.5pt,framearc=0.2]{$h<0$}}
\end{pspicture}
\label{D0t:Diagram1}
\ee
The blue and red line here plot $\fdc(\phi)$ for fixed values of $h$.
The blue line is for the $h>0$ regime, and the solid blue circle at its end
indicates $SU(2)$ restoration in 5D for $\phi=0$.
The red line is for the $h<0$ regimes, and the open red circles indicate
regime changes at $\phi=(-h)$, $\phi=(-h)/3$, and maybe $\phi=0$;
we shall investigate them momentarily.

Let us start with the right circle and take a closer look at the
spectral curve of the deconstructed theory for $\varphi=-H+O(1/\ql a)$.
In this regime, two roots of the curve's discriminant~(\ref{D0t:Disc})
as in eqs.~(\ref{D0t:Roots:posHPhi}) and~(\ref{D0t:Roots:negHPhi}),
\be
x_{1,2}\ \approx\ e^{+\ql a\varphi}\ \pm\ 2i\,e^{-\ql a(H+\varphi)/2} ,
\ee
while the other two roots $x_3,x_4$ satisfy a quadratic equation
\be
\left(e^{\ql a\varphi}\,x\right)^2\
-\ 2\left(1-2e^{-\ql a(\varphi+H)}\right)\times\left(e^{\ql a\varphi}\,x\right)\
+\ 1\ =\ 0
\ee
and collide with each other for $\ql a(\varphi+H)=0$ (modulo $2\pi i$).
This collision creates a pole in $j(\tau)$, indicating a charged particle
becoming massless at this point in the moduli space.
Since the 4D coupling $\tau$ is generally very weak in this area, we conclude
that the massless particle's charge is electric rather than magnetic.
Also, a single pole at a unique (modulo $2\pi$) value of the Wilson line
indicates the charge is $\pm1$ in fundamental units.
In other words, the massless 4D particle is a quark, and it deconstructs
a 5D quark which (in the $h<0$ phase) becomes massless at $\phi=(-h)$. 

Note that the 4D quiver of our model does have quarks, but
{\sl perturbatively} they have no light modes (with $\rm masses\ll(1/a)$)
and thus decouple from the 5D physics.
Apparently, when the quiver theory is strongly coupled
for $h<0$, those quarks somehow become light and show up in 5D.
In other words, the $h<0$ phase of the deconstructed theory
has a quark flavor that the $h>0$ ``ordinary $\rm SYM_5$'' phase does
not know about.

Now consider the left red circle at $\phi=0$.
In the $h<0$ regime, the spectral curve does not degenerate when $\ql a\varphi$
becomes small.
Instead, the four roots of the discriminant~(\ref{D0t:Disc}) form the
same pattern~(\ref{D0t:Roots:negative}) as for $\phi>0$ (but $\phi<(-h)/3$),
and nothing special happens for $\ql a\varphi=O(1)$.
In 5D terms, there is no 5D $SU(2)$ restoration in the $h<0$ phase
for $\phi=0$, nor anything else special at this point.

But after this disappointment,
the middle red circle at $\phi=(-h)/3$ turns out to be very interesting.
Looking at the spectral curve's discriminant for $3\varphi+H=O(1/\ql a)$,
we find one small root $x_4\approx{-1\over4}e^{+\ql aH}$ and three big roots
$x_{1,2,3}$ governed by a cubic polynomial
\be
\left( e^{-\ql a\varphi}\,x\right)^3\ -\ 2\left( e^{-\ql a\varphi}\,x\right)^2\
+\ \left( e^{-\ql a\varphi}\,x\right)\ +\ 4e^{-\ql a(3\varphi+H)}\ =\ 0.
\ee
For large $e^{-\ql a(3\varphi+H)}$ these roots form an equilateral triangle as in
\eq{D0t:Roots:negative}, while for small $e^{-\ql a(3\varphi+H)}$ the $x_3$ root
is much smaller than the $x_{1,2}$ roots as in \eq{D0t:Roots:negHPhi}.
And for an intermediate value of $e^{-\ql a(3\varphi+H)}={-1\over 27}$, two
of the roots collide and the spectral curve degenerates.
In $\ql a\varphi$ terms, the degeneration happens at three distinct points
\be
\ql a\varphi_k\ =\ -{\ql aH\over3}\ +\ \log3\
+\ {2\pi i (k-{1\over2})\over3}\quad
(\mbox{modulo}\ 2\pi i),\qquad k=1,2,3.
\label{D0t:E0pt:degeneration}
\ee
Moreover, at each point the degeneration is due to collision of a different
pair of roots, which leads to distinct, non-commuting monodromies around
each point.
In physical 4D terms, this means singularities due to massless particles
of different charge types: electric, magnetic, and dyonic.
And reconstructing this behavior in terms of a 5D theory compactified on
a large circle calls for a nontrivial superconformal theory in 5D.

Specifically, this pattern of three singularities related by $2\pi i/3$
Wilson lines is characteristic of compactified 5D SCFT known as the $E_0$.
Unlike the other SCFTs in the $E_n$ series which obtain in the $h\to0$
limits of $SU(2)$ gauge theories with $n-1$ massless flavors, the $E_0$
is an isolated SCFT.
It has a dynamical modulus field $\hat\phi$, but it does not
have any non-dynamical parameters (like $h$ or quark masses)
one needs to tune to obtain superconformal behavior for $\hat\phi=0$.
In 5D,  $\hat\phi$ is a real field which takes non-negative values only,
but after compactification to 4D, it becomes a cylindrical complex variable
$\hat\varphi$ whose real part could be either positive or negative.
However, for negative $\Re\hat\varphi$ the 4D gauge coupling
$\tau$ asymptotes to a selfdual point --- which corresponds to infinite
5D coupling --- while for positive $\Re\hat\varphi$
the 4D coupling is weak, $\Im\tau\propto 2\pi R$,
which indicates finite coupling in 5D, specifically $\fdc=9\hat\phi$.
Comparing this behavior to our quiver theory with
$h<0$ and $\phi\approx (-h)/3$, we immediately see that the 5D theory
here is the $E_0$ SCFT whose Coulomb modulus can be identified as
$\hat\phi=\phi-{1\over3}(-h)$.

Altogether, at this point we may complete the figure~(\ref{D0t:Diagram1})
as follows:
\be
\psset{xunit=2cm,yunit=1.5mm,runit=1mm,linewidth=1pt}
\psset{linecolor=black,fillstyle=solid}
\def\jjj#1#2{\rput(#1,0){%
	\psline[linewidth=0.5pt](0,0)(0,-1)
	\rput[t](0,-2){$#2$}
	}}
\begin{pspicture}[](-1.75,-4)(5,42)
\psline[linewidth=0.5pt]{->}(0,0)(4,0)
\rput[l](4.1,0){$\phi$}
\psline[linewidth=0.5pt]{->}(0,0)(0,38)
\rput[b](0,38.5){$\fdc$}
\psline[linecolor=blue](0,6)(4,38)
\pscircle*[linecolor=blue](0,6){1}
\rput[r](-0.05,6){\blue $SU(2)$ restoration $\to$}
\rput[rb](3,31){$\rm\blue slope=8$}
\rput[lb](4,38){\psframebox[linewidth=0.5pt,framearc=0.2]{\blue$h>0$}}
\jjj{0}{0}
\jjj{1}{|h|/3}
\jjj{3}{|h|}
\psline[linecolor=red](1,0)(3,18)(4,26)
\psline[linecolor=red,linestyle=dotted,linewidth=3pt](0,0)(1,0)
\rput[r](-0.05,0){\red no $SU(2)$ restoration $\to$}
\pscircle[linecolor=red,fillcolor=yellow](1,0){1}
\psline[linewidth=0.5pt,arrowscale=2,linecolor=red]{<-}(.94,0.6)(0.8,3)
\rput[br](1.2,4){\red $E_0$ SCFT}
\pscircle[linecolor=red,fillcolor=green](3,18){1}
\psline[linewidth=0.5pt,arrowscale=2,linecolor=red]{<-}(3.04,17.4)(3.3,15)
\rput[lt](3.33,14.5){\red a massless quark}
\rput[lt](3.75,24){$\rm\red slope=8$}
\rput[lt](2,8){$\rm\red slope=9$}
\rput[lb](4,26){\psframebox[linewidth=0.5pt,framearc=0.2]{\red$h<0$}}
\end{pspicture}
\label{D0t:Diagram2}
\ee
Again, the blue line here plots $\fdc(\phi)$ for a fixed $\blue h>0$ while
the red line plots $\fdc(\phi)$ for a fixed $\red h<0$, but now we have
identified all interesting points on both lines.
Also, the red line is now dotted left of the $E_0$ point ($0\le\phi<|h|/3$)
to indicate that this regime does not really exist in five infinite dimensions but
only in the compactified theory.
Indeed, in the decompactification limit $\ql a\to\infty$,
the distance between the $E_0$ point and the $\phi=0$ point disappears in the
field metric
\be
g_{\phi\phi}\ =\ \mbox{via 5D SUSY}\ \ =\ {1\over g_{5D}^2[U(1)]}\
=\ {\sqrt{3}/8\pi\over \ql a}\ \to\ 0,
\label{ZeroDistance}
\ee
and the whole range of $0\le\phi\le(-h/3)$ becomes invisible.

Such disappearance of 4D phases upon decompactification to 5D is well known
in the string theory context.
For example, when a type~IIA superstring on a Calabi--Yau manifold is promoted
to M--theory on the same manifold, the non-geometric phases of the Calabi-Yau
disappear from the 5D physics because the moduli space regions where they live
collapse to to zero volume in the decompactification limit.
The collapse happens due to difference between the 5D ($\NN=1$) and
the 4D ($\NN=2$) supersymmetric geometries and does not depend on any
inherently stringy physics;
any other UV completion allowing $\rm 4D\to 5D$ decompactification of a field
theory should behave in a similar way.

And that's precisely what we see in the spectral curve of
our deconstructed theory.
When compactified on a circle, the theory acquires a ``non-geometric'' phase
occupying the $0\le\phi<(-h)/3$ range of the moduli space, but this phase
disappears in the decompactification limit $\ql a\to\infty$.
Since the deconstruction process requires finite $\ql $ and hence finite fifth
dimension, we have duly deconstructed the non-geometric phase of
the compactified theory.
But we should not try to interpret this phase in 5D terms because it's
an artefact of compactification.

\medskip
\centerline{\large\blue $\star\qquad\star\qquad\star$}
\medskip

Let us briefly compare our deconstructed example to 
a string-theoretical UV completion of the same 5D theory,
$i.\,e.$, $\tilde D_0$.
Specifically, let us use the type~$\rm I'$ superstring
(orientifold of the type~IIA on a circle) where the
$\tilde D_0$ is realized on a D4-brane probe located near and O8
orientifold plane;
the other O8 plane and all 16 D8 branes are far away from the probe.
The 5D scalar field $\phi$ corresponds to the distance between the D4 and the
O8; perturbatively, the gauge theory on the probe is enhanced from $U(1)$ to
$SU(2)$ for $\phi=0$.
The vacuum charge $-8$ of the orientifold plane creates dilaton gradient
in the $\phi$ direction, which makes the gauge coupling on the probe $\phi$
dependent with derivative $\partial \fdc/\partial\phi=+8$.
As long as the dilaton's value at the O8 itself is finite, this
gives us $\fdc=2h+8\phi$ with $h>0$, precisely as for the
``ordinary $\rm SYM_5$'' phase of the deconstructed theory,
{\it cf.}\ the blue line in figure~(\ref{D0t:Diagram2}).

%\par\goodbreak
For $h\to0$ the dilaton value at the O8 blows up and the
type~$\rm I'$ perturbation theory breaks down, but the S-duality between
type~$\rm I'$ and heterotic strings allows analytic continuation into the
non-perturbative $h<0$ phase.
In this phase, the dilaton always blows up at the orientifold plane,
but the vacuum charge of the plane changes from $-8$ to $-9$, and an
extra D8 brane appears out of the O8 to balance the charge;
the distance between the $\rm O8^{-9}$ and the new D8 is proportional
to $-h$.
Putting a D4 probe right on top of the orientifold in this phase leads
to the $E_0$ superconformal theory on the probe \cite{MS}.
And when we move the probe away, we get the Coulomb branch of the $E_0$,
comprised of one massless vector multiplet with Chern--Simons self-coupling
$k=9$, or in terms of the gauge coupling, $\fdc=9\hat\phi$.
Note the Chern--Simons level $k=9$ here, it is characteristic of
the $E_0$ theory.

But besides the vector multiplet living on the D4 probe itself,
there is also a quark hypermultiplet
due to open strings between the D4 and the D8 brane (which was
emitted by the orientifold plane during the phase transition).
This quark is generally massive but becomes massless when the probe reaches
the D8.
And when the D4 probe moves beyond the D8, the quark becomes massive
again but its 5D mass flips sign; consequently, the Chern--Simons level
of the $U(1)$ vector multiplet reduces from $+9$ to $+8$ and hence
$\partial \fdc/\partial\hat\phi=8$ rather than 9.
Altogether, we have
\be
\fdc\ =\ \cases{
	9\hat\phi & for $0\le\hat\phi\le m_e$,\cr
	\noalign{\vskip 5pt plus 5pt}
	8\hat\phi+m_q & for $\hat\phi\ge m_q$.\cr
	}
\label{D0t:string}
\ee
Again, this behavior is in perfect agreement with the deconstructed
theory with $h<0$, {\it cf.}\ the solid part of the red line
in figure~(\ref{D0t:Diagram2}).

Finally, the non-geometric phase of the deconstructed theory --- {\it cf.}\
the dotted part of the red line for $0\le\phi<(-h/3)$ ---
does not have any counterpart in the type~$\rm I'$ string theory
because we haven't compactified it to 4D.
To see this phase in a string implementation of $\tilde D_0$
we would need a 4D $\NN=2$ construction --- for example a D3 probe near
a cluster of four $(p,q)$ 7--branes in type~IIB string, or
a type~IIA string on a singular Calabi--Yau with a collapsed $dP_1$
4--cycle --- but this gets us too deep into string theory
and away from the main subject of this paper.

\par\goodbreak
Instead, let us go back to the deconstructed theory and draw
its phase diagram in the $(h,\phi)$ parameter space:
\be
\psset{unit=1cm,linewidth=0.5pt,arrowscale=2,fillstyle=solid,framearc=0.2}
\begin{pspicture}[0.5](-8,-1)(+6,+8)
\pspolygon[linewidth=0,fillcolor=paleblue](0,0)(+5,0)(+5,7)(-3.5,7)
\rput(+2,3){\psframebox*{SYM phase}}
\pspolygon[linewidth=0,fillcolor=palered](0,0)(-7,3.5)(-7,7)(-3.5,7)
\rput{315}(-4.5,4.5){\psframebox*{$E_0$ Coulomb phase}}
\pspolygon[linewidth=0,fillcolor=yellow](0,0)(-7,0)(-7,3.5)
\rput[l]{347}(-6.0,1.4){\psframebox*{non-geometric phase}}
%\rput[l]{348}(-6.8,1.3){\psframebox*{or deconstruction failure}}
\psline[linewidth=3pt,linecolor=green](0,0)(-3.6,7.2)
\rput[b](-3.5,7.3){\green massless quark}
\psline[linewidth=3pt,linecolor=red](0,0)(-7.2,3.6)
\rput[r](-7.3,3.65){\red $E_0$ SCFT}
\psline[linewidth=3pt,linecolor=blue](0,0)(5,0)
\rput[r](+3.9,-0.6){\blue $SU(2)$ restoration}
\psarc[linecolor=blue]{->}(+4,-0.1){0.5}{270}{360}
\pscircle*[linecolor=red](0,0){0.12}
\rput[r](-0.6,-0.6){\red $\tilde E_1$ SCFT}
\psarc[linecolor=red]{->}(-0.5,-0.1){0.5}{270}{360}
\rput[r](-3.6,-0.6){no $SU(2)$ restoration}
\psarc[linecolor=black]{->}(-3.5,-0.1){0.5}{270}{360}
\psline{<->}(-7.5,0)(+5.5,0)
\rput[l](+5.6,0){$h>0$}
\rput[r](-7.6,0){$h<0$}
\psline{->}(0,0)(0,7.5)
\rput[b](0,7.6){$\phi$}
\end{pspicture}
\label{D0t:Diagram3}
\ee
Note that for fixed $\phi>0$ there is no phase transition across the $h=0$ line.
In 4D, the branching points of the spectral curve follow the same pattern
on both sides of this line --- {\it cf.}\ eqs.~(\ref{D0t:Roots:positive}
and~(\ref{D0t:Roots:posHPhi}) --- and there are no singularities for $H\to0$.
Consequently, the 5D physics also continues unperturbed, and the $h>0$
$\rm SYM_5$ phase continues to negative $h$.
Instead, the transition to a new phase --- which we identify as
a Coulomb branch of an $E_0$ theory (with some massive fields added) ---
happens at  $h=-\phi$.
Along this transition line, a charged hypermultiplet (a quark) becomes
massless while the gauge coupling remains finite.
In M--theory terms, such transition is a flop where a 4--cycle changes
the sign of its area.

When the parameter $h$ becomes more negative and reaches $h=-3\phi$,
the 5D coupling becomes strong and there is a transition from the
Coulomb phase of the $E_0$ to the superconformal phase.
In M--theory terms, this transition corresponds to 4--cycle (shaped as
a $\PP^2$) collapsing to a point rather than flopping the sign of its area.
And left of the transition line $h=-3\phi$ lies the non-geometric phase,
which exist in 4D compactifications of the 5D theory
but not in five infinite dimensions.

The two transition lines intersect at the $(h=0,\phi=0)$ point.
The 5D physics here is superconformal, but the SCFT is $\tilde E_1$
rather than $E_0$.
In the type~$\rm I'$ string construction of this SCFT point, the dilaton
blows up at the O8 orientifold plane which is just about to emit a D8 brane
but has not done it yet, and the D4 probe sits right on top of this
strongly-coupled mess.
The spectral curve of the 4D quiver theory is also rather messy at this
point, or rather its $O(1/\ql a)$ neighborhood in the $(H,\varphi)$ space:
$\tau$ is generally strong here, and there are four singular lines with
non-commuting monodromies around them.
But the general type of singularities agrees with the $\tilde E_1$ SCFT
compactified to 4D \cite{GMS}.

In string theory, the simplest way to produce the phase diagram similar
to~(\ref{D0t:Diagram3}) --- without the non-geometric phase, of course
--- is via the $(p,q)$ 5--brane web construction in type~IIB superstring.
For the Coulomb branch of the SYM phase, the brane web of the $\tilde D_0$
model looks like
\be
\psset{unit=1cm,linewidth=3pt,arrowscale=2}
\begin{pspicture}[](-4,-4)(12,+4)
\psline[linecolor=blue](-4,-4)(-2,-2)(-2,+2)(-4,+4)
\psline[linecolor=blue](+6,-4)(+6,-2)(+10,+2)(+12,+3)
\psline[linecolor=blue](-2,-2)(+6,-2)
\psline[linecolor=blue](-2,+2)(+10,+2)
\psline[linestyle=dotted](-2,-2)(-0.1,-0.1)(+6,-0.1)(+6,-2)
\psline[linestyle=dotted](-2,+2)(-0.1,+0.1)(+6.2,+0.1)(+10,+2)
\psline[linewidth=0.5pt]{<->}(0,-2)(0,+2)
\rput[l](+0.1,+1){$2\phi$}
\psline[linewidth=0.5pt]{<->}(0,-1)(+6,-1)
\rput[b](+3,-0.9){$h$}
\end{pspicture}
\label{D0t:Diagram4}
\ee
Note that the non-dynamical $h$ parameter here corresponds to relative
position of the semi-infinite external lines, while the dynamical modulus
$\phi$ controls the internal lines only.
For $\phi=0$ and $h>0$, two brane segments become coincident (the
dotted lines on the diagram~(\ref{D0t:Diagram4})) over length $h$, and
the string connecting these branes produce an $SU(2)$ SYM with 5D gauge
coupling $g_5^2\propto 1/h$.

For $h<0$ but $\phi>0$, the web flips between two topologies according to the
sign of $\phi+h$:
\be
\psset{unit=0.6cm,linewidth=3pt,arrowscale=2}
\begin{pspicture}[](-5,-5)(22,+5)
\rput(0,0){%
    \psline[linecolor=blue](-5,-5)(-2,-2)(-2,+2)(-5,+5)
    \psline[linecolor=blue](-1,-5)(-1,-2)(+3,+2)(+7,+4)
    \psline[linecolor=blue](-2,-2)(-1,-2)
    \psline[linecolor=blue](-2,+2)(+3,+2)
    }
\rput(17,0){%
    \psline[linecolor=blue](-5,-5)(-3,-3)(-3,-5)
    \psline[linecolor=blue](-3,-3)(-2,-1)(-2,+2)(-5,+5)
    \psline[linecolor=blue](-2,-1)(+1,+2)(+5,+3)
    \psline[linecolor=blue](-2,+2)(+1,+2)
    }
\end{pspicture}
\label{D0t:Diagram5}
\ee
The left web here --- for $h<0$ but $h+\phi>0$ --- has the same topology
as the $h>0$ web~(\ref{D0t:Diagram5}); it corresponds to the extension
of the SYM's Coulomb phase from $h>0$ to $-\phi<h<0$.
The right web --- for $h+\phi<0$ but $h+3\phi>0$ ---
has a different topology and describes a different phase of the theory,
namely the Coulomb phase of $E_0$ (the upper triangle of the web)
with an extra massive hypermultiplet (the lower fork).
The two webs are related by a segment flop; this is dual to a 4--cycle
flop in M--theory.
At the flop transition itself (at $h+\phi=0$),
there is a 4--brane junction which looks like an intersection of two branes.
Here the strings connecting the intersecting branes give rise to a massless
charged hypermultiplet:
\be
\psset{unit=0.6cm,linewidth=3pt,arrowscale=2}
\def\jjj{\pscurve[linecolor=red,linewidth=1pt]%
		(0,0)(.1,+.1)(.2,0)(.3,-.1)(.4,0)(.5,+.1)%
		(.6,0)(.7,-.1)(.8,0)(.9,+.1)(1,0)
		}
\begin{pspicture}[](-5,-5)(8,+5)
\psline[linecolor=blue](-5,-5)(+2,+2)(+7,+4.5)
\psline[linecolor=blue](-2,-5)(-2,+2)(-5,+5)
\psline[linecolor=blue](-2,+2)(+2,+2)
%\pscircle[linewidth=0.5pt](-2,-2){1}
%\rput[l](-0.9,-2){crossing}
\rput(-2,-1){\jjj}
\rput(-3,-3){\jjj}
\end{pspicture}
\label{D0t:Diagram6}
\ee

\par\goodbreak
On the other side of the $E_0$ Coulomb phase, for $h=-3\phi$
the triangle collapses to a point.
In 5D terms, this corresponds to a non-trivial SCFT, in this case $E_0$:
\be
\psset{unit=0.4cm,linewidth=0.25,arrowscale=2}
\begin{pspicture}[](-6,-4.5)(+4,+4.5)
\psline[linecolor=blue](-4.5,+4.5)(-2,+2)(+2,+4)
\psline[linecolor=blue](-2,+2)(-4.5,-3)(-4.5,-4.5)
\psline[linecolor=blue](-4.5,-3)(-6,-4.5)
\pspolygon[linecolor=blue,fillcolor=red,fillstyle=solid]%
	(-2.3,+2.3)(-1.4,+2.3)(-2.3,+1.4)
\rput[lt](-1.5,+1.5){\red $E_0$ SCFT}
\end{pspicture}
\label{D0t:Diagram7}
\ee
In addition, there is a massive hypermultiplet due to the fork in the
lower part of the web.

A different SCFT, namely $\tilde E_1$ obtains at $h=\phi=0$
when the whole web (except for the external legs) collapses to a point:
\be
\psset{unit=0.5cm,linewidth=0.2,arrowscale=2}
\begin{pspicture}[](-3,-4)(+4,+3)
\psline[linecolor=blue](-3,+3)(0,0)(+4,+2)
\psline[linecolor=blue](-3,-3)(0,0)(0,-4)
\pspolygon[linecolor=blue,fillcolor=red,fillstyle=solid]%
	(-.25,-.25)(-.25,+.25)(+.5,+.25)(0,-.25)
\rput[lt](0.3,-0.2){\red $\tilde E_1$ SCFT}
\end{pspicture}
\label{D0t:Diagram8}
\ee
Finally, for $h+3\phi<0$ the web cannot be build;
this impossibility in 5D corresponds to a non-geometric 4D phase of the
deconstructed theory.

%%%%%%%%%%%%%%%%%%%%%%%%%%%%%%%%%%%%%%%%%%%%%%%%%%%%%%%%%%%%%%%%%%%%%
% section 6.2
% D_0 model.
%
\subsection{The $D_0$ Model: $n_c=2$, $n_f=0$, $\Delta F=2$.}
Our second model is also an $SU(2)$ SYM in 5D, but with $\theta=0$
instead of $\theta=\pi$.
In 4D quiver terms, this calls for $n_c=2$, $n_f=0$, and $\Delta F=2$,
hence spectral curve
\be
y^2\ -\ y\times\left(x^2\,-\,2x\,\cosh(\ql a\varphi)\,+\,1\right)\
+\ e^{-\ql aH}\times x^2\ =\ 0.
\label{D0:Curve}
\ee
The discriminant equation~(\ref{FourBP}) for the branching points
of this curve factorizes into two quadratic equations
\be
x^2\ -\ 2\cosh(\ql a\varphi)\times x\ +\ 1\ =\ \pm 2e^{-\ql aH/2}\times x,
\label{D0:Factorize}
\ee
hence
\be
\eqalign{
x_{1,4}\ &=\ \cosh(\ql a\varphi)\,+\,e^{-\ql aH/2}\
	\pm\ \sqrt{\left(\cosh(\ql a\varphi)\,+\,e^{-\ql aH/2}\right)^2\,-\,1\,},\cr
x_{2,3}\ &=\ \cosh(\ql a\varphi)\,-\,e^{-\ql aH/2}\
	\pm\ \sqrt{\left(\cosh(\ql a\varphi)\,-\,e^{-\ql aH/2}\right)^2\,-\,1\,}.\cr
}\label{D0:ExactRoots}
\ee
For $h>0$ these branching points lie approximately at
\be
x_{1,2}\ \approx\ e^{+\ql a\varphi}\,\pm\,2e^{-\ql aH/2},\qquad
x_{3,4}\ \approx\ e^{-\ql a\varphi}\,\pm\,2e^{-\ql a(H+4\varphi)/2},
\label{D0:Roots:positive}
\ee
and although this pattern is somewhat different from \eq{D0t:Roots:positive}
for the previous model, it has a similar crossratio~(\ref{CrossRatio})
$\chi={1\over16}\,e^{-\ql a(H+4\varphi)}$ and therefore leads
to the same gauge coupling
\be
\fdc\ =\ 2h\ +\ 8\phi,
\label{D0:posH}
\ee
{\it cf.}\ Seiberg formula~(\ref{SeibergFla}).
Likewise, there is $SU(2)$ restoration for $\phi\to0$ and $h>0$.
Indeed, the branching points~(\ref{D0:ExactRoots}) degenerate
($x_1=x_4$ or $x_2=x_3$) when
\be
\cosh(\ql a\varphi)\ =\ \pm1\ \pm\ e^{-\ql aH/2},
\label{D0:DegenLocus}
\ee
and for $h>0$ this happen for
\be
(\ql a\varphi)^2\ \approx\ \pm2\,e^{-\ql aH/2}\qquad{\rm or}\qquad
(\ql a\varphi-\pi i)^2\ \approx\ \pm2i\,e^{-\ql aH/2}.
\label{D0:SWpairs}
\ee
Clearly this makes two close pairs of Seiberg--Witten points
according to \eq{SWpairs}, which indicates $SU(2)$ restoration in 5D.

But despite the similarity between the $\Delta F=1$ and $\Delta F=2$ models
for $h>0$, their $h<0$ behaviors are very different.
In the present $\Delta F=2$ model, for $\phi>0$, the branching points
$x_{1,2,3,4}$ follow the pattern~(\ref{D0:Roots:positive}) as long as
%$e^{-\ql aH/2}\ll\cosh(\ql a\varphi)\,\Longleftrightarrow\,$
$h>-2\phi$.
In 5D terms, this means that the SYM phase persist to negative $h$
as long as $h>-2\phi$.
%which is different from the stronger
%$h>-\phi$ condition of the $\Delta F=1$ model.
%
And for $h<-2\phi$ things are getting seriously weird:
the branching points asymptote to
\be
x_{1,2}\ \approx\ \pm2\,e^{-\ql aH/2},\qquad
x_{3,4}\ \approx\ \mp\half\,e^{+\ql aH/2}
\label{D0:Roots:negative}
\ee
regardless of $\varphi$,
the crossratio becomes $\chi\approx-{1\over4}\,e^{-\ql aH}\gg1$,
and all this translates to a finite but $\phi$--independent 5D gauge coupling
\be
\fdc\ =\ 2|h|\ >\ 0\quad\mbox{whenever}\ h<-2\phi<0.
\label{D0:negative}
\ee

Finally, for $h<0$ the $SU(2)$ restoration happens not at $\phi=0$
but at $\phi=(-h)/2$.
Indeed, in this region, the degeneration loci~(\ref{D0:DegenLocus}) become
\be
\ql a\varphi\ \approx\ -\half\ql aH\ +\ \log(2)\
+\ \{0\ \mbox{or}\ \pi i\}\ \pm\ e^{+\ql aH/2} ,
\label{D0:SWpairs:negative}
\ee
which makes two exponentially close pairs (note $e^{+\ql aH/2}\ll1$)
whose centers differ by a Wilson $\rm line=\pi$;
as in \eq{SWpairs}, this is how the $SU(2)$ restoration in 5D looks
to the 4D spectral curve.
On the other hand, the curve does not degenerate for $h<0$ and $\phi\approx0$,
so there is no $SU(2)$ restoration there.

Altogether, we have
\be
\psset{unit=1.2cm,linewidth=3pt,fillstyle=solid,arrowscale=2}
\begin{pspicture}[](-3,-.6)(7.3,6.5)
\psline[linewidth=0.5pt]{->}(0,0)(7,0)
\rput[l](7.05,0){$\phi$}
\psline[linewidth=0.5pt]{->}(0,0)(0,6)
\rput[b](0,6.05){$\fdc$}
\psline[linewidth=0.5pt](3,0)(3,-0.2)
\rput[t](3,-0.3){$|h]/2$}
\psline[linecolor=blue](0,1.5)(6,5.5)
\pscircle*[linecolor=blue](0,1.5){.1}
\rput[lb](6,5.5){\blue\psframebox[linewidth=0.5pt,framearc=0.2]{$h>0$}}
\rput[b]{35.7}(4.5,4.6){$\rm\blue slope=8$}
\psline[linecolor=blue,linewidth=0.5pt]{->}(-0.5,1.5)(-0.1,1.5)
\rput[r](-0.6,1.5){\blue $SU(2)$ restoration}
\psline[linecolor=red](3,1)(6,3)
\rput[lb](6,3){\red\psframebox[linewidth=0.5pt,framearc=0.2]{$h<0$}}
\rput[b]{35.7}(4.5,2.1){$\rm\red slope=8$}
\psline[linecolor=red,linestyle=dotted,linewidth=3pt](0,1)(3,1)
\rput[b](1.5,1.1){$\rm\red slope=0$}
\pscircle*[linecolor=red](3,1){.1}
\psline[linecolor=red,linewidth=0.5pt]{->}(3.5,0.8)(3.1,0.95)
\rput[l](3.6,0.8){\red $SU(2)$ restoration}
\end{pspicture}
\label{D0:Diagram1}
\ee
where the dotted red line for $h<0$ and $0<\phi<|h|/2$ denotes something
is wrong in this regime.
Indeed, a finite but $\phi$--independent 5D gauge coupling would normally
indicate a free $U(1)$ phase, but such a phase cannot possibly connect to
an unbroken--$SU(2)$ point at $\phi=|h|/2$.
Instead, an $SU(2)$ point should be the end-point of the 5D moduli space
because of the $\ZZ_2\subset SU(2)$ reflection of the Coulomb modulus
$\hat\phi\to-\hat\phi$;
in $\phi$ terms, this corresponds to identification
\be
\phi\ =\ {-h\over2}\ +\ {\red|\hat\phi|}\qquad\Longrightarrow\qquad
\phi\ \mbox{always}\ \ge {-h\over2}\,.
\label{D0:NewBottom}
\ee

From the 4D point of view, such premature end of the $\phi$ modulus
indicated sudden divergence between the $\varphi$ coordinate of the
complex moduli space and between the $\NN=2$
superpartner $\cal A$ of the abelian vector field.
Indeed,
\be
\frac{d{\cal A}}{d\varphi}\
=\ \frac{2\sinh(\ql a\varphi)}{2\pi i}\times
\oint\frac{dx}{\sqrt{(x-x_1)(x-x_2)(x-x_3)(x-x_4)}}
\label{D0:superpartner}
\ee
where the integration contour is the electric cycle of the spectral curve,
$i.\,e.$ a loop around a branch cut connecting $x_1$ with $x_2$\footnote{%
	A similar contour integral over the magnetic cycle of the curve
	--- a loop around a cut from $x_1$ to $x_3$ ---
	gives $d{\cal A}_D/d\cosh(\ql a\varphi)$ where ${\cal A}_D$ is
	the superpartner of the magnetic dual of the vector field.
	};
the pre-integral factor $2\sinh(\ql a\varphi)$ here compensates for
the logarithmic definition of the $\varphi$ modulus, {\it cf.}\ \eq{phidef}.
As long as the branching points are as in \eq{D0:Roots:positive}, 
\eq{D0:superpartner} evaluates to
\be
{d{\cal A}\over d\varphi}\ =\ 1\ +\ O\bigl(e^{-\ql a(H+2\varphi)}\bigr),
\ee
thus in the decompactification limit, ${\cal A}=\varphi+\rm const$
with exponentially good accuracy.
However, when the branching point pattern changes from~(\ref{D0:Roots:positive})
to~(\ref{D0:Roots:negative}) for $h+2\phi<0$,
\eq{D0:superpartner} yields
\be
{d{\cal A}\over d\varphi}\
=\ \coeff{i}{2}\,e^{+\ql a(H+2\varphi)/2}\ \ll 1,
\ee
thus the vector's superpartner $\cal A$ no longer tracks $\varphi$.
Instead, it decouples: as long as $\Re\varphi<(-h/2)$,
the actual value of $\varphi$ does not matter anymore.

Consequently, we would like to map $\varphi$ onto a different holomorphic
coordinate which tracks $\varphi$ for $\Re\varphi>(-h/2)$ but bottoms out
at $\Re\varphi=(-h/2)>0$.
From the $\NN=2$ point of view it would be best to use the vector's
superpartner $\cal A$ itself, but since it suffers non-trivial monodromies at the
Seiberg--Witten points~(\ref{D0:SWpairs:negative}), we would rather
use something simpler.
Specifically, we want a holomorphic coordinate $\hat\varphi$ which
lives on a half-cylinder, $i.\,e.\ \hat\varphi\equiv
\hat\varphi+{2\pi i\over\ql a}$ and $\hat\varphi\equiv-\hat\varphi$,
and whose real part $\Re\hat\varphi$ becomes the 5D modulus $\hat\phi$
({\it cf.}\ \eq{D0:NewBottom}) in the decompactification limit.
And since $\varphi$ itself lives on a half-cylinder, the map between
$\varphi$ and $\hat\varphi$ works according to
\be
\cosh(\ql a\hat\varphi)\ =\ e^{+\ql aH/2}\times\cosh(\ql a\varphi).
\label{D0:remap}
\ee
%where the `$-$' sign is for future convenience; for the present purpose
%we may replace it with an arbitrary $O(1)$ constant factor.
As promised, for $\ql a\to\infty$ \eq{D0:remap} reduces to
$\lfloor\hat\varphi\rfloor=\lfloor\varphi\rfloor+{1\over2}H$ and hence
\eq{D0:NewBottom} for the real 5D variables $\hat\phi$ and $\phi$.
But near the $SU(2)$ restoration points, the $\hat\varphi$ variable becomes
double-valued, hence \eq{D0:SWpairs:negative} becomes
\be
(\ql a\hat\varphi)^2\ =\ \pm2\,e^{+\ql aH/2}\qquad{\rm and}\qquad
(\ql a\hat\varphi-\pi i)^2\ =\ \pm2\,e^{+\ql aH/2} .
\label{D0:SWpairs:dual}
\ee
Note Seiberg--Witten's $\tr(({\rm adjoint\ scalar})^2)$ here
corresponds to $(\ql a\hat\varphi)^2$ or $(\ql a\hat\varphi-\pi i)^2$,
similarly to $(\ql a\varphi)^2$ or $(\ql a\varphi-\pi i)^2$
for the $SU(2)$ restoration
at $\phi\to0$ for $h>0$, {\it cf.}\ \eq{D0:SWpairs}.

In fact, this symmetry between the two $SU(2)$ restorations at
$(\phi=0,h>0)$ and $(\hat\phi=0,h<0)$ is an exact symmetry of the
spectral curve of the 4D quiver theory.
To make it manifest, we rewrite the spectral curve~(\ref{D0:Curve}) as
\be
\left(z\,+\,{1\over z}\right)\ 
+\ e^{-\ql aH/2}\times\left( x\,+\,{1\over x}\,
	-\,2\cosh(\ql a\varphi)\right)\
=\ 0,\qquad
z\ =\ -{y\over x}\times -e^{\ql aH/2}.
\label{D0:Zcurve}
\ee
The coordinates $x$ and $z$ appear here in a similar way, and the
curve is symmetric with respect to simultaneous exchanges of
\be
x\,\leftrightarrow\,z ,\qquad H\,\leftrightarrow\,-H ,\qquad
\varphi\,\leftrightarrow\,\hat\varphi
\label{D0:symmetry}
\ee
where $\hat\varphi$ is exactly as in \eq{D0:remap}.

In light of this symmetry, we deconstruct the 5D phase diagram
of our present model --- or rather its Coulomb branch ---
as a single phase bounded by two separate $SU(2)$ restorations:
\be
\psset{unit=1cm,linewidth=0.5pt,arrowscale=2,fillstyle=solid,framearc=0.2}
\begin{pspicture}[.5](-6,-3)(+6,+7)
%\rput{15}(0,0){%
    \pspolygon*[linecolor=magenta](0,0)(+6,6)(+5,7)(-5,7)(-6,6)
    \rput(0,5){\psframebox{common Coulomb phase}}
    \rput[b](0,3.5){\psframebox{strong coupling}}
    \psline[linewidth=1pt]{->}(0,3.5)(0,1.5)
    \psline[linewidth=3pt,linecolor=blue](0,0)(+6,6)
    \rput{45}(+3.3,2.7){\blue $SU(2)$ restoration}
    \psline[linewidth=3pt,linecolor=red](0,0)(-6,6)
    \rput{315}(-3.3,2.7){\red $SU(2)$ restoration}
    \psline{->}(0,0)(+6.5,6.5)
    \rput[lb](+6.6,6.6){$|\hat\phi|$}
    \psline{->}(0,0)(-6.5,6.5)
    \rput[rb](-6.6,6.6){$|\phi|$}
    \psline{<->}(-6,0)(+6,+0)
    \rput[b](+4,0.2){$h=2(|\hat\phi|-|\phi|)>0$}
    \rput[b](-4,0.2){$h=2(|\hat\phi|-|\phi|)<0$}
    \pscircle*[linecolor=green](0,0){0.15}
    \rput(0,0){$\star$}
    \rput[tr](0,-.3){\psframebox[linecolor=green]{$E_1$ SCFT}}
%    }
\rput{340}(0,0){%
    \pspolygon*[linecolor=brown](0,0)(6.08,0)(6,-1)
    \rput[lb](3,.1){\brown Higgs branch}
    }
\end{pspicture}
\label{D0:Diagram2}
\ee
On this diagram, the 5D coupling has a vertical gradient, the higher
the weaker, $\fdc\to\infty$ as one goes up.
And in the opposite direction, the coupling becomes infinite ($\fdc=0$)
in the bottom corner $h=\phi=\hat\phi=0$, so there instead of a SYM we
have a non-trivial superconformal theory.
To identify the SCFT in question, we consider the degeneration
of the spectral curve in this region and notice that all four
degeneration loci~(\ref{D0:DegenLocus}) pass through the $O(1/\ql a)$
neighborhood of the $(H=0,\varphi=0)$ point.
Moreover, for $H=0$ two of the four singularities collide at
$\cosh(\ql a\varphi)=0$ creating a double singularity.
(In Kodaira terms, $I_1+I_1\to I_2$.)
This singularity structure is characteristic of 4D compactification
of the $E_1$ SCFT in 5D, and so we identify the bottom corner of
diagram~(\ref{D0:Diagram2}) as the $E_1$ point.
Also, there is a baryonic branch here corresponding to the Higgs branch
of the $E_1$.
Indeed, at the $I_2$ singularity at $H=\cosh(\ql a\varphi)$
the spectral curve factorizes according to
\be
y^2\ -\ y\times(x^2+1)\ +\ x^2\
=\ (y-x^2)\times(y-1)\ =\ 0,
\ee
and we saw in \S5 that such factorization indicates a baryonic branch
with two $\mu=0$ flavors.

Ideally, to prove the 5D SCFT at $h=\phi=\hat\phi=0$ is indeed the $E_1$,
we would like to see its enhanced ``flavor'' symmetry $E_1=SU(2)$.
Unfortunately, this symmetry does not show up in 4D --- presumably, its
broken by the deconstruction --- and instead,
we have to rely on less transparent signatures such
as singularities of the spectral curve in 4D and the Higgs branches.

\medskip
\centerline{\large\blue $\star\qquad\star\qquad\star$}
\medskip

We conclude this section by comparing the phase diagram~(\ref{D0:Diagram2})
of the deconstructed $n_c=2$, $n_f=0$, $\theta=0$ model
with a stringy implementation of the same 5D theory, $i.\,e.\ D_0$.
Again, we use the type~IIB 5--brane web construction.
On the Coulomb branch of $D_0$, regardless of $h>0$ or $h<0$,
the web contains a rectangular box:
\be
\psset{unit=0.5cm,linewidth=3pt,arrowscale=2}
\begin{pspicture}[](-16,-8.5)(12,+8.5)
\rput(-8,0){%
    \psline[linecolor=blue](-8,-4)(-6,-2)(-6,+2)(-8,+4)
    \psline[linecolor=blue](+8,-4)(+6,-2)(+6,+2)(+8,+4)
    \psline[linecolor=blue](-6,-2)(+6,-2)
    \psline[linecolor=blue](-6,+2)(+6,+2)
    \psline[linestyle=dotted](-6,-2)(-4.1,-0.1)(+4.1,-0.1)(+6,-2)
    \psline[linestyle=dotted](-6,+2)(-4.1,+0.1)(+4.1,+0.1)(+6,+2)
    \psline[linewidth=0.5pt](+6,-2)(+6,-3.5)
    \psline[linewidth=0.5pt](-6,-2)(-6,-3.5)
    \psline[linewidth=0.5pt]{<->}(-6,-3.3)(+6,-3.3)
    \rput[b](0,-3.2){$2\phi$}
    \psline[linewidth=0.5pt](-6,+2)(-7.5,+2)
    \psline[linewidth=0.5pt](-6,-2)(-7.5,-2)
    \psline[linewidth=0.5pt]{<->}(-7.3,+2)(-7.3,-2)
    \rput[l](-7.2,0){$2\hat\phi$}
    \psline[linewidth=0.5pt](+4,-0.2)(+4,+0.8)
    \psline[linewidth=0.5pt](-4,-0.2)(-4,+0.8)
    \psline[linewidth=0.5pt]{<->}(-4,+0.5)(+4,+0.5)
    \rput[b](0,+0.7){$h$}
    }
\rput(+8,0){%
    \psline[linecolor=blue](-4,-8)(-2,-6)(-2,+6)(-4,+8)
    \psline[linecolor=blue](+4,-8)(+2,-6)(+2,+6)(+4,+8)
    \psline[linecolor=blue](-2,-6)(+2,-6)
    \psline[linecolor=blue](-2,+6)(+2,+6)
    \psline[linestyle=dotted](-2,-6)(-0.1,-4.1)(-0.1,+4.1)(-2,+6)
    \psline[linestyle=dotted](+2,-6)(+0.1,-4.1)(+0.1,+4.1)(+2,+6)
    \psline[linewidth=0.5pt](+2,-6)(+2,-7.5)
    \psline[linewidth=0.5pt](-2,-6)(-2,-7.5)
    \psline[linewidth=0.5pt]{<->}(-2,-7.3)(+2,-7.3)
    \rput[b](0,-7.2){$2\phi$}
    \psline[linewidth=0.5pt](-2,+6)(-3.5,+6)
    \psline[linewidth=0.5pt](-2,-6)(-3.5,-6)
    \psline[linewidth=0.5pt]{<->}(-3.3,+6)(-3.3,-6)
    \rput[l](-3.2,0){$2\hat\phi$}
    \psline[linewidth=0.5pt](-0.2,+4)(+0.8,+4)
    \psline[linewidth=0.5pt](-0.2,-4)(+0.8,-4)
    \psline[linewidth=0.5pt]{<->}(+0.5,-4)(+0.5,+4)
    \rput[l](+0.6,0){$-h$}
    }
\end{pspicture}
\label{D0:Diagram3}
\ee
The left web here is for $h>0$ and the right web for $h<0$, and the
only difference between them is which side of the box is longer;
the topology is the same, and there is no flop transition.
For $\phi\to0$ or $\hat\phi\to0$ --- whichever happens first ---
the box collapses to a pair of coincident line segments (the
dotted lines on the diagram~(\ref{D0:Diagram3}), either web);
the strings between those coincident branes give rise to an
$SU(2)$ SYM with 5D gauge coupling $g_5^2\propto1/|h|$.

For $h=0$ and $\phi=\hat\phi\to0$, the box collapses to a point
\be
\psset{unit=0.5cm,linewidth=0.2,linecolor=blue}
\begin{pspicture}[0.5](-3,-3)(+3,+3)
\psline(-3,+3)(0,0)(+3,+3)
\psline(-3,-3)(0,0)(+3,-3)
\pspolygon[fillcolor=red,fillstyle=solid]%
	(-.25,-.25)(-.25,+.25)(+.25,+.25)(+.25,-.25)
\rput[l](0.7,0){\red $E_1$ SCFT}
\end{pspicture}
\label{D0:Diagram4}
\ee
giving rise to the $E_1$ superconformal theory in 5D.
This theory has an $SU(2)=E_1$ global ``flavor'' symmetry, but
it isn't manifest in the brane-web picture.
Likewise, the deconstructed theory does not have an enhanced
flavor symmetry at $H=0$ at the 4D quiver level.
Presumably, in 5D the enhanced symmetry is limited to the marginal
operators of the SCFT, but the irrelevant operators stemming from
a UV completion --- deconstructive or stringy --- break the symmetry.

Finally, the $E_1$ web~(\ref{D0:Diagram4}) can be reconnected as two
intersecting whole branes (infinite in all directions), and then
the two branes can move away from each other (in a direction perpendicular
to both):
\be
\psset{unit=0.5cm,linewidth=0.2,linecolor=blue}
\begin{pspicture}[0.5](-8,-3)(+8,+3)
\rput(-5,0){%
    \psline(-3,+3)(0,0)(+3,+3)
    \psline(-3,-3)(0,0)(+3,-3)
    \pspolygon[fillcolor=red,fillstyle=solid]%
	(-.25,-.25)(-.25,+.25)(+.25,+.25)(+.25,-.25)
    }
\rput(+5,0){%
    \psline(-3,+3)(+3,-3)
    \psline[border=0.1](-3,-3)(+3,+3)
    }
\psline[linecolor=black,linewidth=0.1,arrowscale=2]{->}(-2,0)(+2,0)
\end{pspicture}
\label{D0:Diagram5}
\ee
This is the Higgs branch which connects to the Coulomb branch at
the $E_1$ point; it corresponds to the exotic baryonic branch of
the quiver theory.

To summarize, we have seen that the deconstructed $D_0$ theory
has exactly the same phase structure as the $D_0$ completed via string theory.

\par\vskip 0pt plus 40pt minus 0pt \goodbreak
%%%%%%%%%%%%%%%%%%%%%%%%%%%%%%%%%%%%%%%%%%%%%%%%%%%%%%%%%%%%%%%%%%%%%
% section 6.3
% D_2 model.
%
\subsection{Models with Flavor: $n_c=n_f=2$.}
In this section we explore two models with $n_f=2$,
one model with $\Delta F=1$ and the other with $\Delta F=0$.
Each model has three non-dynamical parameters, namely $h$, $m_1$, and $m_2$,
but for simplicity we restrict our analysis to $|m_1|=|m_2|$ where in 5D
we expect $U(2)$ flavor symmetry for $m\neq0$ and $SO(4)$ for $m=0$.

We begin with the $\Delta F=1$ model with $m_2=-m_1=m\ge0$, or in 4D terms
$\mu_{1,2,3}=(Ve^{-aM},Ve^{+aM},0)$ where $\Re(M)=m$,
hence the spectral curve
\be
y^2\ -\ y\times\Bigl(x^2\,-\,2\cosh(\ql a \varphi)\times x\,+\,1\Bigr)\
-\ e^{-\ql aH}\times x\times\Bigl(x^2\,-\,2\cosh(\ql a M)\times x\,+\,1\Bigr)\
=\ 0.
\label{D2:curve}
\ee
Note that for $M\neq0$, the flavor symmetry of the quiver theory itself
is $U(1)^2$ rather than $U(2)$, but the discrete $\bf C$
symmetry~(\ref{Csym:XY}--\reftail{Csym:S}) of the spectral
curve permutes the two 5D flavors and hence acts as a custodial symmetry:
it assures that the low-derivative operators in 5D are $U(2)_F$ invariant,
although the higher-derivative operators do not have this symmetry.
And for $M=0$ (or $M={\pi i\over\ql a}$) the quiver has flavor symmetry $U(2)$,
but $\bf C$ acts as a custodial symmetry of the $U(2)\to SO(4)$ symmetry
enhancement in the 5D continuum limit.

Along the Coulomb branch, the spectral curve~(\ref{D2:curve}) generally has
four branching points at
\be
x_{1,4}\ =\ a\,\pm\,\sqrt{a^2-1\,},\qquad x_{2,3}\ =\ b\,\pm\,\sqrt{b^2-1\,}
\label{D2:Roots}
\ee
where
\be
a,b\ =\ \cosh(\ql a\varphi)\,-\,e^{-\ql a H}\ \pm\ e^{-\ql aH/2}\times
\sqrt{2\cosh(\ql aM)\,+\,e^{-\ql aH}\,-\,2\cosh(\ql a\varphi)\,}.
\ee
The curve has four simple singularities ($I_1$ in Kodaira terms) at
\be
\sinh^2{\ql a\varphi\over2}\ =\ \pm e^{-\ql a H/2}\sinh{\ql aM\over2}
\qquad{\rm and}\qquad
\cosh^2{\ql a\varphi\over2}\ =\ \pm e^{-\ql a H/2}\cosh{\ql aM\over2}
\label{D2:SingLoci}
\ee
where two of branching points coincide.
For
\be
\cosh(\ql a\varphi)\ =\ e^{-\ql aH}\ =\ e^{\mp\ql aM}
\label{D2:exBB}
\ee
two of the singularities~(\ref{D2:SingLoci}) merge into a double singularity,
$I_1+I_1\to I_2$.
In addition, there is another double singularity at
\be
2\cosh(\ql a\varphi)\ =\ 2\cosh(\ql aM)\ +\ e^{-\ql aH}.
\label{D2:ordBB}
\ee
At all these double singularities the spectral curve factorizes:
\be
\vcenter{\openup 1\jot \ialign{
	for (#\unskip),\quad\hfil &
	$\displaystyle{#}$\cr
	\ref{D2:ordBB} &
	\left(y\ -\ \left(x\,-\,e^{-\ql aM}\right)
			\left(x\,-\,e^{+\ql aM}\right)\right)
		\times\left(y\ +\ e^{-\ql aH}x\right)\
		=\ 0,\cr
	\ref{D2:exBB} &
	\left(y\ -\ x\left(x\,-\,e^{\mp\ql aM}\right)\right)
		\times\left(y\ +\ \left(e^{\mp\ql aM}x\,-\,1\right)\right)\
		=\ 0.\cr
	}}
\ee
According to \S5, this indicates baryonic branches connected to the
Coulomb branch at these points,
namely the ordinary baryonic branch $B_{12}$ at (\ref{D2:ordBB}),
and exotic baryonic branches $B_{13}$ and $B_{23}$ at (\ref{D2:exBB}).
(The subscripts of $B$ here refer to the flavors of the baryonic VEV.)
In the decompactification limit $\ql a\to\infty$, the locations of these
baryonic branches become
\bea
B_{12}\ {\rm branch}:\quad
\phi &=& \max(m,-h),\label{D2:bb5d:12}\\
B_{13}\ {\rm branch}:\quad
\phi &=& 0,\quad h=m,\label{D2:bb5d:13}\\
B_{23}\ {\rm branch}:\quad
\phi &=& m,\quad h=-m.\label{D2:bb5d:23}
\eea
Finally, for $M=0$ or ${\pi i\over \ql a}$, another pair of $I_1$
singularities~(\ref{D2:SingLoci}) merges into an $I_2$ at $\varphi=M$,
but this time the spectral curve does not factorize all the way;
instead, it has one branch cut and one pole (same on both sheets).
Consequently, at this point we have a mesonic branch.
Note that for $h>0$ this branch is located very close to the $B_{12}$
baryonic branch, and in the 5D limit the two Higgs branches are
rooted at the same place $m=\phi=0$.
And indeed, in {\sl classical} \sqcdv\ with $n_c=n_f=2$, the mesons and
the baryons are related by the $O(4)$ flavor symmetry; although in the
quantum theory only $SO(4)\subset O(4)$ is a true symmetry while the
discrete $\ZZ_2=O(4)/SO(4)$ `isoparity' is anomalous,
the anomaly does not affect the moduli space for $h>0$.
In 4D however, the anomaly is more powerful, and the instanton effects
separate the mesonic and the baryonic branches by $\Delta(\ql a\varphi)=
O(e^{-\ql aH})$.
In quiver terms, the instantons here are diagonal, $i.\,e,$ one instanton
in each $SU(2)_\ell$ factor of the $[SU(2)]^\ql$ gauge group, and that's
why the effect is so small for $h>0$.

And now consider the Coulomb branch of the quiver.
In the decompactification limit $\ql a\to\infty$, the branching
points~(\ref{D2:Roots}) of the spectral curve form several distinct patterns
depending on $\phi$, $h$ and $m$;
as usual, this leads to different phases in 5D.
Let us start with the phase structure for $m=0$:

\begin{itemize}
\item[\large\blue$\star$]
For $h>0$, there is only one phase: for any $\phi>0$, the branching points
are
\be
x_{1,2}\ =\ e^{\ql a\varphi}\,\pm\,2i\,e^{\ql a(\varphi-H)/2},\qquad
x_{3,4}\ =\ e^{-\ql a\varphi}\,\pm\,2i\,e^{\ql a(-3\varphi-H)/2},
\label{D2:Roots:positive}
\ee
their crossratio is $\chi={-1\over 16}\,e^{\ql a(H+3\varphi)}$,
and hence in 5D
\be
\fdc\ =\ 2h\ +\ 6\phi,
\label{D2:positive}
\ee
in perfect agreement with the Seiberg formula~(\ref{SeibergFla}) for
$n_f=2$.
For $\phi\to0$, the situation is more complicated and the spectral curve
develops multiple singularities near $\ql a\varphi=0,\ \pi i$.
Specifically, there is {\it one} Seiberg--Witten pair of simple ($I_1$)
singularities at
\be
(\ql a\varphi-\pi i)^2\ \approx\ \pm2\,e^{-\ql aH},
\label{D2:D2:SW0}
\ee
and another pair of double ($I_2$) singularities at
\be
\ql a\varphi\ =\ 0,\ e^{-2\ql H} .
\label{D2:D2:SW2}
\ee
The latter pair is also of Seiberg--Witten type but correspond to $\NN=2$
$SU(2)$ gauge theory with two massless flavors rather than flavorless SYM
as in \eq{D2:D2:SW0}.\footnote{%
	For $\ql a\varphi\to 0$, the spectral curve of the quiver theory
	may be approximated as
	\be
	y^{\prime 2}\ =\ \left(x^{\prime 2}\,-\,(\ql a\varphi)^2\right)^2\
	-\ 4e^{-\ql aH}\times x^{\prime 2}
	\ee
	where $x'= x-\cosh(\ql a\varphi)\approx x-1$ and
	$y'=2y-(x-e^{\ql a\varphi})(x-e^{-\ql a\varphi})$.
	This curve looks exactly like the $\NN=2$ Seiberg--Witten curve of
	$SU(2)$ gauge theory with two massless flavors, where the role of
	$\tr(({\rm adjoint\ scalar})^2)$ is played by $(\ql a\varphi)^2$
	and the role of $\Lambda^2_{\rm SW}$ by $e^{-\ql aH}$.
	}
Thus, above the Seiberg--Witten scale (which is exponentially small for
$\ql a\to\infty$) we have $SU(2)$ restoration for $\ql a\varphi=0$ and $\pi i$,
and for $\ql a\varphi=0$ we also have two massless quarks.
In 5D terms, this means that for $\phi=0$ we have unbroken $SU(2)$ gauge theory
with two massless flavors;
in string constructions, this 5D theory is known as the $D_2$ after its global
symmetry.

\item[\large\green$\star$]
For $h=0$, the branching points of the spectral curve are as
in \eq{D2:Roots:positive} as long as $\phi>0$, and hence the 5D coupling
is as in \eq{D2:positive}.  However, for  $\phi\to0$ we now have $\fdc=0$,
meaning infinitely strong $g_5$ and hence a non-trivial
superconformal theory in 5D.
Since this SCFT obtains in the $h=0$ limit of the $D_2$ theory we expect it
to be $E_3$.
Unfortunately, we cannot directly verify the $E_3=SU(3)\times SU(2)$ global
symmetry of the SCFT because it applies only to the marginal and relevant
operators of the 5D theory, so instead we look at the singularities
of the spectral curve.
The curve of 5D $E_3$ SCFT compactified to 4D should have three singularities,
of respective Kodaira types $I_3$, $I_2$, and $I_1$,
and indeed the quiver's curve~(\ref{D2:curve}) has such singularities
for $H=M=0$: an $I_3$ at $\cosh(\ql a\varphi)=+1$, an $I_2$ at
$\cosh(\ql a\varphi)={3\over2}$, and an $I_1$ at $\cosh(\ql a\varphi)=-3$,
{\it cf.}\ eqs.~(\ref{D2:SingLoci}) and~(\ref{D2:ordBB}).

\item[\large\red$\star$]
Finally, for $h<0$, there are three distinct patterns of the branching
points depending on $\phi$:
for $\phi>-h$, the branching points are as in \eq{D2:Roots:positive}
and $\fdc=2h+6\phi$;
for ${-h\over2}<\phi<-h$, we have
\be
x_1\,\approx\,-4\,e^{-\ql aH},\quad
x_2\,\approx\,-\coeff14\, e^{\ql a(H+2\varphi)},\quad
x_3\,\approx\,-4\, e^{-\ql a(H+2\varphi)},\quad
x_4\,\approx\,-\coeff14\, e^{+\ql aH},
\label{D2:Roots:negHmidPhi}
\ee
with crossratio $\chi={1\over16}e^{\ql a(2H+4\varphi)}$ and hence
$\fdc=4h+8\phi$;
and for $0<\phi<{-h\over2}$ the branching points are
\be
x_1\,\approx\,-4\,e^{-\ql aH},\qquad
x_{2,3}\,\approx\,+1\,\pm\,\half\,e^{\ql a(H+2\varphi)/2},\qquad
x_4\,\approx\,-\coeff14\, e^{+\ql aH},
\label{D2:Roots:negHsmallPhi}
\ee
with crossratio $\chi=e^{+\ql a(H+2\varphi)/2}\ll1$ and hence
$\fdc=-h-2\phi$.
Altogether,
\be
\mbox{for}\ h<0,\qquad \fdc(\phi)\ =\,\cases{
	2h+6\phi & for $\phi>-h$,\cr
	4h+8\phi & for ${-h\over2}<\phi<-h$,\cr
	-h-2\phi & for $\phi<{-h\over2}$.\cr
	}
\label{D2:negative}
\ee
\end{itemize}

The following diagram summarizes the various regimes of the $D_2$
theory for $m=0$:
\be
\psset{xunit=21mm,yunit=3mm,runit=5mm,framearc=0.2}
\psset{arrowscale=2,linewidth=0.5pt,linecolor=black,fillstyle=solid}
\begin{pspicture}[0.4](-1,-2)(+4.5,+26)
\psline{->}(0,0)(3.5,0)
\rput[l](3.5,0){$\,\phi$}
\psline{->}(0,0)(0,25)
\rput[b](0,25.1){$\fdc$}
\psline[linecolor=blue,linewidth=3pt](0,4)(3.5,25)
\rput[lb](3.5,25){\psframebox{$\blue h>0$}}
\rput[b]{40.6}(2.8,21){$\rm\blue slope=6$}
\pscircle*[linecolor=blue](0,4){0.25}
\rput[r](-.1,4){$\blue D_2$}
\psline[linecolor=green,linewidth=3pt](0,0)(3.5,21)
\rput[lb](3.5,21){\psframebox{$\green h=0$}}
\rput[b]{40.6}(2.8,17){$\rm\green slope=6$}
\pscircle[linecolor=green,linewidth=0.12,fillcolor=yellow](0,0){0.25}
\rput[rt](-.1,0){\green $E_3$ SCFT}
\psline[linecolor=red,linewidth=3pt](1,0)(2,8)(3.5,17)
\rput[lb](3.5,17){\psframebox{$\red h<0$}}
\rput[b]{40.6}(2.8,13){$\rm\red slope=6$}
\rput[b]{48.8}(1.42,4){$\rm\red slope=8$}
\pscircle*[linecolor=red](2,8){0.25}
\rput[lt](2,8){\ \red two quarks}
\psline[linecolor=red,linewidth=3pt,linestyle=dotted](0,2)(1,0)
\pscircle[linecolor=red,linewidth=0.12,fillcolor=white](1,0){0.25}
\pscircle[linecolor=red,linewidth=0.12,fillcolor=white](0,2){0.25}
\rput[r](-.1,2){\red\bf ?}
\rput[b](1,.6){\red\bf ?}
\psline(1,-.5)(1,-1)
\rput[t](1,-1.1){$\red -h/2$}
\psline(2,-.5)(2,-1)
\rput[t](2,-1.1){$\red -h$}
\end{pspicture}
\label{D2:Diagram1}
\ee
The solid circle on the red line here (for $h<0$) indicate a double
flop transition at $\phi=-h$.
Two quarks become massless at this point and that's where the $B_{12}$
baryonic branch lives for $h<0$.
Indeed according to \eq{D2:bb5d:12}, the ordinary baryonic branch moves
from its classically expected location at $\phi=0$ (for $m=0$)
to $\phi=-h$.

The physical meaning of the open red circles at $\phi={-h\over2}$
and $\phi=0$ is moot because the whole dotted segment of the red line
is unphysical.
From the 4D, $\NN=2$ point of view, for $0\le\phi<{-h\over2}$
the scalar superpartner $\cal A$ of the massless abelian vector decouples
from the $\varphi$ modulus.
Indeed, for the $x_{1,2,3,4}$ branching points as in \eq{D2:Roots:negHsmallPhi},
\eq{D0:superpartner} yields
\be
{d{\cal A}\over d\varphi}\ \approx\ {-\ql a(H+2\varphi)\over 2\pi}\times
e^{+\ql a(H+2\varphi)/2}\ \ll\ 1.
\label{D2:decoupling}
\ee
To eliminate this unphysical range we change variables from $\varphi$ to
$\hat\varphi$ according to \eq{D0:remap}.
In terms of the $\hat\varphi$, its real part $\hat\phi$ is non-negative,
and in the $\hat\phi\to0$ limit we have $\fdc=0$ and hence a non-trivial
%\eq{D2:negative} yields $\fdc=0$ for $\hat\phi=0$ 
SCFT in 5D.
Also in terms of $\hat\varphi$, the $O(1/\ql a)$ neighborhood of $\hat\varphi=0$
contains three singularities~(\ref{D2:SingLoci}) of the spectral curve, namely
an $I_2$ singularity at $\cosh(\ql a\hat\varphi)\approx0$ (note $M=0$), and a
pair of $I_1$ singularities at $\cosh(\ql a\hat\varphi)\approx\pm2$.
This singularity structure in 4D indicates the 5D SCFT at $\hat\phi=0$ is $E_1$.
And to confirm this identification, we note that
there is a Higgs branch growing out of the superconformal
point $\hat\phi=0$, namely the mesonic branch of the quiver theory.
 
Altogether, we deconstruct the following Coulomb phase diagram of the
$SU(2)$ theory with $n_f=2$, $\Delta F=1$, and $m_1=m_2=0$:
\be
\psset{unit=0.85cm}
\begin{pspicture}[0.5](-9,-1)(+8,+8.5)
\pspolygon*[linecolor=paleblue](0,0)(-5,7)(+7,7)(+7,0)
\pspolygon*[linecolor=palered](0,0)(-5,7)(-7,7)(-7,5)
\pspolygon*[linecolor=gray](0,0)(-7,0)(-7,5)
\psline[linecolor=blue,linewidth=3pt](0,0)(+7,0)
\psline[linecolor=green,doubleline=true,doublecolor=white,linewidth=1.5pt]%
				(0,0)(-5.2,7.28)
\psline[linecolor=red,linewidth=3pt](0,0)(-7.28,5.2)
\psline[linewidth=0.5pt,arrowscale=2]{<->}(-7.5,0)(+7.5,0)
\rput[l](+7.6,0){$h>0$}
\rput[r](-7.6,0){$h<0$}
\psline[linewidth=0.5pt,arrowscale=2]{->}(0,0)(0,7.5)
\rput[b](0,7.6){$\phi$}
\pscircle*[linecolor=yellow](0,0){.2}
\rput(0,0){\large\red $\star$}
\rput*[l](-1,4){$D_2$ Coulomb phase}
\rput*[r]{315}(-3,3){$E_1$ Coulomb phase}
\rput*[r]{352.4}(-3,0.95){unphysical region}
\rput[l](3,-0.7){\blue $D_2$ origin}
\psarc[linecolor=blue,linewidth=1pt]{<-}(2.9,-0.2){.5}{180}{270}
\psarc[linecolor=blue,linewidth=1pt]{->}(5,-0.2){.5}{270}{360}
\rput[r](-0.6,-0.7){\red $E_3$ SCFT}
\psarc[linecolor=red,linewidth=1pt]{->}(-0.5,-0.2){.5}{270}{360}
\rput[bl](-6.5,8){\green double flop}
\rput[bl](-6.5,7.4){\green 2 massless quarks}
\rput[r](-7.4,5.2){\red $E_1$ SCFT}
\end{pspicture}
\label{D2:Diagram2}
\ee
As to the Higgs branches, they are rooted along the {\blue $D_2$ origin},
{\red $E_1$ SCFT}, and {\green double flop} lines
according to
\be
\psset{unit=1mm}
\def\cone[#1]{%
	\pspolygon*[linecolor=#1](-10.125,50)(0,0)(+10.125,50)
	\psellipse[linewidth=0.4pt,fillstyle=solid,fillcolor=#1](0,50)(10,2.5)
	\psline[linewidth=0.4pt](-10,49.4)(0,0)(+10,49.4)
	}
\begin{pspicture}[0.5](-88,-15)(+60,+60)
\rput(-60,0){%
    \rput{20}(0,0){\cone[brown]}
    \rput{340}(0,0){\cone[magenta]}
    \pscircle*[linecolor=blue](0,0){2}
    \rput[b](-18,55){\brown $B_{12}$ baryonic}
    \rput[b](+18,55){\magenta mesonic}
    \rput[t](0,-4){\blue $D_2$ origin}
    \rput[t](0,-11){\blue $h>0$, $\phi=0$}
    }
\rput(0,0){%
    \cone[magenta]
    \pscircle*[linecolor=red](0,0){2}
    \rput[b](0,55){\magenta mesonic}
    \rput[t](0,-4){\red $E_1$ SCFT}
    \rput[t](0,-11){\red $h<0$, $\phi=-h$}
    }
\rput(45,0){%
    \cone[brown]
    \pscircle*[linecolor=green](0,0){2}
    \rput[b](0,55){\brown $B_{12}$ baryonic}
    \rput[t](0,-4){\green double flop}
    \rput[t](0,-11){\green $h<0$, $\phi=-h/2$}
    }
\end{pspicture}
\label{D2:Diagram3}
\ee
and also at the {\red $E_3$ SCFT} point according to
\be
\psset{unit=1mm}
\begin{pspicture}[0.5](-50,-10)(+100,+60)
\rput{20}(0,0){%
    \pspolygon*[linecolor=brown](-10.125,50)(0,0)(+10.125,50)
    \psellipse[linewidth=0.4pt,fillstyle=solid,fillcolor=brown](0,50)(10,2.5)
    \psline[linewidth=0.4pt](-10,49.4)(0,0)(+10,49.4)
    }
\rput[b](-18,55){\brown $B_{12}$ baryonic}
\rput{330}(0,0){%
    \pspolygon*[linecolor=orange](-20.5,50)(0,0)(+20.5,50)
    \psellipse[linewidth=0.4pt,fillstyle=solid,fillcolor=orange](0,50)(20,5)
    \psline[linewidth=0.4pt](-20,48.75)(0,0)(+20,48.75)
    }
\rput[bl](+20,55){\orange mesonic mixed with $B_{13}+B_{23}$ baryonic}
\pscircle[linecolor=red,fillstyle=solid,fillcolor=yellow,linewidth=1](0,0){2}
\rput[t](0,-4.5){\red $E_3$ SCFT, $h=\phi=0$}
\end{pspicture}
\label{D2:Diagram4}
\ee
The mixed branch at the $E_3$ point has hypermultiplet $\rm dimension=2$
(real $\rm dimension=8$); all other Higgs branches have hypermultiplet
$\rm dimension=1$ (real $\rm dimension=4$).

\medskip
\centerline{\large\blue $\star\qquad\star\qquad\star$}
\medskip

And now consider the quiver theory for $m>0$.
Again, we look at patterns of the branching points~(\ref{D2:Roots})
for different values of $\phi$ and $h$.
\begin{itemize}
\item[$\blue\bullet$]
For $h>m$ there are two patterns:
for $\phi>m$ the roots are as in \eq{D2:Roots:positive} and hence
$\fdc=2h+6\phi$,
while for $0<\phi<m$
\be
x_{1,2}\ =\ e^{\ql a\varphi}\,\pm\,2e^{\ql a(H-M)/2},\qquad
x_{3,4}\ =\ e^{-\ql a\varphi}\,\pm\,2e^{\ql a(H-M-4\varphi)/2},
\label{D2:Roots:poslow}
\ee
the crossratio is $\chi=\frac{1}{16}e^{\ql a(4\varphi+H-M)}$,
and $\fdc=2h-2m+8\phi$.
Altogether,
\be
\fdc\ =\ \cases{ 2h\,+\,6\phi & for $\phi>m$,\cr
		2h\,-\,2m\,+\,8\phi & for $\phi<m$,\cr
		}
\label{D2:HbM}
\ee
in perfect agreement with the Seiberg formula for 5D $SU(2)$ with massive
flavors
\be
\fdc\ =\ 2h\ +\ 8\phi\ -\,\sum_f\max(\phi,|m_f|).
\label{SeibergFlaM}
\ee
At $\phi=m$, there is a double flop due to
two quark flavors becoming massless at the same time ---
note $\phi_1=m_1=-m$ and $\phi_2=m_2=+m$.
This point on the
Coulomb branch is the origin of the ordinary baryonic branch~$B_{12}$,
{\it cf.}\ \eq{D2:bb5d:12}.

Finally, for $\phi=0$ we have $SU(2)$ restoration in 5D.
Indeed the singularities~(\ref{D2:SingLoci}) of the spectral curve
form two Seiberg--Witten pairs
\be
(\ql a\varphi)^2\ =\ \pm 2 e^{-\ql a(H-M)/2}
\qquad{\rm and}\qquad
(\ql a\varphi\,-\,\pi i)^2\ =\ \pm 2 e^{-\ql a(H-M)/2}
\label{D2:HbM:SU2}
\ee
separated by Wilson $\rm line=\pi$.
%According to \eq{SWpairs}, such two pairs indicate unbroken $SU(2)$ in 5D.
Note $\Lambda_{\rm SW}^2=e^{-\ql a(H-M)/2}$ here indicates classical 4D
$SU(2)$ coupling ${8\pi^2\over g_4^2}=\ql a(h-m)$ and hence 5D coupling
$\fdc=h-m$.
This agrees with the $\phi\to0$ limit of the abelian coupling:
according to \eq{D2:HbM}, $\fdc(\phi=0)=2\times(h-m)$;
the factor of 2 here is the Clebbsch of the $U(1)\subset SU(2)$.

\item[$\star$]
Another way to understand the $\phi<m$ regime --- including $\phi\to0$ ---
is via effective SYM theory.
Note that for $\phi<m$ the 5D quarks are massive and we may integrate them
out.
The result is an effective $SU(2)$ SYM with no flavors, $\theta=0$
({\it cf.}\ \eq{VacuumAngle}), and inverse coupling $h^{\rm eff}_{}=h-m$
({\it cf.}\ \eq{D2:HbM} for $\phi<m$).
In terms of the spectral curve integration out works by focusing on $x$
being neither too large nor too small, specifically
$e^{-\ql aM}\ll x\ll e^{+\ql a M}$; in this regime,
the curve~(\ref{D2:curve}) may be approximated as
\be
y^2\ -\ y\times\Bigl(x^2\,-\,2\cosh(\ql a\varphi)\times x\,+\,1\Bigr)\
+\ e^{-\ql a(H-M)}\times x^2\,,
\label{D2:effectiveD0}
\ee
which looks exactly like the curve for $n^{\rm eff}_f=0$,
$\Delta F^{\rm eff}_{}=2$, and $H^{\rm eff}_{}=H-M$.
And as in \S6.2, this curve  yields $\fdc=2h^{\rm eff}_{}+8\phi=
2(h-m)+8\phi$ for $\phi>0$, and for $\phi\to0$ and $h^{\rm eff}_{}>0$
it has two pairs of Seiberg--Witten singularities indicating $SU(2)$
restoration in 5D.

\item[$\darkcyan\bullet$]
For $h=m$, the patterns or branching points are similar to the
$h>m$ regimes for $\phi>m$, $\phi=m$, and $0<\phi<m$,
but for $\phi\to0$ there is a difference:
$\fdc\to0$ at this point, which indicates a superconformal theory in 5D.
The nature of this SCFT is clear from the effective theory --- SYM with
$\theta=0$ --- whose superconformal limit at $h^{\rm eff}_{}=h-m=0$
and $\phi=0$ is $E_1$, exactly as in~\S6.2.
Moreover, the $E_1$ has a Higgs branch growing out of the superconformal
point, and the quiver theory does have a Higgs branch at precisely this point,
namely the exotic baryonic branch with flavors 1 and 3,
{\it cf.}\ \eq{D2:bb5d:13}.

\item[$\green\bullet$]
For $-m<h<+m$ there are three regimes:
for $\phi>m$ the branching points are as in \eq{D2:Roots:positive},
for ${m-h\over2}<\phi<m$ they are as in \eq{D2:Roots:poslow}, and
for $0<\phi<{m-h\over2}$ we have a new pattern, namely
\be
x_{1,2}\ =\ \pm 2\,e^{+\ql a(M-H)/2},\qquad
x_{3,4}\ =\ \mp{1\over2}\,e^{-\ql a(M-H)/2},
\label{D2:Roots:midlow}
\ee
with crossratio $\chi=e^{\ql a(M-H)}$ and hence $\fdc=2(m-h)$, regardless
of $\phi$ (as long as $\phi<{m-h\over2}$).
Altogether,
\be
\mbox{for}\ -m<h<+m,\quad \fdc\ =\,\cases{
	2h\,+\,6\phi & for $\phi>m$,\cr
	2h\,-\,2m\,+\,8\phi & for ${m-h\over2}<\phi<m$,\cr
	2m\,-\,2h & for $0<\phi<{m-h\over2}$.\cr
	}
\label{D2:middle}
\ee
However, the third regime here is unphysical because
the $\NN=2$ superpartner $\cal A$ of the vector field
decouples from $\varphi$; indeed, for branching
points as in~(\ref{D2:Roots:midlow}) we have
\bea
\frac{d{\cal A}}{d\varphi} &=& 
\frac{2\sinh(\ql a\varphi)}{2\pi i}\times
    \oint\frac{dx}{\sqrt{(x-x_1)(x-x_2)(x-x_3)(x-x_4)}} &
%\omit\hfill
(\ref{D0:superpartner})\cr
&\approx& \frac{e^{\ql a\varphi}}{\sqrt{x_1\,x_2}}
    \label{D2:superpartner}\\
&\approx& \coeff{i}{2}\,e^{\ql a(2\varphi+H-M)/2}\ \ll\ 1.
\eea
In terms of the effective theory, the decoupling happens for
and $2\phi+h^{\rm eff}_{}<0$, and it works exactly as in \S6.2.
And similar to \S6.2, at the edge of decoupling $\phi=-\half h^{\rm eff}_{}
={m-h\over2}$ there is unbroken $SU(2)$ in 5D --- which manifests in 4D
via singularities~(\ref{D2:SingLoci}) forming
two pairs of Seiberg--Witten points,
\be
\ql a\varphi\ =\ \ql a\,{M-H\over2}\ +\ \log(2)\
+\ \{0\ {\rm or}\ \pi i\}\ \pm\ e^{-\ql a(M-H)/2} .
\label{D2:midlow:SW}
\ee
The proper modulus of $SU(2)$ restoration is $\tilde\phi=\phi+{h-m\over2}$,
or in 5D terms
\be
\cosh(\ql a\tilde\varphi)\ =\ \cosh(\ql a\varphi)\times
{e^{\ql a H/2}\over\sqrt{2\cosh(\ql a M)}}\,,
\label{D2:remap}
\ee
where the hyperbolic cosines indicate that variables $\varphi$, $M$,
and $\tilde\varphi$ all live on half-cylinders.

Finally, at $\phi=m$ there is a double flop due to two quarks
becoming massless at the same time;
the ordinary baryonic branch grows out of this point,
{\it cf.}\ \eq{D2:bb5d:12}.

\item[$\brown\bullet$]
For $h=-m$ the intermediate range of $\phi$ disappears and there are
only two regimes of the branching points: (\ref{D2:Roots:positive})
for $\phi>m$ and (\ref{D2:Roots:midlow}) for $\phi<m$, thus
\be
\mbox{for}\ h=-m,\quad\fdc\ =\ \cases{
	2h\,+\,6\phi & for $\phi>m$,\cr
	4m & for $0\le\phi<m$.\cr
	}
\label{D2:HmM}
\ee
And again, the second regime here is unphysical because $\cal A$
decouples from $\varphi$.
However, the regime boundary at $\phi=m$ is more complicated than
for $h>-m$ because now $SU(2)$ restoration happens at the same
point where two quarks become massless.
In 5D terms, this corresponds to an effective $D_2$ theory with
two $m^{\rm eff}_{}=0$ quarks.
In the continuum 5D limit this effective theory should have a global
$SO(4)$ symmetry, but the 4D quiver theory itself is not $SO(4)$
symmetric.
Instead, we identify the $h=-m$, $\phi=m$ ($i.\,e.,\ \tilde\phi=0$) point
as an effective $D_2$ origin via singularities of
the spectral curve and also via Higgs branches.
Indeed, for $H=-M$ the curve~(\ref{D2:curve})
of our quiver theory has two double ($I_2$) and two simple ($I_1$)
singularities near $\tilde\varphi=0$ or ${\pi i\over\ql a}$, namely
\be
\vcenter{\openup 1\jot \ialign{
	\hfil #\unskip\ at:\quad &
	$\displaystyle{\cosh(\ql a\varphi)\ =\ #}$\quad\hfil &
	$\displaystyle{\Longrightarrow\quad #\ }$\hfil &
	$\displaystyle{\approx\ #}$\hfil\cr
	first $I_2$ & e^{\ql aM}\ +\ \coeff12 e^{-\ql aM} &
	(\ql a\tilde\varphi)^2 & \coeff14 e^{-4\ql a M},\cr
	second $I_2$ & e^{\ql aM} &
	(\ql a\tilde\varphi)^2 & -e^{-2\ql aM},\cr
	two $I_1$'s & -e^{\ql aM}\ \pm 2 &
	(\ql a\tilde\varphi\,-\,\pi i)^2 & \pm\ 4e^{-\ql aM},\cr
	}}
\label{D2:D2:Sing}
\ee
and this is precisely the singularity structure of $D_2$,
{\it cf.}\ \eqrange{D2:D2:SW0}{D2:D2:SW2}.
Moreover, there are two distinct Higgs branches rooted at the
double singularities near $\tilde\varphi=0$, namely
the ordinary baryonic branch $B_{12}$ rooted at the first $I_2$
({\it cf.}\ \eq{D2:ordBB} for $e^{-\ql a H}=e^{+\ql aM}$),
and the exotic baryonic branch $B_{23}$ rooted at the second $I_2$
({\it cf.} \eq{D2:exBB}).
In terms of the effective $D_2$ theory, one of these Higgs branches
corresponds to the mesonic branch and the other to the baryonic branch.
Altogether, the 4D singularities and the Higgs branches confirm our
identification of the $\phi=-h=m$ point as the $D_2$ theory in 5D.

\item[$\red\bullet$]
Finally, for $h<-m$ we have three regimes:
for $\phi>-h$, the branching points of the spectral curve as in
\eq{D2:Roots:positive}; for ${m-h\over2}<\phi<-h$ we have a new pattern
\be
x_1\,\approx\,-4\,e^{-\ql aH},\quad
x_2\,\approx\,-\coeff14\, e^{\ql a(H+2\varphi)},\quad
x_3\,\approx\,-4\, e^{-\ql a(H+2\varphi)},\quad
x_4\,\approx\,-\coeff14\, e^{+\ql aH},
\label{D2:Roots:negmid}
\ee
with crossratio $\chi={1\over16}e^{\ql a(2H+4\varphi)}$;
and for $0\le\phi<{m-h\over2}$ the branching points are
\be
x_1\,\approx\,-4\,e^{-\ql aH},\qquad
x_2\,\approx\,-e^{\ql aM},\qquad
x_3\,\approx\,-e^{-\ql aM},\qquad
x_4\,\approx\,-\coeff14\, e^{+\ql aH},
\label{D2:Roots:neglow}
\ee
with crossratio $\chi=e^{2\ql aM}$, regardless of $\varphi$.
Altogether, this gives us
\be
\mbox{for}\ h<-m,\quad\fdc\ =\ \cases{
	2h+6\phi & for $\phi>-h$,\cr
	4h+8\phi & for ${m-h\over2}<\phi<-h$,\cr
	4m & for $0\le\phi<{m-h\over2}$.\cr
	}
\label{D2:HlmM}
\ee
Note that the third regime here is unphysical because for branching
points~(\ref{D2:Roots:neglow})
\bea
\frac{d{\cal A}}{d\varphi} &\approx&
\frac{e^{\ql a\varphi}}{\sqrt{x_1\,x_2}} & (\ref{D2:superpartner})\cr
\noalign{\vskip 5pt plus 2pt}
&\approx& \half e^{\ql a(H-M+2\varphi)/2}\ \ll\ 1.
\eea
In other words, the proper 5D modulus is not $\phi\ge0$ but $\tilde\phi=
\phi-{m-h\over2}\ge0$.
Moreover, at the endpoint $\tilde\phi=0$ there is unbroken $SU(2)$ in 5D
as evidenced by singularities~(\ref{D2:SingLoci}) of the spectral curve
forming two Seiberg--Witten pairs
\be
\eqalign{
\ql a\varphi\ &
=\ \ql a\, {M-H\over2}\ +\ \log(2)\
+\ \{0\ {\rm or}\ \pi i\}\ \pm\ e^{-\ql aM},\cr
(\ql a\tilde\varphi)^2\quad &
\mbox{or}\quad (\ql a\tilde\varphi\,-\,\pi i)^2\
=\ \pm 2\,e^{-2\ql aM}.\cr
}\label{D2:HlmM:SWW:local}
\ee
Finally, at $\phi=-h$ --- which corresponds to $\tilde\phi={-h-m\over2}>0$ ---
there is a double flop transition due to two quark flavors becoming massless.
As usual, such double flop is root of the ordinary baryonic
branch~$B_{12}$, {\it cf.}\ \eq{D2:bb5d:12} for $h<-m$.

\end{itemize}

Altogether, various regimes of our model are summarized on the following
diagram:
\be
\psset{xunit=28mm,yunit=4mm,runit=5mm,framearc=0.2}
\psset{arrowscale=2,linewidth=0.5pt,linecolor=black,fillstyle=solid}
\begin{pspicture}[0.4](-0.5,-3)(+4.5,+34)
\psline{->}(0,0)(4,0)
\rput[l](4,0){$\,\phi$}
\psline{->}(0,0)(0,32)
\rput[b](0,32.1){$\fdc$}
\psline[linecolor=red,linewidth=3pt](2,4)(3,12)(4,18)
\rput[lb](4,18){\psframebox{$\red h<-m$}}
\pscircle[linecolor=red,linewidth=0.12,fillcolor=white](3,12){0.25}
\pscircle*[linecolor=red](2,4){0.25}
\psline[linecolor=red,linewidth=3pt,linestyle=dotted](2,4)(0,4)
\rput[lt](3.02,11.9){\red\ 2 quarks}
\rput[lt](2.02,3.9){$\red\; SU(2)$}
\psline[linecolor=blue,linewidth=3pt](0,6)(1,14)(4,32)
\rput[lb](4,32){\psframebox{$\blue h>m$}}
\pscircle[linecolor=blue,linewidth=0.12,fillcolor=white](1,14){0.25}
\pscircle*[linecolor=blue](0,6){0.25}
\rput[r](0,6){$\blue SU(2)\,\,$}
%\rput[bl]{41}(0.9,13.7){\blue 2 quarks}
\rput[r](0.92,14){\blue 2 quarks}
\psline[linecolor=darkcyan,linewidth=3pt](0,0)(1,8)(4,26)
\rput[lb](4,26){\psframebox{$\darkcyan h=m$}}
\pscircle[linecolor=darkcyan,linewidth=0.12,fillcolor=white](1,8){0.25}
\pscircle*[linecolor=darkcyan](0,0){0.25}
\rput[r](0,0){\darkcyan $E_1$ SCFT\ }
%\rput[bl]{41}(0.9,7.7){\darkcyan 2 quarks}
\rput[r](0.92,8){\darkcyan 2 quarks}
\psline[linecolor=green,linewidth=3pt](0.5,2)(1,6)(4,24)
\rput[lb](4,24){\psframebox{$\green -m<h<m$}}
\pscircle[linecolor=green,linewidth=0.12,fillcolor=white](1,6){0.25}
\pscircle*[linecolor=green](0.5,2){0.25}
\psline[linecolor=green,linewidth=3pt,linestyle=dotted](0.5,2)(0,2)
\rput[lt](0.5,1.9){$\green\; SU(2)$}
%\rput[bl]{41}(0.9,5.7){\green 2 quarks}
\rput[r](0.90,6){\psframebox*[boxsep=false,framesep=1pt]{\green 2 quarks}}
\psline[linecolor=brown,linewidth=3pt](1,4)(4,22)
\rput[lb](4,22){\psframebox{$\brown h=-m$}}
\psline[linecolor=brown,linewidth=3pt,linestyle=dotted](1,4)(0,4)
\pscircle*[linecolor=brown](1,4){0.35}
\pscircle*[linecolor=black](1,4){0.2}
\rput[lt](1,3.5){\brown\ $D_2$}
\psline[linestyle=dotted](1,-1)(1,20)
\rput[t](1,-1.2){$m$}
\psline[linestyle=dotted](0.5,2)(0.5,-1)
\rput[t](0.5,-1.2){$\darkcyan m-h\over2$}
\psline[linestyle=dotted](2,4)(2,-1)
\rput[t](2,-1.2){$\red m-h\over2$}
\psline[linestyle=dotted](3,12)(3,-1)
\rput[t](3,-1.2){$\red -h$}
\end{pspicture}
\label{D2:Diagram5}
\ee
The colored lines here plot $\fdc(\phi)$ for several fixed values of $h$
(different colors for different $h$); the solid lines correspond to
physical 5D regimes, and the dotted lines to unphysical regimes for
$\phi<{m-h\over2}$ where $\varphi$ decouples from the low-energy
degrees of freedom.
Note the blue, cyan, and green lines are bent at $\phi=m$: the slope
$d\fdc/d\phi$ is 6 for $\phi>m$ and 8 for $\phi<m$.
Likewise, the red line is bend at $\phi=-h$; the slope is 6 for
$\phi>-h$ and 8 for $\phi<-h$.
In terms of the Seiberg formula~(\ref{SeibergFla}), $\rm slope=6$
corresponds to Coulomb branch of the $D_2$ theory, while
$\rm slope=8$ corresponds to Coulomb branch of an effective SYM
theory.
Thus, we deconstruct the following Coulomb phase diagram of
the $SU(2)$ theory with $n_f=2$, $\Delta F=1$, and fixed $m_2=-m_1=m>0$:
\be
\psset{unit=3.7mm,framearc=0.2}
\begin{pspicture}[0.7](-19,-2)(+21,+20)
\pspolygon*[linecolor=paleblue](-5,7)(-12.5,17.5)(+15,17.5)(+15,7)
\pspolygon*[linecolor=palered](-5,7)(+15,7)(+15,0)(+5,0)
\pspolygon*[linecolor=palered](-5,7)(-12.5,17.5)(-15,17.5)(-15,14)
\pspolygon*[linecolor=gray](+5,0)(-15,0)(-15,14)
\psline[linecolor=green,doubleline=true,doublecolor=white,linewidth=1.5pt]%
			(+15.5,7)(-5,7)(-12.8,17.92)
\psline[linecolor=red,linewidth=3pt](+15,0)(+5,0)(-15.4,14.28)
\psline[linewidth=0.5pt,arrowscale=2]{<->}(-16,0)(+16,0)
\rput[l](+16.2,0){$h>0$}
\rput[r](-16.2,0){$h<0$}
\psline[linewidth=0.5pt,arrowscale=2]{->}(0,0)(0,18.5)
\rput[b](0,18.7){$\phi$}
\pscircle*[linecolor=yellow](+5,0){.4}
\rput(+5,0){\large\red $\star$}
\pscircle*[linecolor=blue](-5,7){.4}
\rput(-5,7){\large\white $\star$}
\rput*[l](-3,12){$D_2$ Coulomb phase}
\rput*[l]{345}(+1,5.25){Eff.~SYM Coulomb phase}
\rput*[rb](-4,2.5){unphysical region}
%\rput*(-12,14){$D'_0$}
\rput[r](+4,-1.5){\orange $E_1$ SCFT}
\psarc[linewidth=1pt,linecolor=orange]{->}(+4,-0.5){1}{270}{360}
\rput[r](+11,-1.5){\red $SU(2)$}
\psarc[linewidth=1pt,linecolor=red]{->}(+11,-0.5){1}{270}{360}
\rput[r](-4,-1.5){\red SU(2)}
\psline[linewidth=1pt,linecolor=red,border=0.5pt]{->}(-3.5,-1.5)(-1,4.1)
\rput[r](-15.6,14.42){\red $SU(2)$}
\rput*[rt](-5.6,6.4){\blue $D_2$ point}
\rput[l](+15.7,7.6){\green double flop}
\rput[l](+15.7,6.4){\green 2 massless quarks}
\rput[lb](-15,19.4){\green double flop}
\rput[lb](-15,18.2){\green 2 massless quarks}
\end{pspicture}
\label{D2:Diagram6}
\ee
The effective SYM for $h>-m$ and $\phi<m$ is $D_0$:
it has $\theta=0$ in 5D, and in 4D the effective curve~(\ref{D2:effectiveD0})
has $\Delta F^{\rm eff}_{}=0$.
The other effective SYM for $h<-m$ and $\phi<-h$ is also $D_0$, and in fact
at the spectral curve level, there is a symmetry between the two $D_0$ phases.
To make this symmetry manifest, we change the $y$ variable in \eq{D2:curve}
to $\bar y=-y/(x-e^{-\ql aM})$ and then rewrite the spectral curve as
\be
{\bar y}^2\times\left(x\,-\,e^{-\ql aM}\right)\
+\ \bar y\times\left(x^2\,-\,2\cosh(\ql a\varphi)\times x\,+\,1\right)\
-\ e^{-\ql aH}\times x\times\left(x\,-\,e^{+\ql aM}\right)\ =\ 0,
\label{D2:curve:mod}
\ee
or equivalently
\be
x\bar y(x+\bar y)\
-\ \left(e^{-\ql aH}\times x^2\,+\, e^{-\ql aM}\times {\bar y}^2\,
	+\, 2\cosh(\ql a\varphi)\times x\bar y\right)\
+\ \left( e^{\ql a(M-H)}\times x\,+\,y\right)\ =\ 0.
\label{D2:curve:sym}
\ee
The resulting equation is invariant under
\be
\eqalign{
x\ &\to\ \bar y\times e^{\ql a(M-H)/2},\qquad
\bar y\ \to\ x\times e^{\ql a(M-H)/2},\cr
H\ &\to\ {3M-H\over2},\qquad
M\ \to\ {M+H\over2},\qquad
\cosh(\ql a\varphi)\ \to\ \cosh(\ql a\varphi)\times e^{\ql a(H-M)/2},\cr
}\label{D2:Z2}
\ee
and this symmetry indeed interchanges the two effective $D_0$ phases
on the Coulomb phase diagram~(\ref{D2:Diagram6}).
Note however that this is not a symmetry of the quiver theory itself but
only of its spectral curve.

As to the Higgs branches of the deconstructed
$SU(2)$ theory with $n_f=2$, $\Delta F=1$, and $m_2=-m_1=m>0$, there is
the ordinary baryonic branch $B_{12}$ rooted along the double-flop
line $\phi=\max(m,-h)$, and two exotic baryonic branches:
$B_{13}$ rooted at the $E_1$ SCFT point $h=+m$, $\phi=0$ and $B_{23}$
rooted at the $D_2$ point $\phi=-h=m$.
There are no mesonic or mixed Higgs branches for $m\neq0$.

\medskip
\centerline{\large\blue $\star\qquad\star\qquad\star$}
\medskip

And now we move on to another example of deconstructed \sqcdv\
with $n_c=n_f^{}=2$.
This time, we take $\Delta F=0$ rather than 1, and impose a different
constraint on the quark masses, namely $m_1=m_2=\tilde m$, or in 4D terms
$\mu_1=\mu_2=Ve^{a\tilde M}$.\footnote{%
	For this model, we put tildes on all the variables:
	$\tilde M$, $\tilde H$, $\tilde\varphi$, {\it etc., etc.}
	Such notations help discuss duality~(\ref{D2t:duality})
	betwen spectral curves of this model and the previous model
	with $\Delta F=1$.
	The tildes make clear which variable belongs to which model.
	}
This choice gives us a quiver with manifest $U(2)$ flavor symmetry.
On the other hand, flavor symmetry enhancement $U(2)\to SO(4)$ in 5D for $M=0$
is not protected by the discrete custodial symmetry {\bf C}
(\ref{Csym:XY}--\reftail{Csym:5dp}) because $\Delta F\neq1$ breaks~{\bf C}.

The spectral curve of the $\Delta F=0$ quiver is
\be
\tilde y^2\ -\ \tilde y\times
	(\tilde x^2\,-\,2\cosh(\ql a\tilde \varphi)\times \tilde x\,+\,1)\
+\ e^{-\ql a\tilde S}\times\left(\tilde x\,-\,e^{\ql a\tilde M}\right)^2\ =\ 0
\label{D2t:curve}
\ee
where $\tilde S=\tilde H+\tilde M$ ({\it cf.}\ \eq{Hdef}) and
$\Re\tilde H\ge\Re\tilde M$ ({\it cf.}\ \eq{Hlimit1}.
Consequently, for $\tilde m\equiv\Re\tilde M>0$ we expect no phase transitions
(except for a double flop at $\tilde\phi=\tilde m$),
but for $\tilde m<0$ there should be distinct phases
for $\tilde h\equiv\Re\tilde H>-\tilde m$ and for $\tilde m<\tilde h<-\tilde m$.

The curve~(\ref{D2t:curve}) factorizes for $\tilde\varphi=\pm\tilde M$ where the
quiver has a mesonic Higgs branch, and also for
\be
\cosh(\ql a\tilde\varphi)\ =\ e^{\ql a\tilde M},\qquad
e^{2\ql a\tilde M}\ +\ e^{-\ql a(\tilde H+\tilde M)}\ =\ 1,
\label{D2t:baryonic}
\ee
where the quiver has a baryonic Higgs branch~$B_{12}$.
This is the only baryonic branch for this quiver: because of
$\Delta F=0$ there are no exotic branches.
In fact, there are no Higgs branches other than the mesonic and the (ordinary)
baryonic branches;
in the 5D limit $\ql a\to\infty$ they are located at:
\be
\eqalign{
\omit mesonic:\hfil\quad & \tilde\phi=|\tilde m|,\cr
\omit baryonic:\hfil\quad & \tilde\phi=0,\quad \tilde h\ge 0,\quad
	\tilde m=0\ {\rm or}\ \tilde m=-\tilde h.
}\label{D2t:Higgs:5d}
\ee

We may analyze the Coulomb branch of the present quiver in the usual way,
by studying the branching points of the curve~(\ref{D2t:curve}),
but there is an easier way.
The curve~(\ref{D2t:curve}) happens to be dual to the curve~(\ref{D2:curve})
of the previous model:
\bea
\!y^2\ -\ y\times\Bigl(x^2-2\cosh(\ql a \varphi)\times x+1\Bigr) &-&
e^{-\ql aH}\times x\times\Bigl(x^2-2\cosh(\ql a M)\times x+1\Bigr)\
 =\ 0\!\nonumber\\
&\hbox{\rput{90}(0,0){$\Longleftrightarrow$}} &\nonumber\\
\!\tilde y^2\ -\ \tilde y\times(\tilde x^2\,
	-\,2\cosh(\ql a\tilde\varphi)\times \tilde x\,+\,1) &+&
e^{-\ql a\tilde S}\times\left(\tilde x\,-\,e^{\ql aM}\right)^2\ =\ 0\nonumber\\
\noalign{\vskip 5pt plus 3pt}
\mbox{for}\quad \tilde y\,=\,{y\,+\,e^{-\ql aH}x\over 2\cosh(\ql aM)}\,,&&
\quad \tilde x\,=\,-{y\over x}\times{e^{+\ql aH/2}\over\sqrt{2\cosh(\ql aM)}}\,,
\label{D2t:duality}\\
\noalign{\vskip 5pt plus 3pt}
e^{-\ql a\tilde S}\,=\,{e^{+\ql aH}\over 2\cosh(\ql aM)}\,,&&
\quad e^{\ql a\tilde M}\,=\,{e^{-\ql aH/2}\over\sqrt{2\cosh(\ql aM)}}\,,
\nonumber\\
\noalign{\vskip 5pt plus 3pt}
\mbox{and}\ \cosh(\ql a\tilde\varphi)\ =\ \cosh(\ql a\varphi) &\times&
{e^{\ql a H/2}\over\sqrt{2\cosh(\ql a M)}}\,,\quad
\mbox{{\it cf.}\ \eq{D2:remap}.}\nonumber
\eea
Physically, the two $n_c=n_f=2$ quiver theories
are {\sl not} dual to each other;
they are not even in the same universality class.
Indeed, the Higgs branches of the two theories
are not quite dual to each other:
\bea
n_c=n_f=2,\ \Delta F=1\ \mbox{quiver} &\longleftrightarrow&
n_c=n_f=2,\ \Delta F=0\ \mbox{quiver} \label{D2:HiggsDuality}\\
B_{12}\ \mbox{baryonic branch} &\longleftrightarrow&
\mbox{mesonic branch,}\nonumber\\
B_{13}\ {\rm and}\ B_{23}\ \mbox{baryonic branches} &\longleftrightarrow&
\mbox{baryonic branch(es),}\nonumber\\
\mbox{mesonic branch} &\longleftrightarrow&
\mbox{\red nothing.}\footnotemark\nonumber
\eea
\footnotetext{%
	The spectral curve~(\ref{D2:curve}) of the $\Delta F=1$ quiver
	has an $I_2$ singularity at the mesonic root $\varphi=M=0$,
	and the duality maps it onto a similar $I_2$ singularity of the
	$\Delta F=0$ quiver's spectral curve at $\cosh(\ql a\tilde\varphi)=
	\half e^{-\ql a\tilde M}=2e^{-\ql a\tilde H}$.
	However, despite this singularity, the curve~(\ref{D2t:curve})
	does not factorize.
	Also, at this point, the link resolvent $T(\tilde X)$ of
	the $\Delta F=0$ quiver has no poles at the physical sheet;
	instead, there is a pole on the unphysical sheet with $\rm residue=2$.
	Anyhow, there is no Higgs branch at this point.
	}
However, their spectral curves are dual, and we may use use this duality
to obtain the Coulomb phase diagram of our second $n_c=n_f=2$ example
without too much work.

In the 5D limit $\ql a\to\infty$, the duality map~(\ref{D2t:duality}) becomes
\be
\tilde h\ =\ {3|m|-h\over2}\,,\qquad
\tilde m\ =\ {-|m|-h\over2}\,,\qquad
|\tilde\phi|\ =\ |\phi|\ +\ {h-|m|\over2}\,.
\label{D2t:duality:5d}
\ee
Note that according to this map, there is a lower limit $\tilde h\ge\tilde m$;
remarkably, this limit agrees with \eq{Hlimit1} which follows from very
different physics, namely quantum corrections $V=v+\cdots$ in a quiver
without $\mu=0$ quarks.
This agreement indicates that the map~(\ref{D2t:duality:5d}) covers
all physical Coulomb phases of the two theories
(even though it misses some of the Higgs phases).
Consequently, applying this map to
% the Coulomb phase
diagrams~(\ref{D2:Diagram2}) and~(\ref{D2:Diagram6}) of the $\Delta F=1$
model, we arrive at the following
{\blue Coulomb phase diagram of deconstructed \sqcdv\ with
$n_c=n_f=2$, $\Delta F=0$, and $m_1=m_2=\tilde m$:}
\be
\psset{unit=11mm,linewidth=0.5pt,arrowscale=1.2}
\def\dfl{\psframebox[linecolor=green]{\vbox{%
		\baselineskip=12pt\hbox{double}\hbox{flop}%
		}}}
\begin{pspicture}[0.1](-6.5,-4)(+6.5,+6)
\pspolygon[fillcolor=lightgray,fillstyle=solid](-5,+1)(-5,+5)(+5,+3)(+5,-1)
\pspolygon[fillcolor=darkgray,fillstyle=solid](0,0)(-5,+5)(-5,+1)(-3,-3)
\pspolygon[fillcolor=red,fillstyle=solid](0,0)(-3,-3)(+5,-1)
\pspolygon[fillcolor=red,fillstyle=solid](0,0)(-3,-3)(-5,+5)
\psline[linecolor=yellow,linewidth=4pt](+5.1,-1.02)(0,0)(-3.1,-3.1)
\psset{hatchwidth=1.5pt,hatchsep=3pt}
\pspolygon[hatchcolor=green,fillstyle=crosshatch](0,0)(+5,+3)(+1,-2)
\pspolygon[hatchcolor=green,fillstyle=crosshatch](0,0)(-5,+5)(-3,+1)(+1,-2)
\psline[linecolor=blue,linewidth=4pt](-5.1,+5.1)(0,0)(+1,-2)
\psline(-3,+1)(-3,-3)
\pscircle*[linecolor=yellow](0,0){0.2}
\rput(0,0){\scalebox{2}{\blue$\star$}}
\psset{linewidth=1.5pt,linestyle=dotted}
\psline{>->}(-4.4,-1.1)(+5.4,+1.35)
\rput[lt](+5.5,+1.4){$\tilde m$}
\rput[r](-4.5,-1.1){$\tilde m$}
\psline{->}(0,0)(0,+4.4)
\rput[b](0,+4.5){$\tilde\phi$}
\psline{>->}(-2.4,+4.8)(+1.3,-2.6)
\rput[rt](+1.3,-2.7){$\tilde h$}
\rput[lb](-2.45,+4.9){$\tilde h$}
\psset{linestyle=solid,linewidth=1pt,framearc=0.25}
\rput*[b](0,1.2){$E'_3$ SCFT}
\psline[linecolor=yellow,linewidth=1.5pt]{->}(0,1.2)(0,0.25)
\rput[l](+5.1,-1){\psframebox[linecolor=yellow]{$E'_1$ SCFT}}
\rput[t](-3.1,-3.1){\psframebox[linecolor=yellow]{$E_1$ SCFT}}
\rput[rb](-5.1,+5.1){\psframebox[linecolor=blue]{$D'_2$ origin}}
\rput[l](+2,-2.5){\psframebox[linecolor=blue]{$D_2$ origin}}
\psline[linecolor=blue]{<-}(+1,-2)(+2,-2.5)
\rput[lt](+4,-2){\psframebox[linecolor=red]{$SU(2)$}}
\psline[linecolor=red]{<-}(+3,-1.5)(+4,-2)
\rput[lt](-0.5,-3.2){\psframebox[linecolor=red]{$SU(2)$}}
\psline[linecolor=red]{<-}(-1,-2.5)(-0.5,-3.2)
\rput[r](-5,0){\psframebox[linecolor=red]{$SU(2)$}}
\psline[linecolor=red]{->}(-5,0)(-3.25,0)
\rput[r](-5.5,+2){\dfl}
\psline[linecolor=green]{->}(-5.5,+2)(-3.5,+2)
\rput[l](+5.5,+2.5){\dfl}
\psline[linecolor=green]{->}(+5.5,+2.5)(+4.2,+2)
\rput*[b](0,+2.5){\blue $D_2$ Coulomb}
\rput*[l]{50}(+3,-0.8){\red $D'_0$ Coulomb}
\rput*[r]{320}(-1.5,-2){\red $D_0$ Coulomb}
\end{pspicture}
\label{D2t:CoulombPhases}
\ee
This diagram shows 3D parameter/moduli space spanned by $\tilde m$,
$\tilde h$, and $\tilde\phi$.
There are two unphysical regions --- $\tilde h<\tilde m$ and
$2\tilde\phi+\tilde h+\tilde m<0$ --- and three distinct physical regions
separated by double flop transitions at $\tilde\phi=|\tilde m|$:
\begin{itemize}
\item[(1)]
The $D_2$ Coulomb phase for $\tilde\phi>|\tilde m|$ where
$\fdc=2\tilde h+6\tilde\phi$.

\item[(2)]
The effective $D_0$ (SYM) Coulomb phase for $\tilde m<0$
and $\tilde\phi<|\tilde m|$ where $\fdc=2\tilde h^{\rm eff}_{}+8\tilde\phi$.
In quiver terms, in this phase the quark bare mass $\tilde\mu=Ve^{a\tilde m}$
is effectively $\tilde\mu\approx0$, hence $n_f^{\rm eff}=0$ but
$\Delta F^{\rm eff}_{}=2$, and everything works as in \S6.2.
In particular, there is $SU(2)$ restoration for $\tilde\phi=0$
and $\tilde h^{\rm eff}_{}={\tilde h+\tilde m\over2}>0$,
and also for $\tilde h^{\rm eff}_{}<0$ and
$\tilde\phi=-\tilde h^{\rm eff}_{}/2$ (this is dual to $\phi=0$).
And for $\tilde h^{\rm eff}_{}=0$
% ($i.\,e.,\ \tilde h=-\tilde m$)
and $\tilde\phi=0$ we have $E_1$ SCFT.

\item[(3)]
The $D'_0$ Coulomb phase for $\tilde m>0$ and $\tilde\phi<\tilde m$.
This is another effective SYM Coulomb phase with
$\fdc=2\tilde h^{\rm eff}_{}+8\tilde\phi$ and
$SU(2)$ restoration for $\tilde\phi=0$.
However, this time $\tilde h^{\rm eff}_{}={\tilde h-\tilde m\over2}$
is always nonnegative.
In quiver terms, in this phase $\tilde\mu=Ve^{a\tilde m}$ is so large
the quarks effectively decouple, hence $n_f^{\rm eff}=\Delta F_{}^{\rm eff}=0$;
this effective theory works as in \cite{IK1}.
In particular, for $\tilde h^{\rm eff}_{}=0$
% ($i.\,e.,\ \tilde h=\tilde m>0$)
and $\tilde\phi=0$ there is a 5D SCFT.

We call this superconformal theory $E'_1$ because its spectral curve is dual
to the curve of the $E_1$; in particular, there are two $I_1$ singularities and
one $I_2$.
However, there are major differences between the two SCFTs:
the $E'_1$ does not have a Higgs branch, and its Coulomb branch is limited
to $\tilde h^{\rm eff}_{}\ge0$.
In M~theory, the $E'_1$ SCFT arises from a Calabi--Yau with a $\CC^3/\ZZ_4$
orbifold singularity.
Note that such singular points are not isolated but lie on lines of milder
$A_1$ singularity ($\CC^2/\ZZ_2$), and the line cannot be blown up without
also blowing up the point.
This is unlike $E_1$ which arises from an isolated singular point, namely
${\rm conifold}/\ZZ_2$.

\end{itemize}

\par\noindent
The three Coulomb phases come together at $\tilde\phi=\tilde m=0$, $\tilde h>0$,
where we have a $D_2$ theory at its origin --- unbroken $SU(2)$ and two
massless quarks at the same time.
Likewise, there is unbroken $SU(2)$ and two massless quarks along the
$\tilde\phi=-\tilde h=-\tilde m>0$ line, which we call the `$D'_2$ origin'.
From the spectral curve's point of view, the $D_2$ and the $D'_2$ origins
have similar singularity structures $2I_2+2I_1$.
But beyond the spectral curve level, the two origins origin are different.
In particular, there are three Coulomb phases near the $D_2$ origin
but only two Coulomb phases --- the $D_2$ and the $D_0$ ---
near the $D'_2$ origin; the $D'_0$ Coulomb phase is cut off by
the deconstruction limit $\tilde h\ge\tilde m$.
Likewise, two Higgs branches --- one mesonic and one baryonic ---
have roots at the $D_2$ origin, but the $D'_2$ origin has only
the mesonic Higgs branch.

Indeed, here is the diagram of the Higgs phases~(\ref{D2t:Higgs:5d})
of the $\Delta F=0$ theory:
\be
\psset{unit=0.95mm}
\def\cone[#1]{%
	\pspolygon*[linecolor=#1](-10.125,50)(0,0)(+10.125,50)
	\psellipse[linewidth=0.4pt,fillstyle=solid,fillcolor=#1](0,50)(10,2.5)
	\psline[linewidth=0.4pt](-10,49.4)(0,0)(+10,49.4)
	}
\begin{pspicture}[0.1](-65,-95)(+85,+60)
\rput(-45,0){%
    \rput{20}(0,0){\cone[magenta]}
    \rput[b](-18,55){\magenta mesonic}
    \pscircle*[linecolor=green](0,0){2}
    \rput[t](0,-4){\green double flop}
    \rput[t](0,-11){\green $\tilde\phi=\pm\tilde m$}
    }
\rput(0,0){%
    \rput{340}(0,0){\cone[brown]}
    \rput{20}(0,0){\cone[magenta]}
    \pscircle*[linecolor=blue](0,0){2}
    \rput[b](+18,55){\brown baryonic}
    \rput[b](-18,55){\magenta mesonic}
    \rput[t](0,-4){\blue $D_2$ origin}
    \rput[t](0,-11){\blue $\tilde\phi=\tilde m=0,\ \tilde h>0$}
    }
\rput(+65,0){%
    \rput{340}(0,0){\cone[white]}
    \rput{20}(0,0){\cone[magenta]}
    \pscircle*[linecolor=blue](0,0){2}
    \rput[b](+18,55){\black missing}
    \rput[b](-18,55){\magenta mesonic}
    \rput[t](0,-4){\blue $D'_2$ origin}
    \rput[t](0,-11){\blue $\tilde\phi=-\tilde m=-\tilde h>0$}
    }
\rput(-45,-80){%
    \rput{20}(0,0){\cone[brown]}
    \rput(-18,55){\brown baryonic}
    \pscircle[linecolor=red,fillcolor=yellow,fillstyle=solid](0,0){2}
    \rput[t](0,-4){\orange $E_1$ SCFT}
    \rput[t](0,-11){\orange $\tilde\phi=0,\ \tilde h=-\tilde m>0$}
    }
\rput(0,-80){%
    \rput{20}(0,0){\cone[white]}
    \rput(-18,55){\black missing}
    \pscircle[linecolor=red,fillcolor=yellow,fillstyle=solid](0,0){2}
    \rput[t](0,-4){\orange $E'_1$ SCFT}
    \rput[t](0,-11){\orange $\tilde\phi=0,\ \tilde h=+\tilde m>0$}
    }
\rput(+45,-80){%
    \rput{340}(0,0){\cone[brown]}
    \rput{20}(0,0){\cone[magenta]}
    \rput{330}(0,0){%
    	\psellipse[linewidth=0.5pt,linestyle=dashed](0,45)(18,4.5)
    	\psline[linewidth=0.5pt,linestyle=dashed](-18,44.5)(0,0)(+18,44.5)
    	}
    \pscircle[linecolor=red,fillcolor=yellow,fillstyle=solid](0,0){2.2}
    \rput(0,0){\large\blue$\star$}
    \rput[b](+18,55){\brown baryonic\black, unmixed}
    \rput[b](-18,55){\magenta mesonic}
    \rput[t](0,-4){\orange $E'_3$ SCFT}
    \rput[t](0,-11){\orange $\tilde\phi=\tilde m=\tilde h=0$}
    }
\end{pspicture}
\label{D2t:HiggsPhases}
\ee
where ``missing'' branches do not exist for $\Delta F=0$ but their roots
are dual to roots of mesonic branches of the $\Delta F=1$ theory.
And the dashed lines making an empty wide cone around the last Higgs branch
here indicate that the branch has hypermultiplet $\rm dimension=1$ ---like
all the other Higgs branches of the $\Delta F=0$ theory --- but in the
dual $\Delta F=1$ theory there is a Higgs branch of $\rm dimension=2$,
{\it cf.}\ diagram~(\ref{D2:Diagram4}).

The central point $\tilde\phi=\tilde m=\tilde h=0$
of the Coulomb phase diagram~(\ref{D2t:CoulombPhases}) ---
where the last pair of Higgs branches~(\ref{D2t:HiggsPhases}) are rooted ---
is dual to the $E_3$ SCFT point of the $\Delta F=1$ theory.
For $\Delta F=0$ this point is also superconformal;
we call this SCFT $E'_3$ because of the duality, and also
because it is similar to $E_3$ SCFT in many ways:
(1) the $E'_3$ obtains in the $\tilde h\to0$ limit of $D_2$;
(2) its spectral curve in 4D has $I_3+I_2+I_1$ singularities;
(3) it has two distinct Higgs branches.
However, the dimensions of the Higgs branches are different:
$1+1$ (in hypermultiplet units) for the $E'_3$ {\it versus}
$1+2$ for the $E_3$.
In M theory, the $E_3$ and the $E'_3$ SCFTs are realized on Calabi--Yaus
with different singularity types.
There respective toric diagrams are:
\be
\psset{unit=2cm,linewidth=1.5pt,dotscale=1,dotstyle=*}
\begin{pspicture}[](-4,-1)(+4,+1)
\rput(-2,0){%
    \pspolygon[showpoints=true](+1,0)(+1,+1)(0,+1)(-1,0)(-1,-1)(0,-1)
    \psline[linestyle=dotted](-1,0)(+1,0)
    \psline[linestyle=dotted](-1,-1)(+1,+1)
    \psline[linestyle=dotted](0,-1)(0,+1)
    \psdots[dotstyle=o](0,0)
    \rput[tr](-1,+1){$E_3$}
    }
\rput(+2,0){%
    \pspolygon[showpoints=true](+1,0)(0,+1)(-1,0)(-1,-1)(0,-1)(+1,-1)
    \psline[linestyle=dotted](-1,0)(+1,0)
    \psline[linestyle=dotted](-1,-1)(0,0)(+1,-1)
    \psline[linestyle=dotted](0,-1)(0,+1)
    \psdots[dotstyle=o](0,0)
    \rput[tr](-1,+1){$E'_3$}
    }
\end{pspicture}
\label{D2t:ToricDiagrams}
\ee
Note the lower edge of the $E'_3$ diagram has a middle point:
this indicates that the singularity is not an isolated point
but a more-singular point on a less-singular line, and the line
cannot be blown up without blowing up the point at the same time.

In type~IIB string theory, the brane webs for the $E_3$ and the $E'_3$
are as follows:
for the SCFT points themselves
\be
\psset{unit=1cm,linewidth=3pt,linecolor=blue}
\begin{pspicture}[](-7,-3)(+7,+3)
\rput(-4,0){%
    \psline(-3,0)(+3,0)
    \psline(-3,+3)(+3,-3)
    \psline(0,+3)(0,-3)
    \pscircle[fillstyle=solid,fillcolor=red](0,0){0.2}
    \rput[lb](-3,-3){$E_3$}
    }
\rput(+4,0){%
    \psline(-3,0)(+3,0)
    \psline(-3,+3)(0,0)(+3,+3)
    \psline[doubleline=true,doublecolor=yellow](0,0)(0,-3)
    \pscircle[fillstyle=solid,fillcolor=red](0,0){0.2}
    \rput[lb](-3,-3){$E'_3$}
    }
\end{pspicture}
\label{D2t:e3webs:scft}
\ee
for the respective Coulomb branches ($m=h=0$ but $\phi>0$ or
$\tilde m=\tilde h=0$ but $\tilde\phi>0$)
\be
\psset{unit=1cm,linewidth=3pt,linecolor=blue}
\begin{pspicture}[](-7,-3)(+7,+3)
\rput(-4,0){%
    \pspolygon(+1,0)(0,+1)(-1,+1)(-1,0)(0,-1)(+1,-1)
    \psline(+1,0)(+3,0)
    \psline(0,+1)(0,+3)
    \psline(-1,+1)(-3,+3)
    \psline(-1,0)(-3,0)
    \psline(0,-1)(0,-3)
    \psline(+1,-1)(+3,-3)
    \rput[lb](-3,-3){$E_3$}
    }
\rput(+4,0){%
    \pspolygon(+1,0)(+1,+1)(-1,+1)(-1,0)(0,-1)
    \psline(+1,0)(+3,0)
    \psline(+1,+1)(+3,+3)
    \psline(-1,+1)(-3,+3)
    \psline(-1,0)(-3,0)
    \psline[doubleline=true,doublecolor=yellow](0,-0.9)(0,-3)
    \rput[lb](-3,-3){$E'_3$}
    }
\end{pspicture}
\label{D2t:e3webs:Coulomb}
\ee
for the baryonic Higgs branches
\be
\psset{unit=1cm,linewidth=3pt,linecolor=blue}
\begin{pspicture}[](-7,-3)(+7,+3)
\rput(-4,0){%
    \psline(-2.9,-0.1)(+0.1,-0.1)(+0.1,+2.9)
    \psline(+0.1,-0.1)(+3,-3)
    \psline[border=1.5pt](+2.9,+0.1)(-0.1,+0.1)(-0.1,-2.9)
    \psline(-0.1,+0.1)(-3,+3)
    \rput[lb](-3,-3){$E_3$}
    }
\rput(+4,0){%
    \psline(-2.9,+3)(+0.1,0)(+3,0)
    \psline(+0.1,0)(+0.1,-3)
    \psline[border=1.5pt](+2.9,+3)(-0.1,0)(-3,0)
    \psline(-0.1,0)(-0.1,-3)
    \rput[lb](-3,-3){$E'_3$}
    }
\end{pspicture}
\label{D2t:e3webs:baryonic}
\ee
for the mixed / mesonic Higgs branches
\be
\psset{unit=1cm,linewidth=3pt,linecolor=blue}
\begin{pspicture}[](-7,-3)(+7,+3)
\rput(-4,0){%
    \psline(-3,0)(+3,0)
    \psline[border=1.5pt](-3,+3)(+3,-3)
    \psline[border=1.5pt](0,+3)(0,-3)
    \rput[lb](-3,-3){$E_3$}
    }
\rput(+4,0){%
    \psline(-3,+3)(0,0)(+3,+3)
    \psline[doubleline=true,doublecolor=yellow](0,+0.1)(0,-3)
    \psline[border=1.5pt](-3,0)(+3,0)
    \rput[lb](-3,-3){$E'_3$}
    }
\end{pspicture}
\label{D2t:e3webs:mesonic}
\ee
Note the web for the mixed branch of $E_3$ has three
disconnected lines: this corresponds to hypermultiplet $\rm dimension=2$.
In comparison, the webs for baryonic branches of both SCFTs and also for
the mesonic branch of $E'_3$ have only two disconnected pieces each:
this corresponds to $\rm dimension=1$.

And of course there are many more webs for non-conformal values of the
Coulomb parameters $m\neq0$ and/or $h\neq0$ (or $\tilde m\neq0$
and/or $\tilde h\neq0$).
In fact, there too many webs, so we don't diagram them here.
Instead, let us simply state the main result:
The brane webs for the $E_3$ and its resolutions and deformations have
precisely the same physical phases --- both Coulomb and Higgs ---
as the deconstructed \sqcdv\ with $n_c=n_f^{}=2$ and $\Delta F=1$.
Likewise, the webs for the $E'_3$ and its resolutions and deformations have
precisely the same physical phases as the deconstructed \sqcdv\ with
$n_c=n_f^{}=2$ and $\Delta F=0$.
And in both cases, the webs corresponding to the unphysical phases
cannot be built.
In other words, in both cases, {\blue the brane webs in string theory are
in perfect agreement with the dimensional deconstruction}.

%% file: chapter7.tex
%auto-ignore
%
% chapter 7 of the DESQCD paper
%
\section{Deconstruction / String Universality}
In the last section we saw four examples of non-perturbative phase diagrams,
and in all four cases dimensional deconstruction and type~IIB brane web
implementation of the same 5D theory yielded identical diagrams (except for
the non-geometric phases of unphysical regions of the moduli/parameter space).
In this section, we shall see that such deconstruction / string universality
is general and holds for any $n_c$, $n_f$, and $\kcs$.
Specifically, deconstructive and brane-web completions of the same \sqcdv\
are in the same universality class and have similar moduli/parameter spaces and
similar prepotentials ${\cal F}(\phi_1,\ldots,\phi_{n_c};h;m_1,\ldots,m_{n_f})$.
However, the two completions are {\it not dual} to each other and become
dissimilar outside the zero-energy limit.
This is similar to the universality between the 4D SQCD and the MQCD
\cite{witten97, HSZ, BIKSY}: they are not dual
to each other but are in the same universality class and have similar
holomorphic properties.

In fact, the 5D universality between deconstruction and brane webs is
{\it based} on the 4D universality between SQCD and MQCD, or rather its
generalization to more complicated 4D theories.
Specifically, we start a deconstructed \sqcdv, treat it as a 4D
$[SU(n_c)]^\ql$ quiver theory, and take its M--theory counterpart:
an M5 brane spanning the 4D Minkowski space and the
quiver's spectral curve~(\ref{SW}).
We are going to take the large $\ql$ limit of this correspondence, so instead
of identifying the $x$ and $y$ coordinates of the spectral curve with some
of the 7 extra dimensions of the M theory, we embed them in a non-linear
manner based on eqs.~(\ref{XiEta}), namely
\be
\eqalign{
x\ =\ \exp(\ql a\times\xi),&\qquad
\xi\ =\ {X^5+iX^9\over C}\,,\cr
y\ =\ \exp(\ql a\times\eta),&\qquad
\eta\ =\ {X^6+iX^{10}\over C}\,,\cr
}\label{U:embed}
\ee
where $C$ is a constant parameter, to be determined later in this section.
Consequently, the induced metric on the M5 brane itself is
\be
ds^2\
=\ dX_{0123}^2\ +\ C^2\left( d\bar\xi\,d\xi\,+\,d\bar\eta\,d\eta\right)\
=\ dX_{0123}^2\ +\ {C^2\over (\ql a)^2}\,\left(
	{d\bar x\,dx\over|x|^2}\ +\ {d\bar y\,dy\over|y|^2}\right)
\label{U:M5metric}
\ee
where $x$ and $y$ are related according to \eq{SW}.

We claim that the $\ql a\to\infty$ limit of this M5 brane is dual to a
5D brane web, and moreover this web implements the very \sqcdv\
we have started from.
Combining this duality with the generalized SQCD/MQCD duality in 4D, we
arrive at the following diagram:
\be
\psset{unit=5mm,linewidth=2pt,linecolor=blue}
\begin{pspicture}(-15,-7)(+15,+7)
\rput(0,+6){\ovalnode{deco}{Deconstructed \sqcdv}}
\rput(0,0){\ovalnode{mqcd}{M Theory of the Quiver, $\ql a\to\infty$}}
\rput(0,-6){\ovalnode{web}{Brane-web Engineered \sqcdv}}
\psset{linecolor=red,framearc=0.2}
\ncline{<->}{deco}{mqcd}
\ncline{<->}{mqcd}{web}
\rput*(0,+3){\psframebox{Universality}}
\rput*(0,-3){\psframebox{Duality}}
\end{pspicture}
\label{U:Summary}
\ee
And since duality implies universality (but not the other way around),
we find that
{\blue the dimensional deconstruction and the brane-web engineering
of the same \sqcdv\  are in the same universality class.}

To prove the duality part of the diagram~(\ref{U:Summary})
we will show the following:
\begin{itemize}

\item[(1)]
For $\ql a\to\infty$, the spectral curve of the quiver becomes a union of
linear segments $\alpha\xi-\beta\eta=\rm const$ with
integer $\alpha$ and $\beta$.
The joints between the segments are infinitesimal
but have $\delta$--like curvature,
which allows different $(\alpha,\beta)$ for different segments.

\item[(2)]
The coordinates $X^9=C\Im\xi$ and $X^{10}=C\Im\eta$  are periodic.
Together they form a $T^2$ torus, and M theory on this torus is dual to
the type~IIB string theory on a circle $S^1$.
Under this duality, the M5 brane's part spanning an $\alpha\xi-\beta\eta=\rm const$
piece of the spectral curve (times the $\RR^{3,1}$ Minkowski space) becomes
the $(p=\alpha,q=\beta)$ 5--brane spanning $\RR^{3,1}\times S^1\times{}$a
real line segment $\alpha X^5-\beta X^6=\rm const$.
And the M5 spanning the whole spectral curve is dual to a $(p,q)$ brane web
made of such segments.

\item[(3)]
This brane web turns out to be precisely the web implementing the \sqcdv\
with appropriate $n_c$, $n_f$, and $\kcs$.
(Except that one of the 5 dimensions is compactified on the $S^1$.)
In particular, for positive enough coupling $h$, the web forms a ladder
with $n_c$ parallel rungs; this implements the SQCD Coulomb phase of
the 5D theory.
For negative or low enough $h$, the web flips to a different topology, and
this happens precisely when the deconstructed theory has a phase transition.

\item[(4)]
Finally, comparing the semiclassical gauge boson masses in
the M theory and in the deconstructed \sqcdv\ yields $C=\sqrt{\ql a/2\pi t_3}$
where $t_3$ is the M2 brane tension.
This gives us the area of the $T^2$ torus in the M theory and hence the radius
of the $S^1$ circle in the dual string theory~\cite{witten95}.
That radius turns out to be $R=\ql a/2\pi$ ---
which is precisely the radius of the deconstructed dimension.
Hence, in the decompactification limit of the deconstructed theory,
the brane web also decompactifies to 5D.

\end{itemize}

So let us start with part (1) of our argument.
Consider the spectral curve~(\ref{SW}) as a quadratic equation for the $y(x)$
and let us take the $\ql a\to\infty$ limit for fixed $\xi$ and $\eta$.
Similar to \eqrange{phipprox}{quivertauii}, the polynomials $p(x)$ and $b(x)$
become in this limit
\bea
p\left(x=e^{\ql a\xi}\right) & \longrightarrow &
\pm\exp\left[ \ql a\times\sum_{i=1}^{n_c}\max(\xi,\varphi)\right] \label{U:p}\\
&& \mbox{as long as}\ \Re\xi\neq\mbox{any of the}\ \phi_i\,,
\vrule width 0pt height 10 pt depth 10pt \nonumber\\
b\left(x=e^{\ql a\xi}\right) & \longrightarrow &
\pm\exp\left[ \ql a\times\left(
        \Delta F\times\xi\ +\,\sum_{f=1}^{n_f}\max(\xi,m_f)\right)
	\right] \label{U:b}\\
&& \mbox{as long as}\ \Re\xi\neq\mbox{any of the}\ \Re m_f\,.\nonumber
\eea
(By abuse of notations, here max of two complex numbers denotes the number
with the larger real part.)
And as in \eq{BPratio}, for almost all $\xi$ either $p^2(x)\gg e^{-\ql a S}b(x)$
or else $p^2(x)\ll e^{-\ql a S}b(x)$.
When $p^2(x)\gg e^{-\ql a S}b(x)$,
\eq{SW} for the $y(x)$ has two very different solutions, namely
\be
y_1(x)\ =\ p(x),\qquad
y_2(x)\ =\ {e^{-\ql aS}\,b(x)\over p(x)}\,,\qquad
y_1\ \gg\ y_2;
\label{U:Ydiff}
\ee
in terms of $\eta(\xi)$, these solutions translate to
\bea
\eta_1(\xi) &=& \sum_{i=1}^{n_c}\max(\xi,\varphi), \label{U:eta1}\\
\eta_2(\xi) &=& -S\ -\,\sum_{i=1}^{n_c}\max(\xi,\varphi)\
    +\ \Delta F\times\xi\ +\,\sum_{f=1}^{n_f}\max(\xi,m_f). \label{U:eta2}
\eea
Note that both $\eta_1$ and $\eta_2$ are piecewise-linear functions of $\xi$,
and their derivatives are integer-valued.
(That is, for each piece $d\eta/d\xi$ is a constant integer, but its value
jumps from piece to piece.)
And when $p^2(x)\ll e^{-\ql a S}b(x)$, the two solutions of \eq{SW} become
\be
\hat y_{1,2}(x)\ = \pm i\,e^{-\ql aS/2}\,\sqrt{b(x)},
\label{U:Ysame}
\ee
or in terms of $\xi$ and $\eta$,
\be
\hat\eta_{1,2}(\xi)\
=\ -{S\over2}\ +\ {\Delta F\over2}\times\xi\
+\ {1\over2}\sum_{f=1}^{n_f}\max(\xi,m_f)\ \pm\ {\pi i\over2\ql a}\,.
\label{U:etahat}
\ee
Again, $\hat\eta_{1,2}$ are piecewise-linear functions of $\xi$,
and for each piece the derivative $d\hat\eta_{1,2}/d\xi$ is
integer or half-integer.
Consequently, the whole spectral curve consists of a bunch of linear pieces,
and each piece is rational, $i.\,e.$ satisfies $\alpha\xi-\beta\eta=\rm const$
for some integers $\alpha$ and $\beta$.

Actually, \eqrange{U:eta1}{U:eta2} and (\ref{U:etahat}) miss some of the pieces
of the spectral curve, but the missing pieces are also linear and rational.
These missing pieces are located at $\xi=\varphi_i$ or $\xi=m_f$ where
the limits (\ref{U:p}--\reftail{U:b}) do not work.
Instead, for $\xi$ in a $O(1/\ql a)$ neighborhood of a modulus $\varphi_i$
the value of the $p(x)$ polynomial can be anywhere between zero and the
right hand side of \eq{U:p};
likewise, for $\xi$ in a $O(1/\ql a)$ neighborhood of a mass $m_f$ the value
of $b(x)$ can be anywhere between zero and the right hand side of \eq{U:b}.
Consequently, in the decompactification limit, a fixed value of $\xi$ which
happens to coincide with a modulus $\varphi_i$ or a mass $m_f$
agrees with a wide range of values of $\eta$.\footnote{%
	Specifically, for $\xi=\varphi_i$, $\eta$ can be anywhere between
	$\eta_1(\varphi_i)$ and $\eta_2(\varphi_i)$, or rather
	$\Re\eta_1(\varphi_i)>\Re\eta>\Re\eta_2(\varphi_i)$.
	Likewise, for $\xi=m_f$, $\eta$ may vary from $\eta_2(m_f)$
	or $\hat\eta_2(m_f)$ (whichever applies) all the way to $-\infty$.
	}
Thus, the spectral curve has several $\xi=\rm const$ pieces ---
which are linear and rational with $(\alpha,\beta)=(1,0)$.

Altogether, the decompactification limit of the spectral curve~(\ref{SW})
of the quiver is a union of segments of rational straight lines 
$\alpha\xi-\beta\eta=\rm const$.
For large but finite $\ql a$, the joints between adjacent segments have small
but finite sizes of the order $O(1/\ql a)$.
Such joints are strongly curved --- curvature${}=O(\ql a)$ --- which
allows for finite differences between directions $\alpha/\beta$ of segments
they connect.
For $\ql a\to\infty$, the joints' sizes become infinitesimal while the
curvature becomes $\delta$--like.
This completes part (1) of our proof.

Next, consider the spectral~curve~(\ref{SW}) as a Riemann surface.
This surface is a double cover of the complex $x$ sphere, or in other words
$x$ spans $\CC^*$, twice.
The map $x=\exp(\ql a\times\xi)$ takes out the $x=0$ and $x=\infty$ points,
which turns the $x$ sphere into a complex cylinder:
$\Re\xi$ is single valued and spans the real line $\RR$ while
$\Im\xi$ is periodic modulo~$(2\pi/\ql a)$.
Likewise, the map $y=\exp(\ql a\times\eta)$ turns the $y$ sphere into a
cylinder: $\Re\eta$ is single valued and spans $\RR$
while $\Im\eta$ is periodic modulo~$(2\pi/\ql a)$.
In terms of eqs.~(\ref{U:embed}), this means that $(X^5,X^6)$ span an infinite
plain $\RR^2$ while $(X^9,X^{10})$ span a torus $T^2$
of area $A=(2\pi C/\ql a)^2$.

M theory compactified on this torus is dual to the type~IIB
string theory\footnote{%
	The string coupling $\tau_s=ie^{-2\Phi}+\rm axion$ follows from the shape
	of the torus; in our case it has two $\perp$ periods of equal length, hence
	$\tau_s=i$.
	We may choose a different $\tau_s$ by modifying
	eqs.~(\ref{U:embed}) according to
	\be
	X^5+iX^9\ =\ (\xi\,+\,\mbox{axion}\times\eta)\times Ce^{+\Phi},\qquad
	X^6+iX^{10}\ =\ \eta\times Ce^{-\Phi}\,.
	\ee
	However, in brane-web engineering the actual value of
	the $\tau_s$ is not important, and so
	we stick with eqs.~(\ref{U:embed}) as they are and hence $\tau_s=i$.
	}
compactified on a circle $S^1$ whose radius is inversely proportional
to the torus's area,
\be
{1\over R[S^1]}\ =\ t_3\times\mathop{\rm Area}[T^2]\
\equiv\ t_3\times(2\pi C/\ql a)^2
\label{U:Radius}
\ee
where $t_3$ is the membrane (M2) tension in M theory.
Under this duality, an M5 brane which wraps one circle of the $T^2$
is dual to a $(p,q)$ 5--brane which wraps the $S^1$ circle.
%(times the remaining $4+1$ dimensions of the M5 brane).
The Neveu--Schwarz--Ramond charges $(p,q)$ of this dual 5--brane
depend on a particular circle of the $T^2$ wrapped by the M5:
for a circle $\alpha X^9-\beta X^{10}=
\rm const$ with integer $\alpha$ and $\beta$, the Ramond charge $p=\alpha$ and
the Neveu--Schwarz charge $q=\beta$.

According to eqs.~(\ref{U:embed}), a linear piece $\alpha\xi-\beta\eta=
\rm const$ of the spectral curve is the direct product of a real line segment
$\alpha X^5-\beta X^6=\rm const$ in the $X^{5,6}$ plane, times a circle
$\alpha X^9-\beta X^{10}=\rm const$ in the $T^2$ torus.
Hence, the M5 brane spanning this piece (times the 4D Minkowski space $\RR^{3,1}$)
is dual to the $(p=\alpha,q=\beta)$ 5--brane which spans
$\RR^{3,1}\times S^1\times{}$the real line segment
$\alpha X^5-\beta X^6=\rm const$.
And the M5 brane spanning the entire spectral curve of the quiver theory
is dual to the $(p,q)$ brane web made of such segments.
This completes part (2) of our proof.

Now consider the web's geometry for different phases of the deconstructed 5D
theory.
Let us start with the {\sl ordinary SQCD${}_5$} phase where
branching points  of the 4D spectral curve over the $x$ plane
form $n$ very close pairs~(\ref{pairs}--\reftail{dformula}).
This pattern requires $\Re S=h+\half\sum m_f$ large enough to assure that
$p^2(x)\gg e^{-\ql a S}b(x)$ for all $x=e^{\ql a\xi}$
(except when $\xi$ equals one of the $\varphi_i$).
Consequently, the spectral curve follows eqs.~(\ref{U:Ydiff}) rather than
(\ref{U:Ysame}) for all $x$, or in terms of $\xi$ and $\eta$, it comprises
complex lines $\eta_1(\xi)$ and $\eta_2(\xi)$ according to
\eqrange{U:eta1}{U:eta2} for all $\xi$.
And there are also $\xi=\rm const$ complex lines for $\xi=\varphi_i$ and $\xi=m_f$.
The brane web dual to this curve spans the real parts of all these complex
lines, thus:
\be
\psset{unit=4mm,linewidth=3pt}
\begin{pspicture}(-16,-6)(+23,+7.5)
\psset{linecolor=blue}
\psline(-12,-4)(+2,-4)
\psline(-9,-2)(+4,-2)
\psline(-5,+2)(+12,+2)
\psline(-4,+4)(+18,+4)
\rput[b](+7,+4.3){$\blue\xi=\varphi_{n_c}$}
\rput[b](+3.5,+2.3){$\blue\xi=\varphi_{n_c-1}$}
\rput[b](-2.5,-1.7){$\blue\xi=\varphi_2$}
\rput[b](-5,-3.7){$\blue\xi=\varphi_1$}
\psline[linestyle=dotted](-1,-0.5)(+2,+1)
\psset{linecolor=green}
\psline(-16,-3)(-11,-3)
\psline(-16,+3)(-5,+3)
\rput[bl](-15,+3.3){$\green\xi=m_{n_f}$}
\rput[bl](-15,-2.7){$\green\xi=m_1$}
\psline[linestyle=dotted](-13,-1.5)(-13,+2.5)
\psset{linecolor=red}
\psline(2,-6)(2,-4)(4,-2)
\psline[linestyle=dashed](4,-2)(12,+2)
\psline(12,+2)(18,+4)(22,+5)
\psline(-16,-6)(-12,-4)(-11,-3)(-9,-2)
\psline[linestyle=dashed](-9,-2)(-5,+2)
\psline(-5,+2)(-5,+3)(-4,+4)(-4,+6)
\rput[l](2.3,-5){$\red\eta=\eta_1(\xi)$}
\rput[rb](22,+5.2){$\red\eta=\eta_1(\xi)$}
\rput[r](-4.3,+5){$\red\eta=\eta_2(\xi)$}
\rput[lt](-14,-5.2){$\red\eta=\eta_2(\xi)$}
\psset{linewidth=0.5pt,linecolor=black,arrowscale=2}
\psline{->}(-16,0)(+22,0)
\rput[l](22.1,0){$\Re\eta$}
\psline{->}(0,-6)(0,+6.5)
\rput[b](0,6.6){$\Re\xi$}
\end{pspicture}
\label{U:Web1}
\ee
As promised, this web looks like a ladder with $n_c$ horizontal rungs
(blue lines for $\xi=\varphi_i$) between two multiply-bent sides
(red lines for $\eta=\eta_1(\xi)$ and $\eta=\eta_2(\xi)$),
but there also are $n_f$ semi-infinite horizontal branes attached to
the left side of the ladder (green lines for $\xi=m_f$).
In 5D (after eventual decompactification of the $S^1$ circle), this web
obviously gives rise to the Coulomb phase of an \sqcdv\ with
$n_c$ colors and $n_f$ flavors.\footnote{%
	Indeed, strings between the `blue' branes produce vector multiplets in
	the adjoint of $U(n_c)$ (spontaneously broken to $U(1)^{n_c}$ by
	distances $\phi_i-\phi_j$ between the branes),
	with the overall $U(1)$ frozen due to interactions with
	the `red' branes (the sides of the ladder).
	Likewise, strings between the `green' branes try to produce $SU(n_f)$ gauge
	fields, but they decouple because the `green' branes are infinitely long;
	hence, the $SU(n_f)$ symmetry is flavor rather than gauge.
	Finally, $n_c\times n_f$ quark hypermultiplets arise from strings connecting
	the blue and the green branes to each other.
	}
The Chern--Simons level of this \sqcdv\ is less obvious, but it is related
to the asymmetry between the top and the bottom ends of the ladder:
At the top of the ladder ($\Re\xi\to+\infty$), its two sides
separate from each other at the rate
\be
{\rm rate}_{\rm\, top}\
=\ \left[ {d\eta_1\over d\xi}\,-\,{d\eta_2\over d\xi}\right]
	_{\Re\xi\to+\infty}\
=\ (n_c)\,-\,(n_f+\Delta F-n_c),
\ee
while at the bottom of the ladder they separate at a different rate
\be
{\rm rate}_{\rm\, bottom}
=\ \left[ {d\eta_2\over d\xi}\,-\,{d\eta_1\over d\xi}\right]
	_{\Re\xi\to-\infty}\
=\ (\Delta F)\,-\,(0).
\ee
The Chern--Simons level is one half of the difference between these rates,
\be
\kcs\ =\ \half\Bigl[ {\rm rate}_{\rm\, top}\ -\ {\rm rate}_{\rm\, bottom}\Bigr]\
=\ n_c\ -\ \half n_f\ -\ \Delta F;
\label{U:KCS}
\ee
note that it comes out exactly as in dimensional deconstruction, {\it cf.}\
\eq{Flavors}.
Thus, the web~(\ref{U:Web1}) indeed gives rise to the same \sqcdv\
as the deconstructed theory we have started from,
at least in the {\sl ordinary SQCD${}_5$} Coulomb phase.

In terms of the web~(\ref{U:Web1}),
lowering the $h$ parameter of the 5D theory makes the left side of the
ladder move right, {\it cf.} \eq{U:eta2}.
Eventually, for some critical $h=h_c$,
the left side collides with the right side, and then the
web switches to a different configuration.
The details of this transition depend on whether the two sides of the
ladder collide at a single point or over some length of parallel segments.
For a point collision, the web has a flop transition
\be
\psset{unit=8mm,linewidth=2pt}
\begin{pspicture}(-11,-2)(+8,+3)
\psline[linewidth=1pt,linestyle=dotted](-11,0)(+8,0)
\rput*[l](-11,0){a $\phi_i$}
\rput(-7,0){%
	\psline[linecolor=blue](-2,0)(0,0)
	\psline[linecolor=red](-4,-2)(-2,0)(-2,+2)
	\psline[linecolor=red](0,-2)(0,0)(+2,+2)
	\rput[b](0,2.2){$h>h_c$}
	}
\rput(0,0){%
	\psline[linecolor=red](0,-2)(0,+2)
	\psline[linecolor=red](-2,-2)(+2,+2)
%	\pscircle*[linecolor=blue](0,0){0.1}
	\rput[b](1,2.2){$h=h_c$}
	}
\rput(+6,0){%
	\psline[linecolor=brown](0,-1)(+1,+1)
	\psline[linecolor=red](-1,-2)(0,-1)(0,-2)
	\psline[linecolor=red](+1,+2)(+1,+1)(+2,+2)
	\rput[b](0.5,2.2){$h<h_c$}
	}
\psline[doubleline=true]{->}(-4,+1)(-2,+1)
\psline[doubleline=true]{->}(-5,-1)(-3,-1)
\psline[doubleline=true]{->}(+3,+1)(+5,+1)
\psline[doubleline=true]{->}(+2,-1)(+4,-1)
\end{pspicture}
\label{U:Flop1}
\ee
(Note that this picture shows only a part of the web.)
For a parallel-line collision, we have a more complicated picture:
\be
\psset{unit=7.5mm,linewidth=2pt}
\begin{pspicture}(-13,-4)(+7,+5)
\psline[linewidth=1pt,linestyle=dotted](-13,+2)(+7,+2)
\rput*[l](-13,+2){$\phi_{i+1}$}
\psline[linewidth=1pt,linestyle=dotted](-13,-2)(+7,-2)
\rput*[l](-13,-2){$\phi_{i}$}
\rput(-9,0){%
	\psline[linecolor=blue](-2,-2)(+2,-2)
	\psline[linecolor=blue](-2,+2)(+2,+2)
	\psline[linecolor=red](-4,-4)(-2,-2)(-2,+2)(-4,+4)
	\psline[linecolor=red](+4,-4)(+2,-2)(+2,+2)(+4,+4)
	\rput[b](0,4.2){$h>h_c$}
	}
\rput(0,0){%
	\psline[linecolor=blue](-0.1,-2.1)(+0.1,-2.1)
	\psline[linecolor=blue](-0.1,+2)(+0.1,+2.1)
	\psline[linecolor=red](-2,-4)(-0.1,-2.1)(-0.1,+2.1)(-2,+4)
	\psline[linecolor=red](+2,-4)(+0.1,-2.1)(+0.1,+2.1)(+2,+4)
	\rput[b](0,4.2){$h=h_c$}
	}
\rput(+6,0){%
	\psline[linecolor=red](-1,-4)(0,-3)(+1,-4)
	\psline[linecolor=red](-1,+4)(0,+3)(+1,+4)
	\psline[doubleline=true,linecolor=brown](0,-3.05)(0,+3.05)
	\rput[b](0,4.2){$h<h_c$}
	}
\psline[doubleline=true]{->}(-4.5,0)(-2.5,0)
\psline[doubleline=true]{->}(+2,0)(+4,0)
\end{pspicture}
\label{U:Flop2}
\ee
(again, only a part of the web is shown).
This time, the web does not change topology for $h<h_c$; instead, the
two coincident segments (colored brown in the above picture) are
no longer bounded by $\phi_i<\Re\xi<\phi_{i+1}$ but become longer and longer
with decreasing $h$.

From the spectral curve's point of view, the brown segments of the webs
(\ref{U:Flop1}) and (\ref{U:Flop2}) for $h<h_c$ correspond to $\eta(\xi)$
following eqs.~(\ref{U:etahat}) instead of \eqrange{U:eta1}{U:eta2}.
Pictorially, we have
\be
\psset{unit=1cm,linewidth=3pt,arrowscale=2}
\begin{pspicture}(-4,-4)(+4,+4.5)
	\psline[linecolor=red](-3,-4)(-1,-2)(-1,-4)
	\psline[linecolor=red](+3,+4)(+1,+2)(+1,+4)
	\pspolygon[linecolor=red,linestyle=dotted](-1,-2)(-1,0)(+1,+2)(+1,0)
	\psline[linecolor=brown](-1,-2)(+1,+2)
	\rput[l](-0.9,-3){$\eta_1(\xi)$}
	\rput[lt](+2.1,+2.9){$\eta_1(\xi)$}
	\rput[rb](-1.1,+0.1){$\eta_1(\xi)$}
	\rput[r](+0.9,+3){$\eta_2(\xi)$}
	\rput[rb](-2.1,-2.9){$\eta_2(\xi)$}
	\rput[lt](+1.1,-0.1){$\eta_2(\xi)$}
	\rput*(0.2,0.4){$\hat\eta_{1,2}(\xi)$}
	\psline[linewidth=0.5pt]{->}(-4,-1)(+3,-1)
	\rput[l](+3.1,-1){$\Re\eta$}
	\psline[linewidth=0.5pt]{->}(-3.5,-4)(-3.5,+4)
	\rput[b](-3.5,+4.1){$\Re\xi$}
	\psline[linewidth=1pt,linestyle=dotted](-3.5,0)(+3,0)
	\rput[r](-3.6,0){$\phi_i$}
\end{pspicture}
\label{U:Switch1}
\ee
for the right web (\ref{U:Flop1}), and
\be
\psset{unit=1cm,linewidth=3pt,arrowscale=2}
\begin{pspicture}(-3,-4)(+3,+4.5)
	\psline[linecolor=red](-1,-4)(0,-3)(+1,-4)
	\psline[linecolor=red](-1,+4)(0,+3)(+1,+4)
	\pspolygon[linecolor=red,linestyle=dotted](0,+3)(+1,+2)(+1,-2)(0,-3)(-1,-2)(-1,+2)
	\psline[linecolor=brown,doubleline=true,doublesep=2pt](0,-3)(0,+3)
	\rput[lb](+1.1,-3.9){$\eta_1(\xi)$}
	\rput[lt](+1.1,+3.9){$\eta_1(\xi)$}
	\rput[r](-1.1,0){$\eta_1(\xi)$}
	\rput[rb](-1.1,-3.9){$\eta_2(\xi)$}
	\rput[rt](-1.1,+3.9){$\eta_2(\xi)$}
	\rput[l](+1.1,0){$\eta_2(\xi)$}
	\rput*(0,0){$\hat\eta_{1,2}(\xi)$}
	\psline[linewidth=0.5pt]{->}(-3,-1)(+3,-1)
	\rput[l](+3.1,-1){$\Re\eta$}
	\psline[linewidth=0.5pt]{->}(-2.5,-4)(-2.5,+4)
	\rput[b](-2.5,+4.1){$\Re\xi$}
	\psline[linewidth=1pt,linestyle=dotted](-2.5,-2)(+2,-2)
	\psline[linewidth=1pt,linestyle=dotted](-2.5,+2)(+2,+2)
	\rput[r](-2.6,-2){$\phi_i$}
	\rput[r](-2.6,+2){$\phi_{i+1}$}
\end{pspicture}
\label{U:Switch2}
\ee
for the right web (\ref{U:Flop2});
in both figures, the dotted red lines plot  \eqrange{U:eta1}{U:eta2}
in the range of $\Re\xi$ where those equations do not apply.
Indeed, the $\ql a\to\infty$ limit of the spectral curve follows
\eqrange{U:eta1}{U:eta2} {\it only} when $p^2(x)\gg e^{-\ql a S}b(x)$;
in terms of \eqrange{U:eta1}{U:eta2} themselves, this corresponds to
$\Re\eta_1(\xi)>\Re\eta_2(\xi)$.
But for $h<h_c$, there is a range of $\Re\xi$ for which
$\Re\eta_1(\xi)<\Re\eta_2(\xi)$ --- {\it cf.}\ the dotted red lines
in figures~(\ref{U:Switch1}--\reftail{U:Switch2}) --- and in this range
$p^2(x)\ll e^{-\ql a S}b(x)$ and the spectral curve follows eqs.~(\ref{U:etahat})
instead of \eqrange{U:eta1}{U:eta2}.
Note that the switchover is continuous because
\be
\hat\eta_{1,2}(\xi)\
=\ {\eta_1(\xi)+\eta_2(\xi)\over2}\ \pm {\pi i\over 2\ql a}\,.
\label{U:etarel}
\ee

The $(p,q)$ 5--branes dual to the $\hat\eta_{1,2}(\xi)$ segments of the curve
can be single as in figure~(\ref{U:Switch1})
or double (two coincident branes) as in figure~(\ref{U:Switch2}).
The factor deciding between single or double branes is the derivative
$d\hat\eta_{1,2}/d\xi$, which is quantized in units of $\half$:
if it is integer the brane is double,
and if it is half-integer the brane is single.
To see this, consider the imaginary parts of the $\hat\eta_1(\xi)$ and
$\hat\eta_2(\xi)$ segments of the spectral curve.
For an integer $d\hat\eta_{1,2}/d\xi$, the line $(\Im\xi,\Im\hat\eta_1(\xi))$
on the torus $T^2$ is a complete circle,
and the line $(\Im\xi,\Im\hat\eta_2(\xi))$ is a separate complete circle;
the two circles are parallel but separated from each other by half-a-period
in the $\Im\eta$ direction, {\it cf.}\ \eq{U:etarel}.
Together, the M5 branes wrapping these two circles are dual to two
5--branes with similar $(p,q)$ charges, and the positions of these two
branes coincide because $\Re\hat\eta_1(\xi)=\Re\hat\eta_2(\xi)$.
On the other hand, when $d\hat\eta_{1,2}/d\xi$ is a half-integer,
the lines $(\Im\xi,\Im\hat\eta_1(\xi))$ and $(\Im\xi,\Im\hat\eta_2(\xi))$
on the torus are two halves of the same circle.
Consequently, the M5 brane wrapping this circle is dual to a single
$(p,q)$ 5--brane.

In any case, the very existence of branes dual to $\hat\eta_{1,2}(\xi)$
instead of un-hatted $\eta_{1,2}(\xi)$ indicates that
there is a range of $\xi$ for which $p^2(x)\ll e^{-\ql a S}b(x)$.
By reasons of concavity this range must include at least one
modulus~$\varphi_i$, and consequently
some of the branching points of the spectral curve over the $x$
plane do not form close pairs (\ref{pairs}--\reftail{dformula}).
Therefore, some of the deconstructed 5D gauge couplings deviate form
\eqrange{deconGij}{deconGii}, which means that
the deconstructed theory is no longer in the {\sl ordinary SQCD${}_5$}
Coulomb phase.
Instead, we have a Coulomb phase of an exotic 5D theory such as
the $E_0$ Coulomb phase of the $\tilde D_0$ model of \S6.1, or
perhaps an unphysical phase such as in the $D_0$ model of \S6.2.

In general, distinct Coulomb phases of the deconstructed theory
correspond to distinct patterns of the spectral curve's branching points
$x_1,\ldots,x_{2n_c}\,$.
The duality translates these branching points into specific features
of the brane web:
a rung of the ladder
\be
\psset{unit=1cm,linewidth=3pt}
\begin{pspicture}(-3,-1)(+3,+1)
\psline[linecolor=blue](-2,0)(+2,0)
\psline[linecolor=red](-3,-1)(-2,0)(-2,+1)
\psline[linecolor=red](+2,-1)(+2,0)(+3,+1)
\rput[b](0,+0.1){$\xi=\varphi_i$}
\psline[linewidth=0.5pt](-2,0)(-2,-0.5)
\psline[linewidth=0.5pt,arrowscale=2]{<->}(-2,-0.4)(+1.95,-0.4)
\rput[t](0,-0.5){$\Delta\eta$}
\end{pspicture}
\label{U:Rung}
\ee
corresponds to a close pair
\be
x_{2i-1},x_{2i}\ =\ e^{\ql a\varphi_i}
\times\left(1\,\pm\,e^{-\ql a\Delta\eta}\right);
\label{U:Pair}
\ee
a joint where two sides of the ladder merge into a single brane dual to
$\hat\eta_{1,2}(\xi)$
\be
\psset{unit=1cm,linewidth=3pt}
\begin{pspicture}(-5,-2)(+5,+2)
\rput(-4,0){%
	\psline[linecolor=brown](0,0)(-1,-2)
	\psline[linecolor=red](0,+2)(0,0)(+2,+2)
	\rput[lt](+1.1,-1.1){$\xi_j$}
	\psline[linewidth=0.5pt,arrowscale=2]{->}(+1,-1)(+0.1,-0.1)
	}
\rput(+4,0){%
	\psline[linecolor=brown](0,0)(+1,+2)
	\psline[linecolor=red](0,-2)(0,0)(-2,-2)
	\rput[rb](-1.1,+1.1){$\xi_j$}
	\psline[linewidth=0.5pt,arrowscale=2]{->}(-1,+1)(-0.1,+0.1)
	}
\rput(0,0){\bf or}
\end{pspicture}
\label{U:Joint}
\ee
corresponds to an un-paired branching point at $x=e^{\ql a\xi_j}$;
a joint involving a double brane
\be
\psset{unit=1cm,linewidth=3pt,doublesep=2pt}
\begin{pspicture}(-6,-2)(+6,+2)
\rput(-4,0){%
	\psline[linecolor=brown,doubleline=true](0,+0.05)(0,-2)
	\psline[linecolor=red](-2,+2)(0,0)(+2,+2)
	\rput[b](0,+1.5){$\xi_j$}
	\psline[linewidth=0.5pt,arrowscale=2]{->}(0,+1.4)(0,+0.15)
	}
\rput(+4,0){%
	\psline[linecolor=brown,doubleline=true](0,-0.05)(0,+2)
	\psline[linecolor=red](-2,-2)(0,0)(+2,-2)
	\rput[t](0,-1.5){$\xi_j$}
	\psline[linewidth=0.5pt,arrowscale=2]{->}(0,-1.4)(0,-0.15)
	}
\rput(0,0){\bf or}
\end{pspicture}
\label{U:Joint2}
\ee
corresponds to two branching points at $x=\pm e^{\ql a\xi_j}$;
and finally, a joint hiding collapsed cycles of the web such as
\be
\psset{unit=7mm,linewidth=3pt}
\begin{pspicture}(-11,-3)(+11,+2)
\rput(-8,0){%
	\psline[linecolor=blue](-1,+1)(+3,+1)
	\psline[linecolor=blue](0,0)(+1,0)
	\psline[linecolor=red](-3,+2)(-1,+1)(0,0)(0,-1)
	\psline[linecolor=red](+6,+2)(+3,+1)(+1,0)(0,-1)
	\psline[linecolor=brown](-1,-3)(0,-1)
	}
\psline[doubleline=true]{->}(-2,0)(+2,0)
\rput(+5,0){%
	\psline[linecolor=red](-3,+2)(+0.6,+0.2)(+6,+2)
	\psline[linecolor=brown](-1,-3)(+0.6,+0.2)
	\pscircle*[linecolor=blue](+0.6,+0.2){0.2}
	\rput[lt](+2.8,-2){$\xi_j$}
	\psline[linewidth=0.5pt,arrowscale=2]{->}(+2.7,-1.9)(+0.7,+0.1)
	}
\end{pspicture}
\label{U:Hidden}
\ee
corresponds to $K$ branching points
\be
x_\nu\ =\ e^{\ql a\xi_j}\times e^{2\pi i\nu/K},\quad \nu=1,2,\ldots,K
\label{U:Cluster}
\ee
where
\be
K\ =\ 2\times\#\{\mbox{hidden cycles}\}\ +\,\cases{
	1 & for a single $\hat\eta_{1,2}(\xi)$ brane,\cr
	2 & for a double $\hat\eta_{1,2}(\xi)$ brane.\cr
}\label{U:Kpoints}
\ee
Although only the real parts of  $\Delta\eta$ or $\xi_j$ are visible
in the brane web, this gives enough information to identify the pattern
of branching points --- and hence the phase of the deconstructed theory ---
and even to calculate the matrix of 5D abelian gauge couplings.
Thus, the phases of the deconstructed theory are in perfect correspondence
with the phases of the brane web, and the transition between those phase
happen at exactly the same $h_c$.

This completes part (3) of our proof.

Finally, let us calculate the $C$ parameter in eqs.~(\ref{U:embed}).
This parameter does not depend on the moduli, quark masses, or gauge coupling
of the 5D theory, so let us make all the $\phi_i$ distinct to
break $SU(n_c)\to U(1)^{n_c-1}$ and take the $h\to+\infty$ limit.
This makes the quiver theory weakly coupled and allows semiclassical analysis.
Consequently, the universality between the 4D gauge theory and the M theory
on the M5 spanning the quiver's spectral curve should extend beyond the
purely holomorphic data to other low-energy properties such as
masses of light particles.
In particular, M theory should reproduce the semiclassical mass $M=|\phi_i-\phi_j|$
of a non-abelian gauge boson $A^\mu_{ij}$.

In string theory on the brane web~(\ref{U:Web1}), this vector field
arises from the string connecting appropriate rungs of the ladder,
\be
\psset{unit=1cm,linewidth=3pt,linecolor=blue}
\begin{pspicture}(-4,-2)(+3,+2)
\psline(-3,+1)(+2,+1)
\psline[linestyle=dashed](-3,+1)(-4,+1)
\psline[linestyle=dashed](+2,+1)(+3,+1)
\psline(-3,-1)(+2,-1)
\psline[linestyle=dashed](-3,-1)(-4,-1)
\psline[linestyle=dashed](+2,-1)(+3,-1)
\rput[b](-2,+1.1){$\Re\xi=\phi_j$}
\rput[b](-2,-0.9){$\Re\xi=\phi_i$}
\psline[linestyle=dotted](0,-0.7)(0,+0.7)
\psline[linestyle=dotted](0,+1.3)(0,+2)
\psline[linestyle=dotted](0,-1.3)(0,-2)
\pscurve[linecolor=red,linewidth=1pt]%
	(1.1,+1)(1,+.9)(0.9,+.8)(1,+.7)(1.1,+.6)(1,+0.5)(0.9,+.4)(1,+.3)%
	(1.1,+.2)(1,+.1)(0.9,0)(1,-.1)(1.1,-.2)%
	(1,-.3)(0.9,-.4)(1,-.5)(1.1,-.6)(1,-.7)(0.9,-.8)(1,-.9)(1.1,-1)
\end{pspicture}
\label{U:Joint3}
\ee
In M theory, this string is dual to M2 brane forming a cylinder:
its long dimension is $X^5=C\times\Re\xi$ and the circular dimension
is $X^{10}=C\times\Im\eta$.
The mass of this tube is
\be
M\ =\ t_3\times\mbox{Area}\
=\ t_3\times C|\phi_i-\phi_j|\times {2\pi C\over \ql a}
\ee
where $t_3$ is the membrane tension;
equating this mass to the field-theoretical mass $M=|\phi_i-\phi_j|$
of the vector field we arrive at
\be
C\ =\ \sqrt{\ql a\over 2\pi t_3}\,.
\label{U:Cvalue}
\ee

Given this value of $C$, \eq{U:Radius} tells us that
the string theory dual to M theory
is compactified on the circle $S^1$ of radius
\be
R\ =\ {1\over t_3}\times\left({\ql a\over 2\pi C}\right)^2\
=\ {\ql a\over 2\pi}
\label{U:UltimateRadius}
\ee
--- which is precisely the radius of the deconstructed dimension
for finite~$\ql$.
And this completes the last part (4) of our proof.

To summarize, we have established that {\it
dimensional deconstruction and brane-web engineering
of the same \sqcdv\ are always in the same universality class}.
%and have exactly similar phase diagrams.}
And of course, other string/M implementations of \sqcdv\
are also in the same universality class because they are
dual to the brane-web engineering.
Which means that {\it\blue the phase diagram and other zero-energy features
of a 5D gauge theory are inherent properties of the theory itself,
regardless of its UV completion}.

\medskip
\centerline{\large\blue $\star\qquad\star\qquad\star$}
\medskip

And this completes our analysis of quantum deconstruction.
We showed how  to use dimensional deconstruction
as a UV completion of a 5D SUSY gauge theory such as \sqcdv.
We showed how to extract 5D quantum effects such as loop corrections to
the prepotential from the 4D loop and instantonic effects --- which can
be calculated exactly thanks to the unbroken $\NN=1$ SUSY in 4D.
We showed how to deconstruct the 5D phase diagrams, including the
non-classical $h<0$ phases.
And at the end of the story, the dimensional deconstruction turned
out to be in the same universality class as the string-theoretical
UV completions of the same 5D theory.

\smallskip
\noindent
{\bf ACKNOWLEDGEMENTS:}
The authors thank Amer Iqbal and Jacob Sonnenschein for many interesting
discussions.
V.~K. thanks the HEP theory group at Tel Aviv University for hospitality
during his many visits there.

%We apologize for stylistic differences between sections of this paper.
%We were writing it for a very long time, and our writing styles have
%evolved.

The research of V.~K. is supported in part by the US National Science
Foundation under grant  PHY--0455649.
The research of E.~dN. is supported by the US Department of Energy
under grant DE--FG02--06ER41418.

%% file: biblio.tex
%auto-ignore
%
% References for the DESQCD paper
%

%% file: desqcd.bbl
\begin{thebibliography}{99}

\bibitem{seiberg}
Nathan Seiberg, {\it``Five Dimensional SUSY Field Theories, Non--Trivial
Fixed Points, and String Dynamics'', \sl Phys. Lett. \bf B 388} (1996)
753 {\tt arXiv:hep-th/9608111}.

\bibitem{MS}
David~R.\ Morrison and Nathan Seiberg, {\it``Extremal Transitions
and Five--Dimensional Supersymmetric Field Theories'',
\sl Nucl. Phys. \bf B 483} (1997) 229, {\tt arXiv:hep-th/9609070}.

\bibitem{CCAF}
A.~C. Cadavid, A.~Ceresole, R.~D'Auria, and S.~Ferrara,
{\it``11-Dimensional Supergravity Compactified on Calabi-Yau
Threefolds'', \sl Phys. Lett. \bf B357} (1995) 76, {\tt arXiv:hep-th/9506144}

\bibitem{FMS}
Sergio Ferrara, Ruben Minasian, and Augusto Sagnotti,
{\it``Low--Energy Analysis of $M$ and $F$ Theories on Calabi--Yau Threefolds'',
\sl Nucl. Phys. \bf  B 474} (1996) 323, {\tt arXiv:hep-th/9604097}  

\bibitem{IMS}
K.~Intriligator, D.~R.\ Morrison and N.~Seiberg,
{\it``Five--Dimensional Supersymmetric Gauge Theories
and Degeneration of Calabi--Yau Spaces'',
\sl Nucl. Phys. \bf B 497} (1997) 56, {\tt arXiv:hep-th/9702198}.

\bibitem{FKM}
Sergio Ferrara, Ramzi~R.\ Khuri, and Ruben Minasian,
{\it``M theory on a Calabi--Yau manifold'',
\sl Phys. Lett. \bf B 375} (1996) 81, {\tt  arXiv:hep-th/9602102}.

\bibitem{GMS}
Ori~J.\ Ganor, David~R.\ Morrison and Nathan Seiberg,
{\it``Branes, Calabi--Yau Spaces, and Toroidal Compactifications of the
$\NN=1$ Six--Dimensional $E_8$ Theory'', \sl Nucl. Phys. \bf B 487}
(1997) 93, {\tt arXiv:hep-th/9610251}.

\bibitem{AH}
Ofer Aharony and Amihay Hanany,
{\it``Branes, Superpotentials, and Superconformal Fixed Points'',
\sl Nucl. Phys. \bf B 504} (1997) 239, {\tt arXiv:hep-th/9704170}. 

\bibitem{AHK}
Ofer Aharony, Amihay Hanany, and Barak Kol,
{\it``Webs of $(p,q)$ 5--Branes, Five Dimensional Field Theories,
and Grid Diagrams'', \sl JHEP \bf 9801} (1998) 002,
{\tt arXiv:hep-th/9710116}.

\bibitem{DHIK}
Oliver DeWolfe, Amihay Hanany, Amer Iqbal, Emanuel Katz,
{\it``Five-branes, Seven-branes, and Five-dimensional $E_n$ Field Theories'',
\sl JHEP \bf 9903} (1999) 006, {\tt arXiv:hep-th/9902179}.

\bibitem{LV}
N.~C.\ Leung and C.~Vafa, {\it``Branes and Toric Geometry'',
\sl Adv. Theor. Math. Phys. \bf 2} (1998) 91, {\tt arXiv:hep-th/9711013}.

\bibitem{KR2}
Barak Kol and J.~Rahmfeld,
{\it``BPS Spectrum of 5 Dimensional Field Theories,
$(p,q)$ Webs, and Curve Counting'',
\sl JHEP \bf 9808} (1998) 006, {\tt arXiv:hep-th/9801067}.

\bibitem{DKV}
M.~R.\ Douglas, S.~Katz and C.~Vafa,
{\it``Small Instantons, del Pezzo Surfaces, and Type~$I'$ Theory'',
\sl Nucl. Phys. \bf B 497} (1997) 155, {\tt arXiv:hep-th/9609071}.

\bibitem{PhasesW}
Edward Witten, {\it``Phase Transitions in M--Theory and F--Theory'',
\sl Nucl. Phys. \bf B 471} (1996) 195, {\tt arXiv:hep-th/9603150}.

\bibitem{ACG}
Nima Arkani--Hamed, Andrew~G.\ Cohen, and Howard Georgi,
{\it``(De)Constructing Dimensions'',
\sl Phys. Rev. Lett. \bf 86} (2001) 4757, {\tt arXiv:hep-th/0104005}.

\bibitem{HPW}
Christopher~T.\ Hill, Stefan Pokorski, and Jing Wang,
{\it``Gauge Invariant Effective Lagrangian for Kaluza--Klein Modes'',
\it Phys. Rev. \bf D 64} (2001), 105005, {\tt arXiv:hep-th/0104035}.

\bibitem{ACG1}
Nima Arkani--Hamed, Andrew~G.\ Cohen, and Howard Georgi,
{\it``Electroweak Symmetry Breaking from Dimensional Deconstruction'',
\sl Phys. Lett. \bf B 513} (2001) 232,  {\tt arXiv:hep-ph/0105239}

\bibitem{CTW}
Hsin--Chia Cheng, Christopher~T.\ Hill, and Jing Wang,
{\it``Dynamical Electroweak Breaking and Latticized Extra Dimensions'',
\sl Phys. Rev. \bf D 64} (2001) 095003, {\tt arXiv:hep-ph/0105323}

\bibitem{csaki}
C.~Csaki, J.~Erlich, C.~Grojean, and G.~Kribs,
{\it``4D Constructions of Supersymmetric Extra
Dimensions and Gaugino Mediation,'' \sl Phys. Rev. \bf D 65} (2002) 015003,
{\tt arXiv:hep-ph/0106044}.

\bibitem{IK1}
Amer Iqbal and Vadim~S.\ Kaplunovsky,
{\it ``Quantum Deconstruction of a 5D SYM and its Moduli Space,''
\sl JHEP \bf 0405 \rm (2004) 013,  \tt arXiv:hep-th/0212098}.\nextline
\S3.2 of this paper --- deconstructing Chern--Simons --- was done in
collaboration with Edoardo Di~Napoli.

\bibitem{seiberg1}
N. Seiberg, {\it ``Naturalness versus Supersymmetric Non--Renormalization
Theorems'', \sl Phys. Lett. \bf B 318} (1993) 469, {\tt arXiv:hep-ph/9309335}

\bibitem{APS}
P.~C.\ Argyres, M.~R.\ Plesser, and  N.~Seiberg,
{\it``The Moduli Space of Vacua of $\NN=2$ SUSY QCD
and Duality in $\NN=1$ SUSY QCD'',
\sl Nucl. Phys. \bf B 471} (1996) 159, {\tt arXiv:hep-th/9603042}.

\bibitem{csaki2}
C.~Csaki, J.~Erlich, V.~V.\ Khoze, E.~Poppitz, Y.~Shadmi, and Y.~Shirman,
{\it``Exact Results in 5D from Instantons and Deconstruction,''
\sl Phys. Rev. \bf D 65} (2002) 085033, {\tt arXiv:hep-th/0110188}.

\bibitem{paper1}
Csaba Csaki, Joshua Erlich, Daniel Freedman, and Witold Skiba,
{\it``$\NN=1$ Supersymmetric Product Group Theories in the Coulomb Phase'',
\sl Phys. Rev. \bf D 56} (1997) 5209, {\tt arXiv:hep-th/9704067}.

\bibitem{KSdN}
Edoardo Di~Napoli, Vadim~S.\ Kaplunovsky, and Jacob Sonnenschein,
{\it ``Chiral Rings of Deconstructive $[SU(n_c)]^N$ Quivers,''
\sl JHEP \bf 0406 \rm (2004) 060, \tt arXiv:hep-th/0406122}.

\bibitem{witten97}
Edward Witten, {\it ``Solutions Of Four-Dimensional Field Theories Via 
M Theory'', \sl Nucl. Phys. \bf B 500} (1997) 3, {\tt arXiv:hep-th/9703166}.

\bibitem{HSZ}
Amihay Hanany, Mathew~J.\ Strassler, and Alberto Zaffaroni,
{\it``Confinement and Strings in MQCD'',
\sl Nucl. Phys. \bf B 513} (1998) 87, {\tt arXiv:hep-th/9707244}

\bibitem{BIKSY}
A. Brandhuber, N. Itzhaki, V. Kaplunovsky,J. Sonnenschein, and S. Yankielowicz,
{\it``Comments on the M Theory Approach to $\NN=1$ SQCD and Brane Dynamics'',
\sl Phys. Lett. \bf B 41} (1997) 27, {\tt arXiv:hep-th/9706127}

\bibitem{DM}
Michael~R.\ Douglas and Gregory Moore,
{\it\\D-branes, Quivers, and ALE Instantons''}, {\tt arXiv:hep-th/9603167}

\bibitem{AHCKKM}
Nima Arkani--Hamed, Andrew~G.\ Cohen, David~B.\ Kaplan,
Andreas Karch, and Lubo\v{s} Motl,
{\it``Deconstructing $(2,0)$ and Little String Theories'',
\sl JHEP \bf 0301} (2003) 083, {\tt arXiv:hep-th/0110146}.

\bibitem{GP}
Eric~G.\ Gimon and Joseph Polchinski,
{\it``Consistency Conditions for Orientifolds and D--Manifolds,''
\sl  Phys.\ Rev. \bf D 54 \rm (1996) 1667 \tt  arXiv:hep-th/9601038}.

\bibitem{IRU}
L.~E. Ibanez, R.~Rabadan, and A.~M.\ Uranga,
{\it``Anomalous $U(1)$'s in Type I and Type IIB $D=4$, $\NN=1$ String Vacua,''
\sl Nucl.\ Phys. \bf B542 \rm (1999) 112, \tt arXiv:hep-th/9808139}.

\bibitem{SkibaSmith}
Witold Skiba and David Smith,
{\it``Localized Fermions and Anomaly Inflow via Deconstruction,''
\sl Phys. Lett. \bf D65} (2002) 095, {\tt arXiv:hep-ph/0201056}.

\bibitem{GMSW}
Jerome~P.\ Gauntlett, Dario Martelli, James Sparks, and Daniel Waldram,
{\it``Supersymmetric $\rm AdS_5$ Solutions of M--Theory'',
Class. Quant. Grav. \bf 21} (2004) 4335, {\tt arXiv:hep-th/0402153};
{\it``Sasaki--Einstein Metrics on $S^2\times S^3$'',
\sl Adv. Theor. Math. Phys. \bf 8} (2004) 711, {\tt arXiv:hep-th/0403002}.

\bibitem{HKW}
Amihay Hanany, Pavlos Kazakopoulos, Brian Wecht,
{\it``A New Infinite Class of Quiver Gauge Theories'',
\sl JHEP \bf 0508} (2005) 054, {\tt arXiv:hep-th/0503177}.

\bibitem{witten95}
Edward Witten, {\it ``String Theory Dynamics in Various Dimensions'',
\sl Nucl. Phys. \bf B 443} (1995) 85, {\tt arXiv:hep-th/9503124}


\end{thebibliography}
